\documentclass[twocolumn]{aastex62}

\newcommand{\del}{\mathbf{\nabla}}

\published{2020 March 16}

\submitjournal{ApJ}

\shorttitle{The Role of the Parker Instability}
\shortauthors{Heintz, Bustard and Zweibel}

\usepackage{amsmath}
\usepackage{natbib}
\usepackage{xcolor}
\usepackage{bm}

\begin{document}

\title{The Role of the Parker Instability in Structuring the Interstellar Medium}

\correspondingauthor{Evan Heintz}
\email{eheintz@wisc.edu, bustard@wisc.edu, zweibel@astro.wisc.edu}

\author{Evan Heintz}
\affil{Department of Physics, University of Wisconsin - Madison, 475 North Charter Street, Madison, WI 53706, USA}

\author{Chad Bustard}
\affil{Department of Physics, University of Wisconsin - Madison, 475 North Charter Street, Madison, WI 53706, USA}

\author{Ellen G. Zweibel}
\affiliation{Department of Physics, University of Wisconsin - Madison, 475 North Charter Street, Madison, WI 53706, USA}
\affiliation{Department of Astronomy, University of Wisconsin - Madison, 475 North Charter Street, Madison, WI 53706, USA}

\begin{abstract}
The Parker instability, a Rayleigh-Taylor like instability of thermal gas supported against gravity by magnetic fields and cosmic rays, is thought to be dynamically important for galaxy evolution, possibly promoting molecular cloud formation and the galactic dynamo. In previous work, we examined the effect of three different cosmic ray transport models on the Parker instability: decoupled ($\gamma_c = 0$), locked to the thermal gas ($\gamma_c = 4/3$) and coupled to the gas with streaming by self-confinement. We expand upon that work here by considering radiative cooling, a smooth gravitational potential, and simulations into the nonlinear regime. We determine that cosmic ray transport away from compression points, whether by diffusion or streaming, is the largest driver of the instability. Heating due to cosmic ray streaming is also destabilizing and especially affects the nonlinear regime. While cooling de-pressurizes the dense gas, streaming cosmic rays heat and inflate the diffuse extraplanar gas, greatly modifying the phase structure of the medium. In 3D, we find that the fastest growth favors short wavelength modes in the horizontal direction perpendicular to the background magnetic field; this is imprinted on Faraday rotation measure maps that may be used to detect the Parker instability. The modifications to the Parker instability that we observe in this work have large implications for the structure and evolution of galaxies, and they highlight the major role that cosmic rays play in shaping their environments. 
\end{abstract}

\keywords{cosmic rays, gravitational instability, interstellar medium, interstellar plasma, interstellar dynamics, interstellar magnetic fields} 

\section{Introduction}
\label{sec:intro}

The Parker instability \citep{parkerinstability1966} is a key example of the effect that magnetic fields and cosmic rays can have on the structure of the interstellar medium (ISM). In the Parker instability, a perturbation to the magnetic field causes the field lines to bend, allowing the gas to fall down into the valleys of the magnetic field due to the force of gravity. This releases gravitational potential energy, some of which is used to compress the gas into the valleys and resist magnetic tension. The instability, then, is a constant battle between gravity, compression, and magnetic tension. While Parker originally employed the instability as a mechanism by which molecular clouds could form, a concept further investigated by \cite{KosinskiParkerCooling2006} \& \cite{MouschoviasParkerCooling2009}, many additional astrophysical phenomena have been pinned on the Parker instability, including disk stability \citep{KimParkerDiskStability2000, RodriguesParker2015, HeintzParker2018} and the magnetic dynamo \citep{HanaszParkerDynamo1997, MachidaParkerDynamo2013}. 

The stability of stratified media was first analyzed by \cite{newcombinstability1961} in a system with a horizontal magnetic field but without cosmic rays. In the limit of infinitely long parallel wavelength and infinitely short perpendicular wavelength, the instability criterion reduces to the Schwarzschild criterion:
\begin{equation}
    -\frac{g}{\rho}\frac{d\rho}{dz} - \frac{\rho g^2}{\gamma_g P_g} > 0
\end{equation}
where $g$ is the gravitational acceleration, $\rho$ is the thermal gas density, $\gamma_g$ is the adiabatic index of the gas and $P_g$ is the thermal gas pressure. While the magnetic field does not explicitly appear in that criterion, it affects the stability of the system through its effect on the stratification.

\cite{parkerinstability1966} adapted 
this system to galactic disks with the addition of cosmic rays. These highly energetic particles comprise about a third of the total energy density in interstellar gas and can help drive galactic winds \citep{breitschwerdtwinds1991, everettwinds2008, girichidiswinds2016, ruszkowskiwinds2017, WienerCRTransport2017}, ionize the interstellar medium \citep{GrenierReview2015}, and contribute to the formation of the Fermi Bubbles \citep{guofermi2012,yangfermi2012} and the galactic dynamo \citep{HanaszDynamo2009}. 

In Parker's original analysis, cosmic rays were assumed to provide equilibrium pressure support but behave as a $\gamma=0$ fluid when perturbed, thereby significantly destabilizing the medium.

Since that seminal paper, our understanding of cosmic ray coupling to the thermal gas has significiantly expanded. For GeV cosmic rays, which make up the bulk of the cosmic ray pressure, cosmic ray streaming, described by the self-confinement model, is the dominant transport mode \citep{KulsrudSelfConfine1969, KulsrudSelfConfine1971, zweibelreview2017}. In this model, cosmic rays that are in gyro-resonance with Alfv\'{e}n waves can exchange energy and momentum with these waves. If their bulk speed is larger than the Alfv\'{e}n speed, $v_A$, the cosmic rays will scatter off the waves until they reach marginal stability, namely isotropy in the wave frame, in the process amplifying the waves \citep{KulsrudSelfConfine1969}. In a steady state, the waves will then transfer the energy to the surrounding thermal gas in the form of heating, and the bulk flow of cosmic rays proceeds at the local Alfven speed down the cosmic ray pressure gradient directed along the magnetic field. 

The extrinsic turbulence model is an alternative model of cosmic ray transport, according to which the cosmic rays scatter off waves generated by a turbulent cascade. While they still interact resonantly with the turbulence, their gains and losses cancel due to the waves propagating in both directions with equal intensity. Therefore, in this model, there is no cosmic ray heating of the gas \citep{zweibelreview2017}. This model effectively reduces to cosmic ray advection with the thermal gas. Both the self confinement and extrinsic turbulence models admit magnetic field aligned diffusion, at a rate which depends on the amplitude of the scattering waves.

With these modern cosmic ray transport theories in hand, \cite{HeintzParker2018} (hereafter HZ18) revisited the Parker instability. The results of our linear stability analysis, which we outline in \S\ref{subsec:pastresults}, show significant dependence on the transport model. For streaming, in particular, we found that the range of unstable wavelengths is greatly expanded and the growth rate is significantly increased. These changes to the canonical Parker instability picture promise wide-ranging implications for the structure the ISM; therefore, we are motivated to explore these effects further with an expanded linear stability analysis and numerical simulations. 




In this paper, we analyze the Parker instability under three different cosmic ray transport models, using both a linear stability analysis and magnetohydrodynamic (MHD) simulations. The paper is outlined as follows. In \S\ref{subsec:pastresults}, we summarize the results of \cite{HeintzParker2018}. We then discuss the advances we have made to that linear theory through the addition of radiative cooling in \S\ref{subsec:cooling}. The remaining bulk of the paper centers around numerical simulations of the Parker instability, focusing on both the linear and nonlinear development of the instability with different cosmic ray transport models. We outline the methods for these simulations in \S\ref{sec:sim_method}, as well as cover the addition of a smooth gravitational potential to the system in \S\ref{subsec:gravity}. We then proceed to discuss our results from 2D simulations in \S\ref{subsec:evol_nonlin}, focusing on the effect of cosmic ray heating in both the linear and nonlinear regimes of the instability. We then investigate the effect of cosmic ray heating on the system in \S\S \ref{subsec:diff_vs_streaming} and \ref{subsec:heatingUnstable}, including the competing role of radiative cooling in \S\ref{sec:coolingSection}.  We perform the same analysis for our 3D simulations in \S\ref{sec:nonlinear3D} and use those results to make some mock observations of the Parker instability. We then summarize the main results, conclusions, and implications of our work in \S\ref{sec:summary}.

\section{Linear Theory}\label{sec:Linear}
\subsection{Basic Equations and Summary of Past Results} \label{subsec:pastresults}
Here, we outline the linear stability analysis and main results from HZ18. For completeness, our equations include the additional terms for radiative cooling. These are explained in \S\ref{subsec:cooling} and derived in more detail in the Appendix. Following our setup from HZ18, we assume a 2D stratified system in magnetohydrostatic equilibrium where the cosmic ray pressure ($P_c$), gas pressure ($P_g$), density ($\rho$), and magnetic field ($\mathbf{B} = B \hat{x}$)
are all functions of $y$. We also assume as \cite{parkerinstability1966} did that the gravitational acceleration ($\mathbf{g} = -g \hat{y}$) is constant, and that the ratios of magnetic pressure and cosmic ray pressure to thermal pressure are constant. Under these conditions, $\rho$, $P_g$, $P_c$, and $P_m\equiv B^2/(8\pi)$ all depend on $y$
as $e^{-y/H}$, where
$H\equiv(P_g+P_c+P_m)/\rho_gg$ is the (constant) scale height.

The equations for the background and perturbed (denoted by $\delta$) quantities are (HZ18):
\begin{gather}
\label{eq:contin}
\frac{\partial\delta\rho}{\partial t} = -\mathbf{\delta u} \cdot \del\rho - \rho\del \cdot \mathbf{\delta u} 
\end{gather}
\begin{gather}
\label{eq:momconsv}
\rho\frac{\partial\mathbf{\delta u}}{\partial t} = -\del(\delta P_c + \delta P_g) + \frac{\mathbf{J}\times\delta\mathbf{B}}{c} + \frac{\delta\mathbf{J} \times \mathbf{B}}{c} + \delta\rho\mathbf{g} 
\\
\frac{\partial\delta\mathbf{B}}{\partial t} = \del \times (\delta\mathbf{u} \times \mathbf{B}) 
\end{gather}
\begin{equation}
\begin{split} \label{eq:crenergy}
\frac{\partial\delta P_c}{\partial t} + \mathbf{u}_A\cdot\del\delta P_c = -(\mathbf{\delta u} + \mathbf{\delta u}_A)\cdot\del P_c \\ 
-\gamma_c P_c \del \cdot (\mathbf{\delta u} + \mathbf{\delta u}_A) + \nabla \cdot (\Bar{\Bar{\kappa}}\cdot\nabla\delta P_c)
\end{split}
\end{equation}
\begin{gather}
\begin{split} \label{eq:thermenergy}
\Big(\frac{\partial}{\partial t} + (\gamma_g - 1)\frac{n^2T}{P_g}\frac{d\Lambda(T)}{dT}\Big)\delta P_g &= - \mathbf{\delta u}\cdot\mathbf{\del}P_g \\ 
- \gamma_g P_g \mathbf{\del}\cdot\mathbf{\delta u} -\Big(\gamma_g - 1\Big)\Big(\frac{n^2 T}{\rho} \frac{d\Lambda(T)}{dT} &- \frac{n^2\Lambda(T)}{\rho}\Big)\delta\rho \\
  - (\gamma_g - 1)(\mathbf{u}_A \cdot \del\delta P_c + \mathbf{\delta u}_A &\cdot \del P_c)
\end{split}
\end{gather}
where $T$ is the temperature, $n$ is the gas density, $\Lambda(T)$ is the optically thin cooling function, $\kappa$ is the diffusion tensor ($\Lambda$ and $\kappa$ were omitted from HZ18), $\mathbf{u}$ is the velocity, $\mathbf{u_A}$ is the Alfv\'{e}n velocity, $\mathbf{B}/\sqrt{4\pi\rho_i}$, and $\gamma_g$ and $\gamma_c$ are the adiabatic indices of the gas and cosmic rays respectively. We have written $\rho_i$ to denote the plasma density, which differs from the total thermal gas density $\rho$ in regions that are weakly ionized; we will ignore ionization effects throughout. These five equations, in order, describe mass continuity, momentum conservation, magnetic induction, and the energy equations for the cosmic rays and thermal gas. All terms associated with the self-confinement model appear in the cosmic ray and thermal gas equations (Equations \ref{eq:crenergy} and \ref{eq:thermenergy}). 

Following the same procedure as in HZ18, we assume a Fourier decomposition for the perturbation quantities in terms of $x$ and $t$; $\delta \propto e^{ik_x x - i\omega t}$, and then assume the background quantities and perturbations depend on $y$ as:
\begin{equation}
\begin{split}
    \rho, P_c, P_g&, B^2 \propto e^{-y/H} \\
    \delta u \propto e^{(ik_y y+\frac{y}{2H})}, \hspace{2pt} \delta P &\propto  e^{(ik_y y-\frac{y}{2H})}, \hspace{2pt} \delta B \propto e^{ik_y y}
\end{split}
\end{equation}
Again, following HZ18, we introduce the dimensionless variables:
\begin{equation}
\label{eq:dimensionless}
    \begin{split}
        H = \frac{H_0}{q}&, \qquad q = \frac{\gamma_g}{1+\frac{\gamma_g}{2}m^2 + \frac{\gamma_g}{\gamma_c}c^2}, \\
        \hat{\omega} = \frac{\omega H_0}{a_g}&, \qquad \hat{k} = kH_0,\\
        \hat{\delta\rho} = \frac{\delta\rho}{\rho}&, \qquad \hat{\delta B} = \frac{\delta B}{B}, \qquad \hat{\delta u} = \frac{\delta u}{a_g}\\
        \hat{\delta P_c} = \frac{\delta P_c}{\gamma_c P_c}&, \qquad \hspace{4pt} \hat{\delta P_g} = \frac{\delta P_g}{\gamma_g P_g}, \\
        \hat{\Lambda}(T) &= \frac{n^2 H_0}{\rho a_g^3}\Lambda(T) \\
        H_0 \equiv \frac{a_g^2}{g_0}&, \qquad a_{g,c} \equiv \sqrt{\frac{\gamma_{g,c} P_{g,c}}{\rho}}, \\
        m \equiv \frac{u_A}{a_g}&, \qquad c \equiv \frac{a_c}{a_g}
    \end{split}
\end{equation}
where $u_A$ is the Alfv\'{e}n speed, $a_g\equiv\sqrt{\gamma_gP_g/\rho}$ is the thermal gas sound speed, $a_c\equiv\sqrt{\gamma_cP_c/\rho}$ is the cosmic ray sound speed, $H_0$ would be the scale height if there were no pressure support from magnetic fields or cosmic rays, $H$ is the actual scale height, and $g_0$ is the gravitational acceleration constant. The m and c values used here relate to $\alpha$ and $\beta$ used in \cite{parkerinstability1966}:
\begin{equation}
    \begin{split}
        \alpha = \frac{\gamma_{g} m^{2}}{2}, \qquad \beta = \frac{\gamma_{g}}{\gamma_{c}} c^{2} 
    \end{split}
\end{equation}

Plugging all of these substitutions into Eqs. \ref{eq:contin} - \ref{eq:thermenergy}, we get our linearized perturbed equations:
\begin{gather}
\label{eq:parkstreamcont}
i\hat{\omega}\hat{\delta\rho}-i\hat{k_x}\hat{\delta u_x}-(i\hat{k_y}-\frac{q}{2})\hat{\delta u_y} = 0 \\
i\hat{\omega}\hat{\delta u_x}-i\hat{k_x}(c^2\hat{\delta P_c}+\hat{\delta P_g})-\frac{qm^2}{2}\hat{\delta B_y} = 0 \\
\begin{split}
i\hat{\omega}\hat{\delta u_y}-(i\hat{k_y}-\frac{q}{2})(c^2\hat{\delta P_c}+\hat{\delta P_g}+m^2\hat{\delta B_x}) \\
+im^2\hat{k_x}\hat{\delta B_y}-\hat{\delta\rho} = 0
\end{split} \\
\hat{\omega}\hat{\delta B_x}-\hat{k_y}\hat{\delta u_y} = 0 \\
\hat{\omega}\hat{\delta B_y}+\hat{k_x}\hat{\delta u_y} = 0 \\
\begin{split}
(i\hat{\omega}-i\hat{k_x}m)\hat{\delta P_c}+\frac{i\hat{k_x}m}{2}\hat{\delta\rho}-i\hat{k_x}(\hat{\delta u_x}+m\hat{\delta B_x}) \\
+(\frac{q}{\gamma_c}-i\hat{k_y}-\frac{q}{2})(\hat{\delta u_x}+m\hat{\delta B_y}) = 0 
\end{split}
\end{gather}
\begin{gather}
\begin{split}
\label{eq:parkstreamgas}
\Big(i\hat{\omega} - (\gamma_g - 1)T\frac{d\hat{\Lambda}(T)}{dT}\Big)\hat{\delta P_g} &+(\frac{q}{\gamma_g}-i\hat{k_y}-\frac{q}{2})\hat{\delta u_y} \\
 - i\hat{k_x}(\hat{\delta u_x} + qmc^2(\gamma_g-1)\hat{\delta P_c}) 
 &+ \frac{qmc^2}{\gamma_c}(\gamma_g-1)\hat{\delta B_y} \\
 + \Big(\gamma_g - 1\Big)\Big(T\frac{d\hat{\Lambda}(T)}{dT} &- \hat{\Lambda}(T)\Big)\hat{\delta\rho} = 0
\end{split}
\end{gather}

\begin{figure*}[]
\centering
\includegraphics[width=0.3\textwidth]{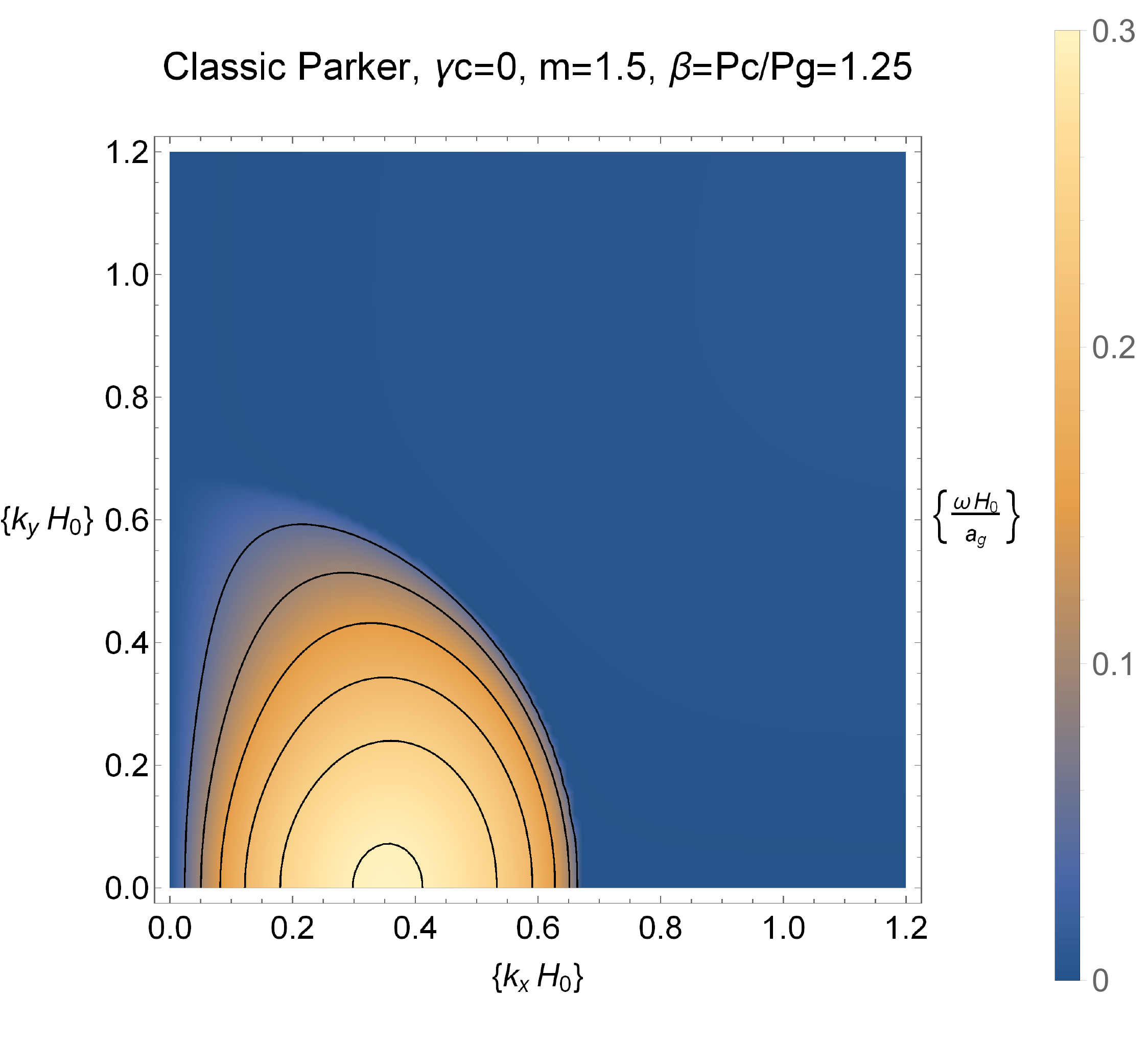}{(a)}
\includegraphics[width=0.3\textwidth]{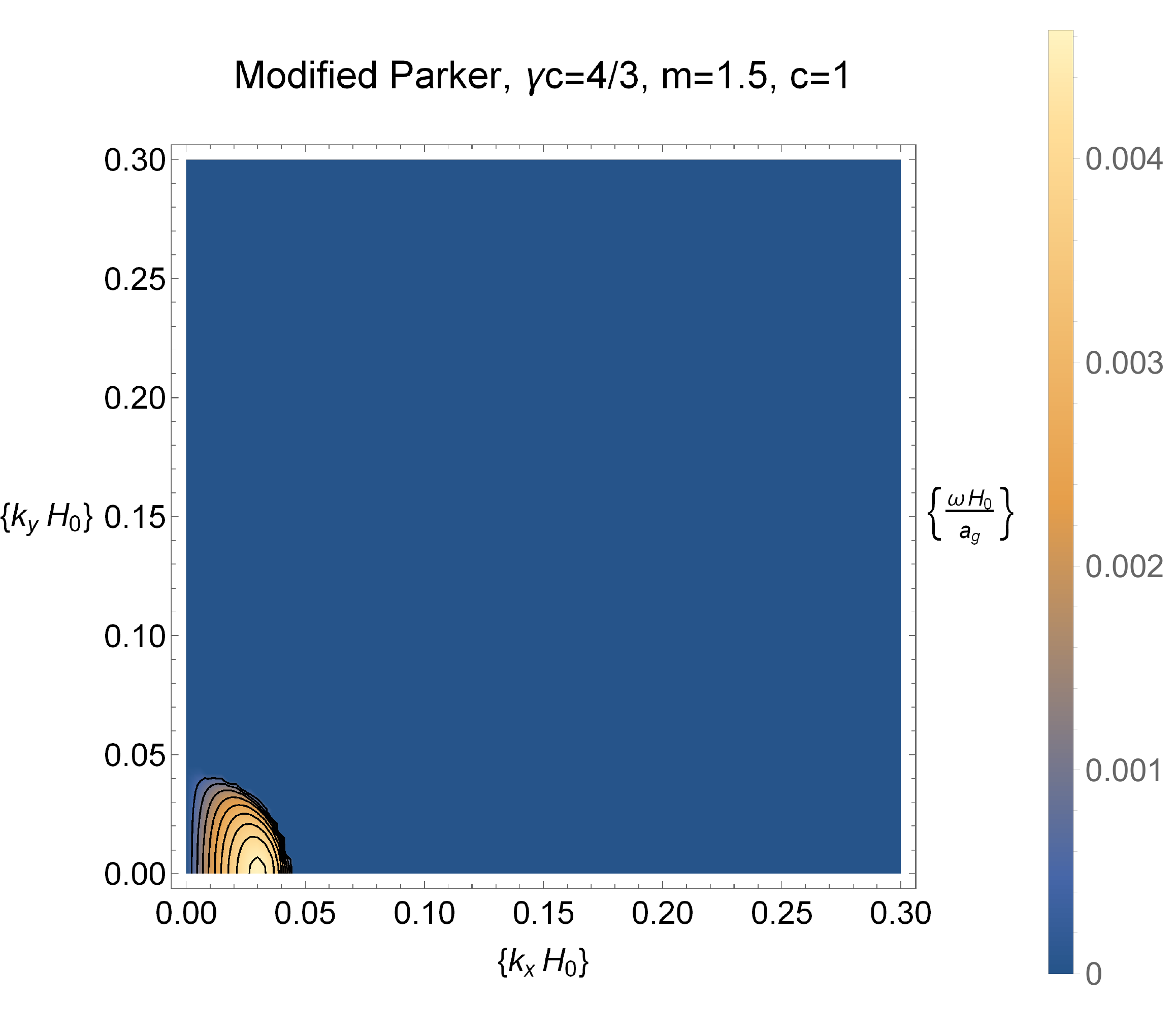}{(b)}
\includegraphics[width=0.3\textwidth]{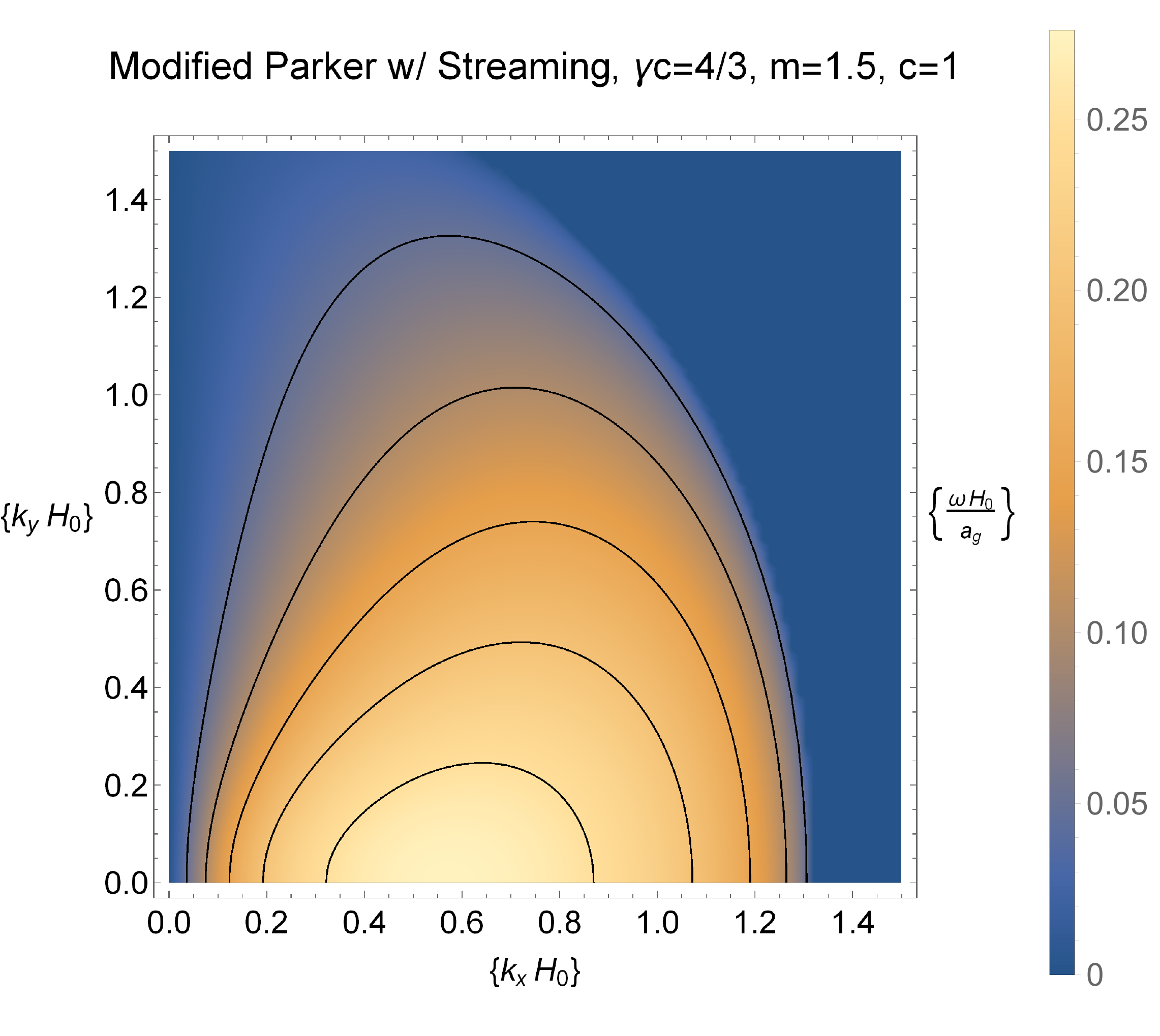}{(c)}
\includegraphics[width=0.3\textwidth]{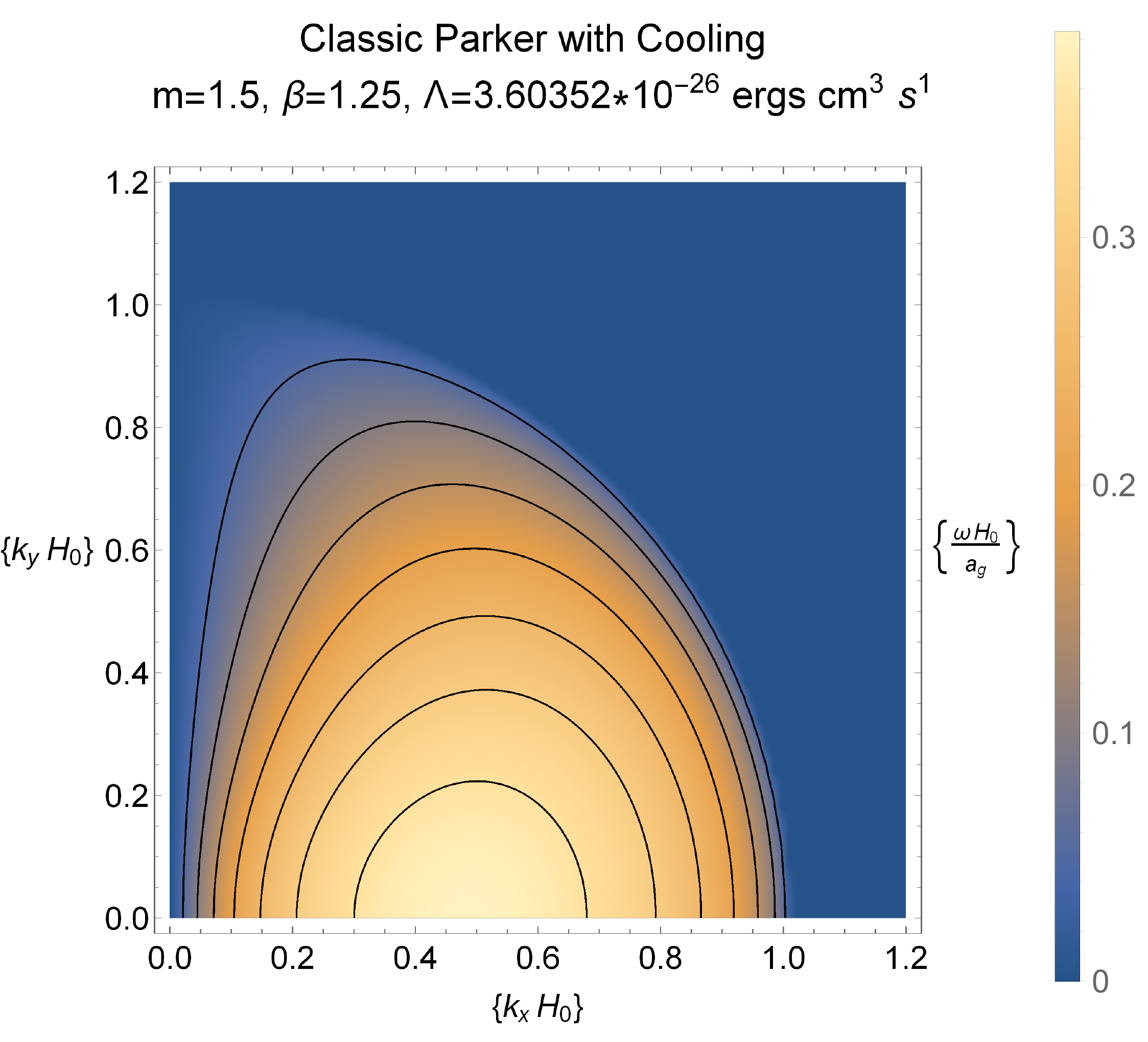}{(d)}
\includegraphics[width=0.3\textwidth]{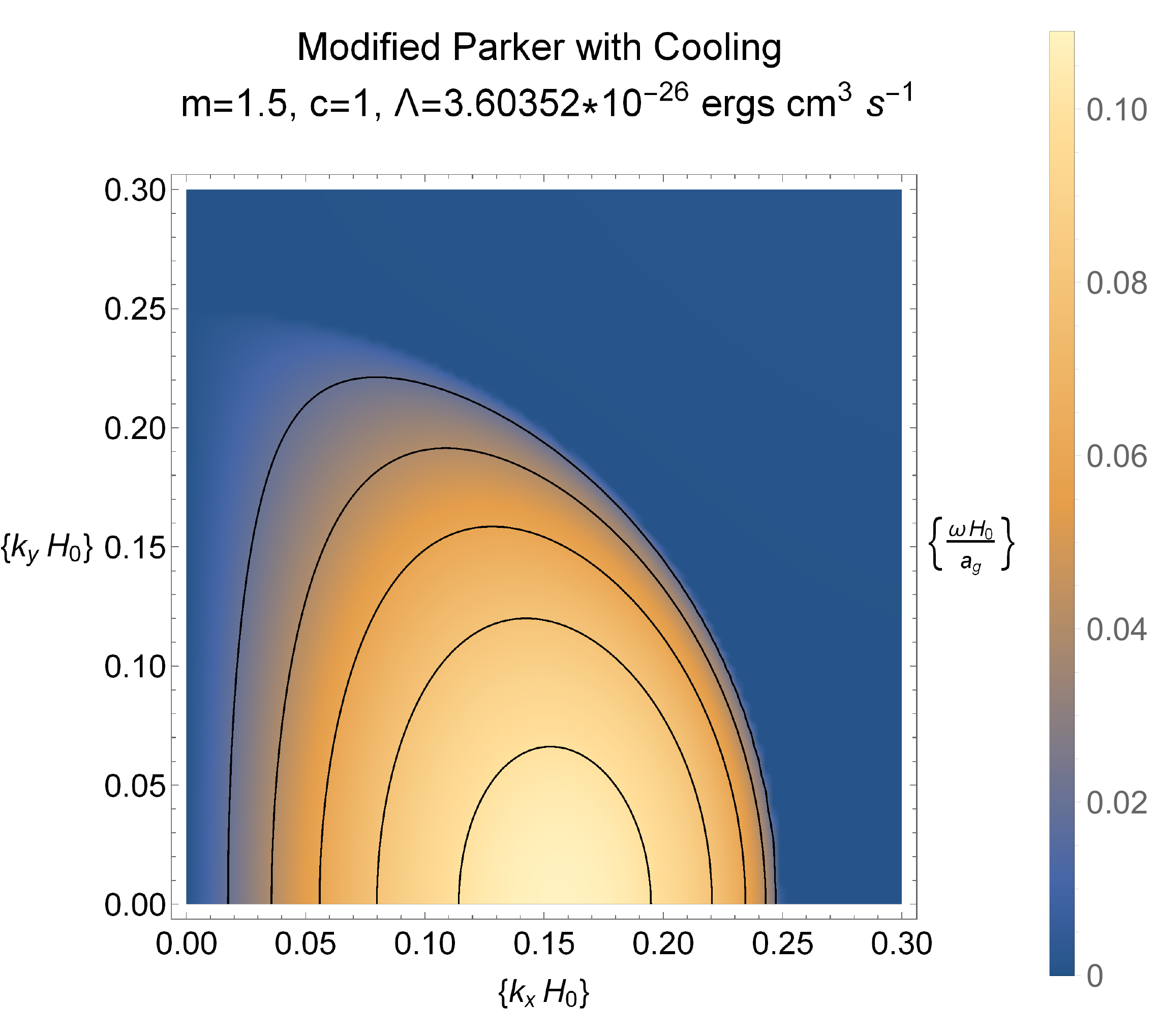}{(e)}
\includegraphics[width=0.3\textwidth]{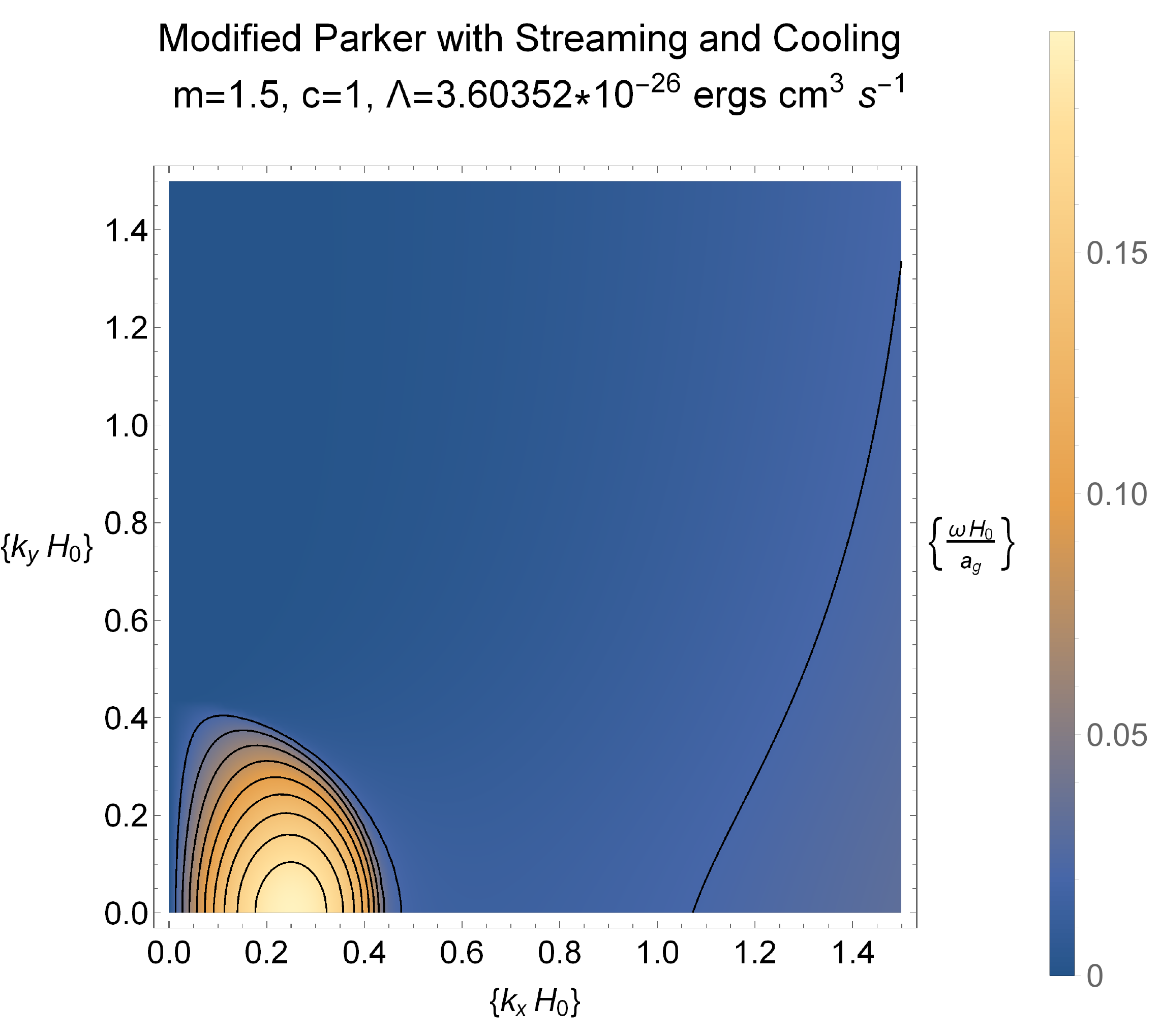}{(f)}
\caption{\emph{Top: no cooling, Bottom: with cooling}. The contour plots of growth rates for the three Parker cases we investigated in HZ18. Note the different domains of ($k_x H_0$) and ($k_y H_0$) in the graphs as well as the different ranges for ($\omega H_0/ a_g$) in the bar legend. In all three cases, $\gamma_g=5/3$. While any dependence on temperature is excluded for the non-cooling case, for the ones with cooling, we do have a temperature dependence and assume here that T=8000K. This gives H $\approx$ 194 pc. We see without streaming that cooling further destabilizes the system by allowing gas to be easier to compress. But with streaming, the cosmic ray heating is less efficient when cooling is turned on and so that system becomes more stable.}
\label{fig:cooling_contourPlots}
\end{figure*}

We considered three transport models: (1) ``Classic Parker," where the cosmic rays are assumed to behave as a $\gamma_c = 0$ fluid but still affect the equilibrium system through their pressure gradient. 
(2) The extrinsic turbulence model (Modified Parker) with $\gamma_c=4/3$, and (3) the self-confinement model (Modified Parker with Streaming). We solve Equations \ref{eq:parkstreamcont} - \ref{eq:parkstreamgas} for a dispersion relation and obtain the contour plots shown in Figure \ref{fig:cooling_contourPlots}. A comparison between the instability found in Classic Parker, Modified Parker, and Modified Parker with Streaming is shown in the top row of Figure \ref{fig:cooling_contourPlots} (from HZ18). In Modified Parker \citep{zweibelparker1975, boettcherEDIG2016}, we found that the increase in adiabatic index led to a decrease in the compressibility of the cosmic rays, resulting in a much more stable system than Classic Parker due to the energy needed to compress the  cosmic rays in the magnetic valleys. In Modified Parker with Streaming, we kept $\gamma_c = 4/3$ and inserted the cosmic ray streaming terms into Equations \ref{eq:crenergy} and \ref{eq:thermenergy}. We found this system to be much more unstable than Modified Parker and more unstable than Classic Parker as well. Based upon the work contributions in the three different systems, we concluded that cosmic ray heating in the self-confinement model was responsible for the increased instability. We revisit this diagnosis in Section \ref{subsec:heatingUnstable} with new information from our MHD simulations.

In all three cases, we found that increasing $m$ increases instability. The picture for $c$ is more complex: while increasing $c$ is destabilizing in the Classic Parker model,  increasing $c$ for 
Modified Parker is stabilizing, as the cosmic rays become more difficult to compress into the magnetic valleys. With streaming, increasing $c$ initially destabilizes the system more until reaching a threshold value of $c$ ($\approx 3$, usually) at which the growth rates begin to decrease again.

We also performed a 2D vs. 3D comparison of Modified Parker and Modified Parker with Streaming. In 2D, we assumed $k_z \rightarrow 0$, while in 3D, we assumed that $k_z \rightarrow \infty$, allowing a comparison of these two to give us boundaries on the range of instability for a 3D case with a finite $k_z$. 
We found that in general, the 2D case is always unstable over a smaller range of wavelengths than the 3D case. While in Modified Parker, the 3D case always reaches higher growth rates than the 2D case, for Modified Parker with Streaming, the 2D case peaks at larger growth rates than the 3D case, especially at larger values of $c$. 

\subsection{Radiative Cooling}\label{subsec:cooling}
In many studies of the Parker instability, strong cooling is effectively assumed by varying $\gamma_{g}$ to be less than 5/3. As shown in \cite{KosinskiParkerCooling2006} and \cite{MouschoviasParkerCooling2009}, explicit implementations of low-temperature, optically-thin radiative cooling give decent but not exact agreement with simply using $\gamma_{g} = 1$; however, the compressibility of the gas, which works to stabilize the system, is altered between the two cases, and the further collapse of over-dense regions due to thermal instability cannot be described by simply changing the adiabatic index of the gas. Additionally, HZ18 previously implicated cosmic ray heating as the probable destabilizing effect, due in part to lower growth rates observed when setting $\gamma_{g} = 1$, which negates the cosmic ray heating term (the last line of Equation \ref{eq:thermenergy}). An analysis of work contributions similarly indicated that cosmic ray heating was destabilizing. Given these differences between explicitly including cooling and simply assuming $\gamma_{g} < 5/3$, a linear stability analysis with explicit radiative cooling is necessary.

\begin{figure}[]
\centering
\includegraphics[width = 0.41\textwidth]{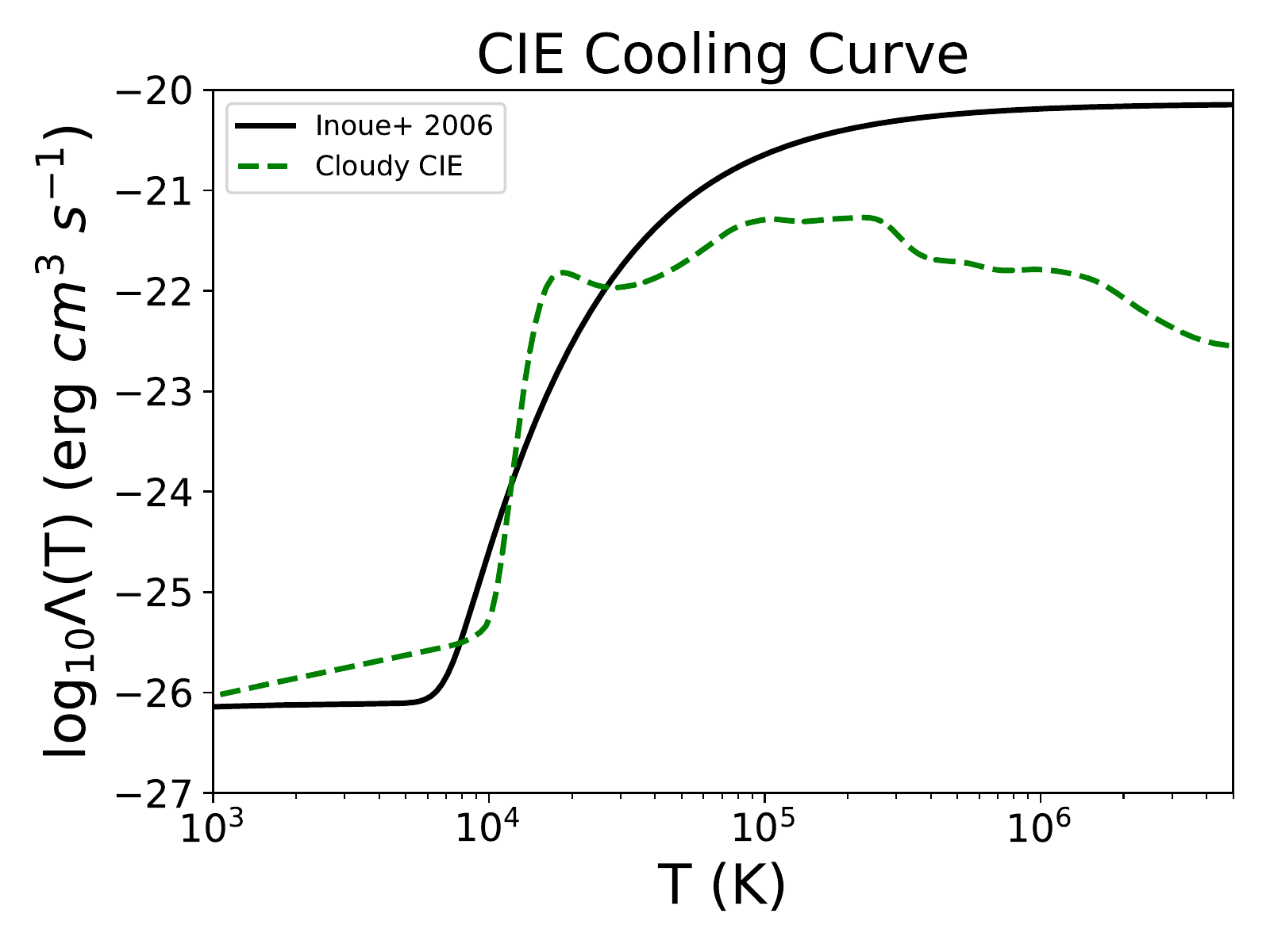}
\caption{The low-temperature cooling curve \citep{InoueCooling2006} used in this work, which has contributions from Ly$\alpha$ and CII fine structure emission. The full CIE cooling curve generated from CLOUDY models \citep{wiersma2009} is also plotted for comparison, showing a reasonable fit at lower temperatures but discrepancies near the peak of the cooling curve.}
\label{fig:coolingCurve}
\end{figure}

To add radiative cooling to our linear stability analysis, we use an analytic approximation to the optically thin cooling curve 
\citep{InoueCooling2006}:
\begin{align}
\label{coolingEqn}
    \Lambda(T) = 7.3 \times 10^{-21} (e^{-118400/(T + 1500.0)}) \nonumber \\
    + 7.9 \times 10^{-27} (e^{-92/T}){\rm{ erg \hspace{2pt }cm^3 \hspace{2pt} s^{-1}}}
\end{align}
where the first term represents cooling from Ly$\alpha$ emission, and the second term comes from CII fine structure emission. A plot of this cooling function, with a comparison to the CIE cooling curve from CLOUDY, is given in Figure \ref{fig:coolingCurve}. While the \cite{InoueCooling2006} cooling curve is a reasonable quantitative fit and produces a medium with two stable phases, there is a larger discrepancy at high temperatures. The \cite{InoueCooling2006} cooling curve levels off at high temperatures and misses the prominent O VI peak that triggers the warm unstable phase. This is not important for our linear stability analysis, but led us to choose a different cooling model for our MHD simulations in the nonlinear regime.

As in \cite{KosinskiParkerCooling2006}, a temporally constant but spatially varying heating function $\Gamma = n_{0}(y)\Lambda(T_{0})$ is applied to keep the initial setup in equilibrium, where $n_{0}(y)$ and $T_{0}$ are the unperturbed density and temperature. Note that while the heating function is constant with temperature, it does depend on $y$ to satisfy thermal equilibrium since the initial density $n_{0}$ is a function of $y$. We ignore cosmic ray heating in this equilibrium state, despite assuming that the cosmic rays are already coupled to the gas. This rests on the assumption, also made by \cite{begelmanacoustic1994}, that the equilibrium horizontal cosmic ray pressure gradient is too weak to contribute significantly to heating. 

For our chosen cooling function, the cooling rate is $\dot{E} = -n_{H}^{2}\Lambda(T) =  -0.49(\rho/m_H)^{2}\Lambda(T)$, where we assume hydrogen is 90\% by number. Therefore, the radiative loss and corresponding gain terms in eq. (\ref{eq:thermenergy}) are:
\begin{align}
\label{eq:coolingrate}
    \dot{E} = -0.49(\rho/m_H)^{2}\Lambda(T) &+ 0.7(\rho/m_H)\Gamma \nonumber \\ = -0.49(\rho/m_H)^{2}\Lambda(T) &+ 0.49 (\rho/m_H) (\rho_{0}(y)/m_H) \Lambda(T_{0}(y)).
\end{align}

Inserting the terms in eq. \ref{eq:coolingrate} into eq. \ref{eq:thermenergy}, we derive a dispersion relation and create the contours in the bottom panel of Figure \ref{fig:cooling_contourPlots}. For these plots, we assumed a base value of $n=0.1$ c$m^{-3}$, $\Bar{m} = 10^{-24}$ g, $T = 8000$K, $a_g = 10^{6}$ cm/s, and $g \approx 4\times 10^{-9}$ cm s$^{-2}$, which are all reasonable for the Milky Way warm neutral medium. These parameter values give an $H_0 \approx 79$ pc and an $H \approx 194$ pc.

After adding radiative cooling for these parameter values, we now see that Classic Parker is the most unstable case of the three, while Modified Parker and Modified Parker with Streaming are now much closer to each other in levels of instability. When compared with the top row of \ref{fig:cooling_contourPlots}, we see that both Classic Parker and Modified Parker are more unstable than their non-cooling counterparts. We believe this is because cooling the gas makes it easier to compress into the valleys of the magnetic field. However, for Modified Parker with Streaming, we see that radiative cooling makes the system becomes more stable than without cooling. The cooling of the gas makes cosmic ray heating effect less efficient and therefore makes the system more stable.  But even with cooling, Modified Parker with Streaming is still more unstable than Modified Parker for this set of parameter values.

\begin{figure*}
\centering
\includegraphics[width=0.3\textwidth]{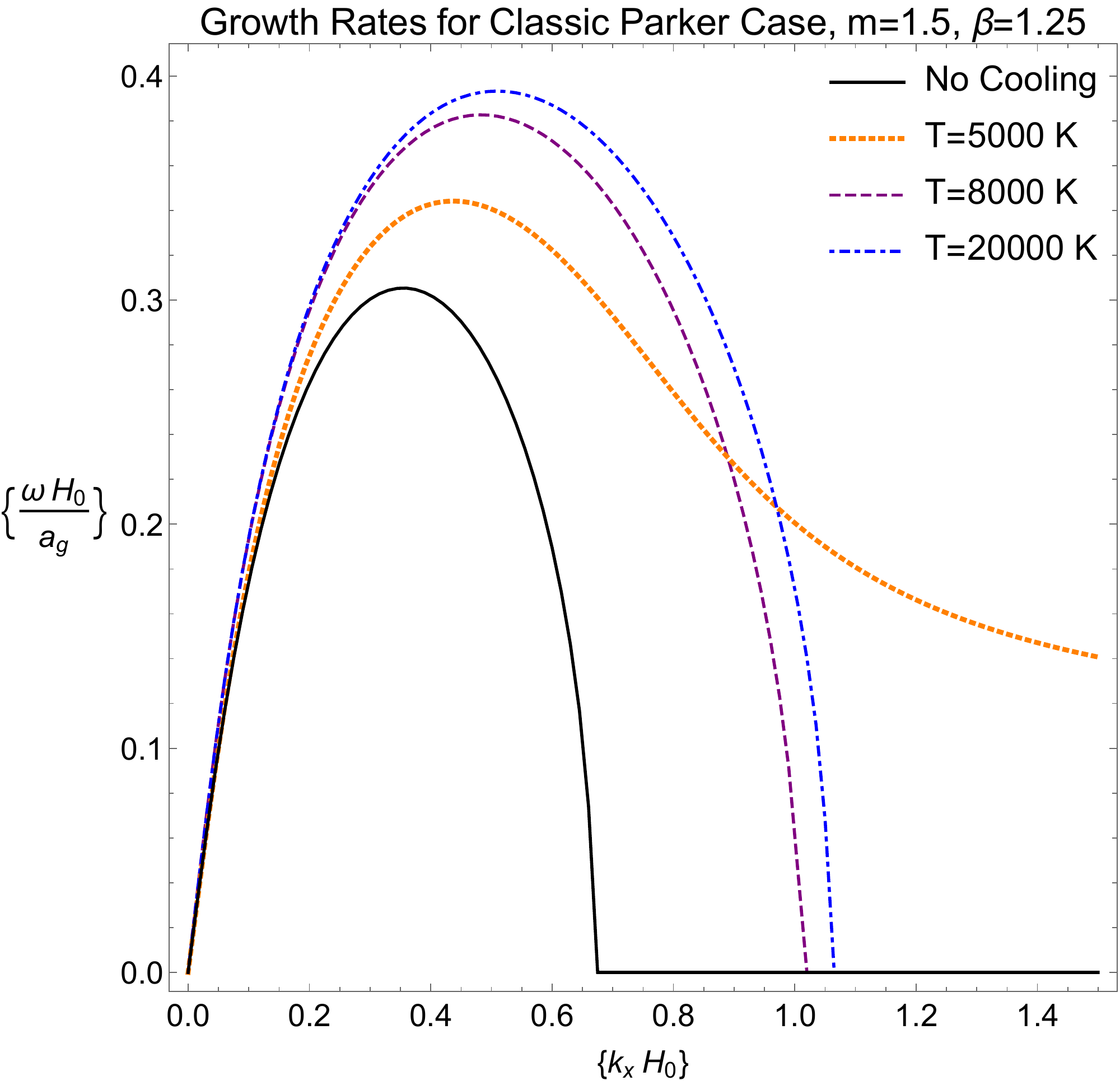}{(a)}
\includegraphics[width=0.3\textwidth]{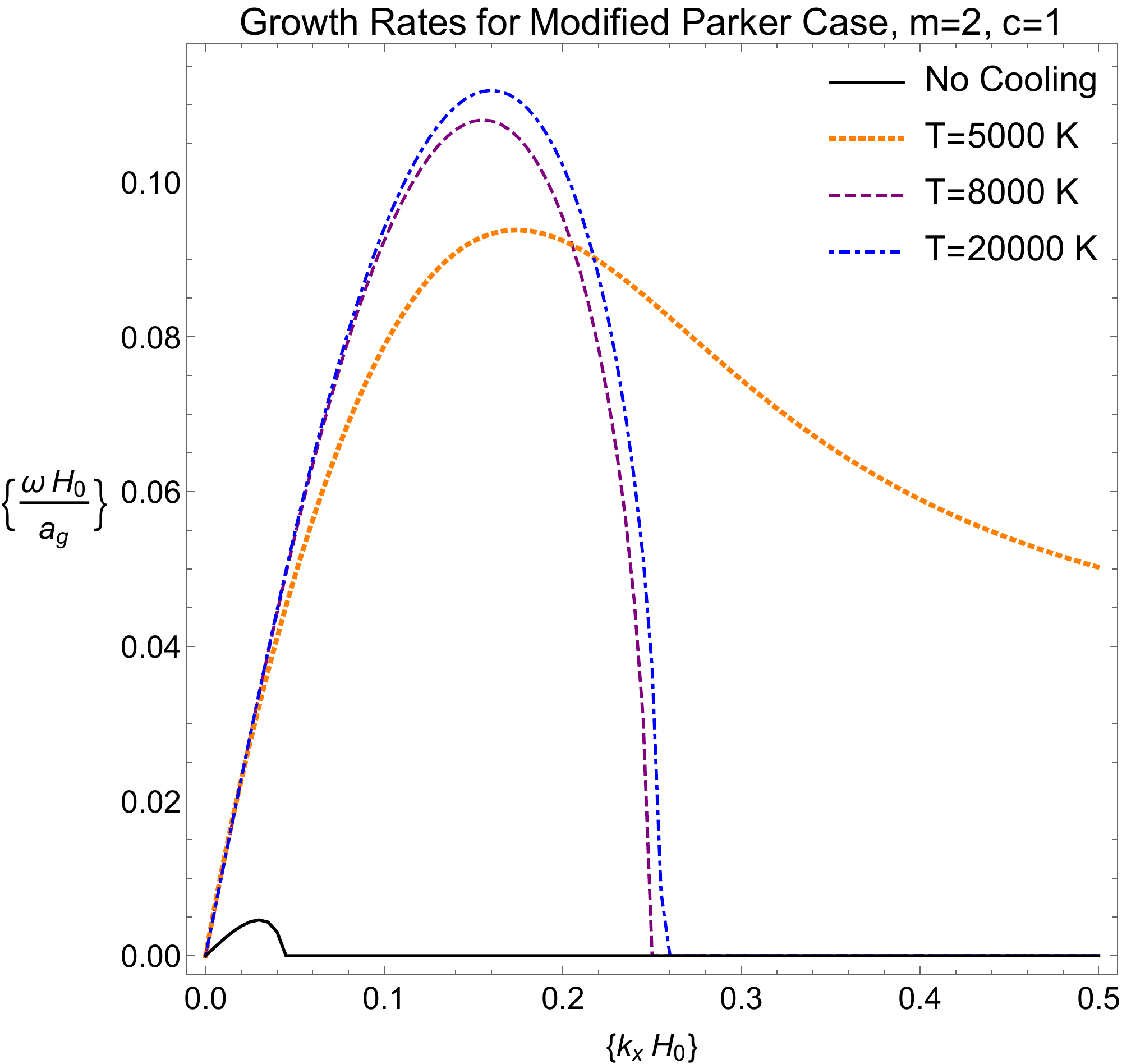}{(b)}
\includegraphics[width=0.3\textwidth]{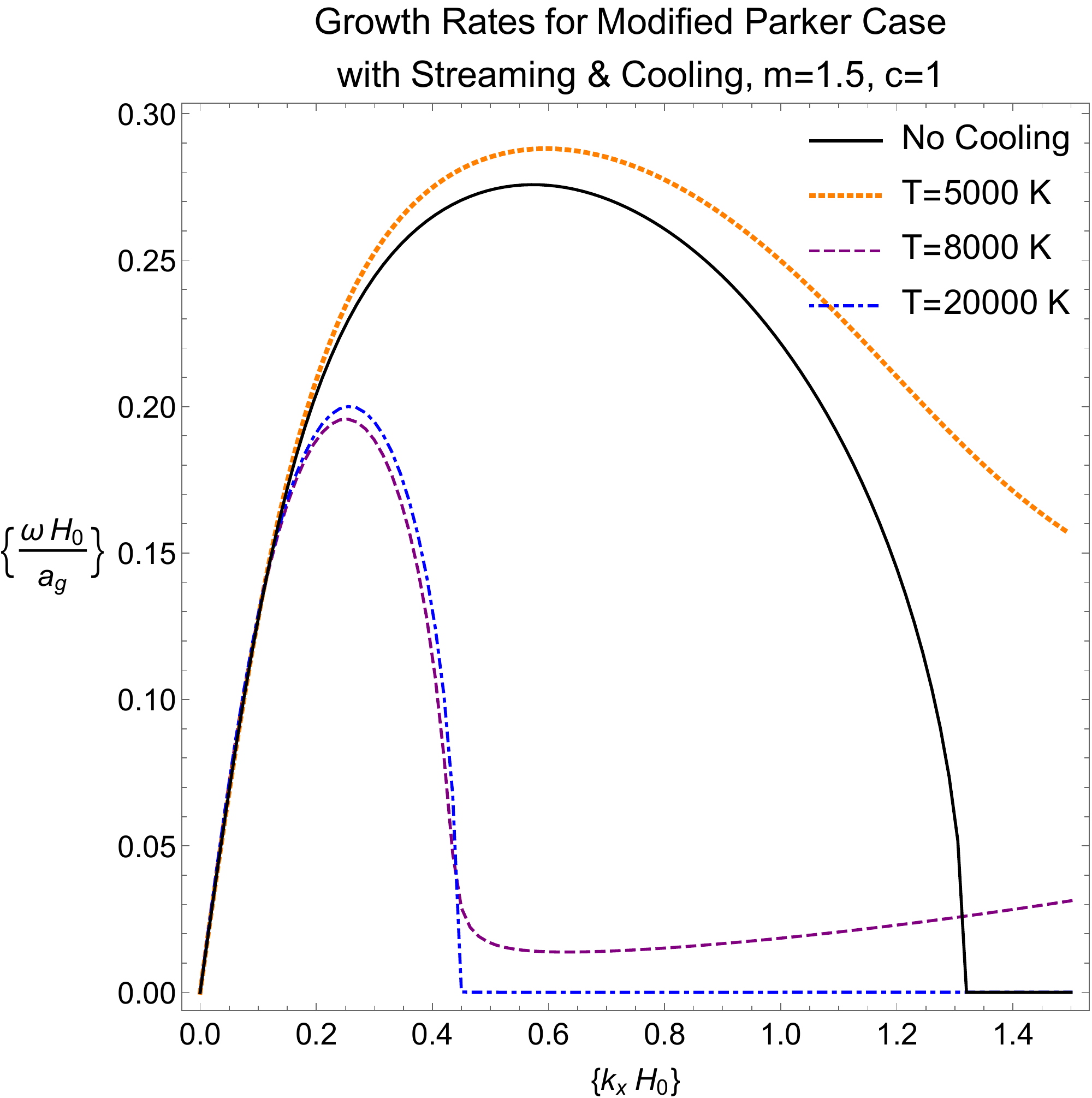}{(c)}
\caption{Plots of $\hat{\omega}$ vs. $\rm \hat{k_x}$ for the three Parker cases with radiative cooling with $k_y=0$. For each case, we have plot three different curves with three different temperatures (meaning different cooling functions). No cooling is the black curve, $\rm T=8000K$ is the purple curve, $\rm T=5000 K$ is the orange curve, and $\rm T=20000K$ is the blue curve. Again, note the different domains of ($k_x H_0$) and ($\omega H_0/a_g$). Similar to the contour plots, the more cooling without streaming, the more unstable the system, while the opposite holds true with streaming. The behavior at large values of $\hat{k_x}$ is due to thermal instability.}
\label{fig:difftempPlots}
\end{figure*}

In Figure \ref{fig:difftempPlots}, we  plot the instability curves for three different values of equilibrium temperature $T_0$ for all three cosmic ray transport models. Increasing $T_0$ from $\rm 8000K$ to $\rm 20000K$  creates negligible changes between the instability curves. However, lowering $T_0$ to $\rm 5000K$ again shows the stark differences between the cases with and without streaming. In Classic Parker and Modified Parker, lowering the temperature causes the peak growth rate to decrease but results in a much larger range of instability that appears to asymptote to a certain growth rate. With Modified Parker with Streaming, the range of instability is still increased but now we instead see the peak growth rate is larger when the temperature is decreased to $\rm 5000K$. Furthermore, at a certain value of $k_x$, even the $\rm T=8000K$ curve begins to rise back up into a more unstable regime.

Since the Parker instability turns off at short wavelengths, it seems likely that the increased ranges of instability are due to thermal instability. For the analytic cooling curve of \cite{InoueCooling2006}, we  calculate that our initial setup is unstable to pure condensation modes (satisfying $\delta n / n \propto -\delta T / T$) when $92 < T < 5800$ K. This corresponds to the relatively flat part of the cooling curve with $d \Lambda /dT < \Lambda / T$ \citep{FieldCriterion1965}. 
Inspection of the eigenfunctions derived in the linear stability analysis supports this conclusion. For an isobaric system (similar to our system with cooling), one expects the temperature and density eigenfunctions to differ in phase by $\pi$. We find for all temperatures where the instability extends out and asymptotes at large values of $\hat{k_x}$, that this relationships seems to hold at those values. However, as we decrease back to smaller values of $\hat{k_x}$, the two eigenfunctions align in phase, as one would expect for an adiabatic system where thermal instability is not the dominant instability. In general, we find that somewhere around $\hat{k_x}\approx 0.3$ for the lower temperature systems and $\hat{k_x}\approx 0.4$ for the higher temperature systems the instability switches over from the Parker instability to thermal instability.

For further study of the wealth of effects uncovered here, we now turn to nonlinear simulations.

\section{Simulation Method}
\label{sec:sim_method}
We use the FLASH v4.2 \citep{FryxellFLASH2000} magnetohydrodynamics (MHD) code to carry out our simulations. To solve the ideal MHD equations, we use the unsplit staggered mesh MHD solver \citep{LeeMeshMHD2009, LeeMeshMHD2013}, which is based on a finite-volume, high-order Godunov scheme and uses a constrained transport scheme to ensure divergence-free magnetic fields. We use a modified version of FLASH that includes an additional cosmic ray module \citep{yangfermi2012, ruszkowskiwinds2017}, which evolves cosmic rays as a second fluid and includes a fluid description of the kinetic cosmic ray streaming process. In practice, the cosmic ray fluid is defined as a mass scalar in FLASH, and it obeys a relativistic equation of state, as well as a separate evolution equation that depends on whether streaming is desired or not. We refer the reader to \citealt{ruszkowskiwinds2017} for the full set of CR-modified equations evolved in FLASH. 

Because, in the streaming picture, the direction of flow along field lines is always directed down the cosmic ray pressure gradient, numerical issues arise near extrema in cosmic ray pressure, where the gradient changes sign. To counteract this, \citealt{ruszkowskiwinds2017} implement a regularization method \citep{SharmaRegularization2009}, for which one must choose a characteristic cosmic ray scale length, L. For our simulations, we consistently use $L = 5$ kpc. We show convergence with respect to this parameter in Appendix \ref{sec:convergence}. 

Our simulation setup assumed periodic boundary conditions along the direction of the magnetic field ($\hat{x}$), with diode boundary conditions (allowing outflow but not inflow) set along the direction of gravity ($\hat{y}$). We also tested outflow boundaries and found consistent growth rates and system evolution away from the immediate boundary cells. In 3D simulations, we also assume periodic boundary conditions along the third direction ($z$). For our fiducial 2D simulations, we used a uniform 512x512 grid, resulting in a resolution of 31.25 pc or 16.125 pc depending on box size (8 kpc x 8 kpc for Solar Neighborhood parameters, 16 kpc x 16 kpc for \cite{RodriguesParker2015} parameters). See Appendix B for how the simulations converge for different resolutions. Our more computationally-expensive 3D simulations were only run using the set of \cite{RodriguesParker2015} parameters on a 16 kpc x 16 kpc x 8 kpc grid (with gravity again in the $\hat{y}$ direction) with a resolution of (62.5 pc, 62.5 pc, 31.25 pc). We shortened the box length and cell size in the $\hat{z}$ direction because we expected (and corroborated) a prevalence of short wavelength perpendicular modes in the $\hat{z}$ direction. For all simulations, we used the HLLD Riemann solver but didn't find any differences when using the HLLC solver.

\subsection{Seeding the Parker Instability}
One typically seeds the Parker instability by applying one of two types of perturbations\footnote{A third method, waiting for the instability to grow from numerical noise, is typically too slow to be efficient.}: to isolate a specific wavelength $\lambda$, one would perturb the vertical component of the velocity with a harmonic function of $2\pi x/\lambda$ -- we refer to this as ``leading the horse to water"; \emph{or} one can simultaneously perturb many wavelengths and apply all those velocity perturbations at once -- we refer to this as the ``horse race". In this work, we consistently perturb the first 100 wavelengths that fit within the box width along the background magnetic field. In our coordinate system, the initial magnetic field is in the $\hat{x}$-direction, while the vertical stratification is in the $\hat{y}$-direction. Our vertical velocity perturbation is then:

\begin{equation}
\label{seedPerturbEqn}
    \delta v_{y} = \sum_{n_{x}=1}^{100} \sum_{n_{z}=1}^{100} A \rm sin\Big(\frac{2 \pi x}{ \lambda_{n_{x}}} - \theta_{n_{x}}\Big) sin\Big(\frac{2 \pi z}{ \lambda_{n_{z}}} - \theta_{n_{z}}\Big)
\end{equation}
where $\lambda_{n_{x}} = (x_{\rm max} - x_{\rm min})/n$ and $\lambda_{n_{z}} = (z_{\rm max} - z_{\rm min})/n$ are the wavelengths that fit within the box width in the $\hat{x}$-direction (parallel to the background magnetic field) and the z-direction (perpendicular to the field). $\theta_{n_{x}}$ and $\theta_{n_{z}}$ are phases randomly drawn between 0 and $2 \pi$. Note that the summation over z-terms is excluded for 2D simulations. The above perturbation also has no y-dependence; we found no difference between simulations with velocity perturbations constant in y and ones following an exponential profile falling off with height.

The variable $A$ defines the initial perturbation amplitude. For 2D simulations, we set the fiducial value of $A$ to  10$^{-4}$ cm/s, which gives a highly subsonic velocity perturbation, resulting in vertical magnetic field perturbations growing by 4 or more orders of magnitude throughout the instability evolution. Changing this value of A to $10^{-6} \rm cm/s$ makes no significant difference in the resulting growth rates (see Appendix \ref{sec:convergence}). In 3D, since we now apply 100 more perturbations, we find that $A = 10^{-4} \rm cm/s$ is too high; $10^{-6} \rm cm/s$ gives a similar velocity kick as in the 2D case and promotes a long linear phase from which we can compute growth rates, so we choose this to be our fiducial 3D value.  

To get a growth rate, we Fourier transform the y-component of the magnetic field across the simulation box, which gives a Fourier amplitude $A(k)$ for each mode. We then back out the growth rate by fitting a line to a section of log($A(k)$) vs time.\footnote{Choosing the time interval is not an exact science. To isolate linear growth, one wants to choose an interval when initially oscillatory/spurious modes are clearly not excited, but the time interval should capture enough output times to get an accurate line fit. We varied the interval start and end times for every growth curve shown in this paper, and we find our presented results to be robust.} How the Fourier amplitudes are scaled does not matter for our growth rate calculations, and the amplitudes are \emph{not} to be confused with the value of $B_{y}$ for each mode. While ``leading the horse to water" avoids mode coupling to give the most direct growth rate comparison for a given wavenumber, it is computationally expensive to run such simulations over and over again.  The ``horse race" allows us to run only one simulation instead of many simulations with one wavelength each. While it does so at the expense of possible mode coupling, it is also more representative of the ISM, which would naturally seed many wavenumbers at once. 

\begin{figure*}[]
\centering
\includegraphics[width = 0.43\textwidth]{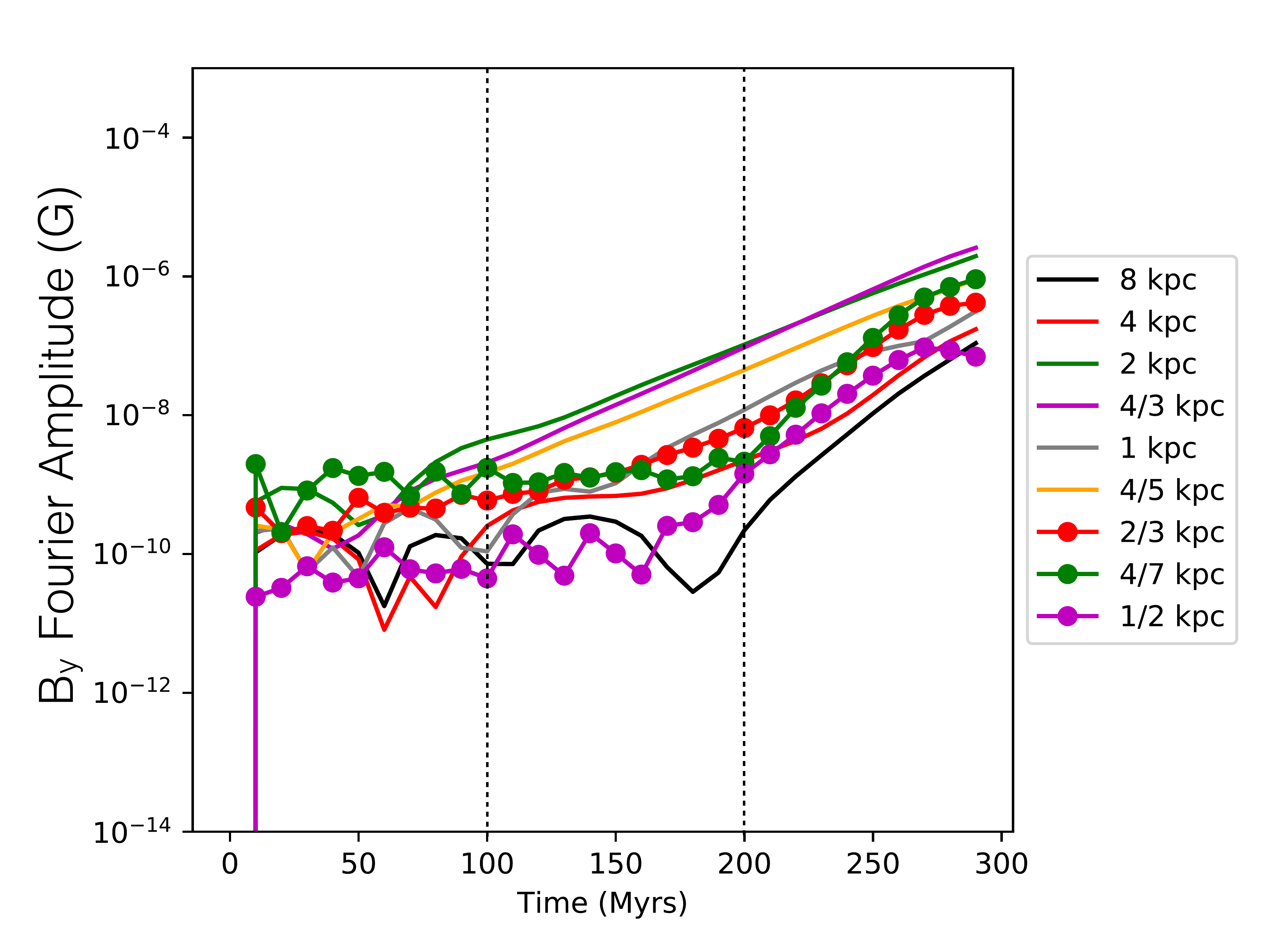}
\includegraphics[width = 0.43\textwidth]{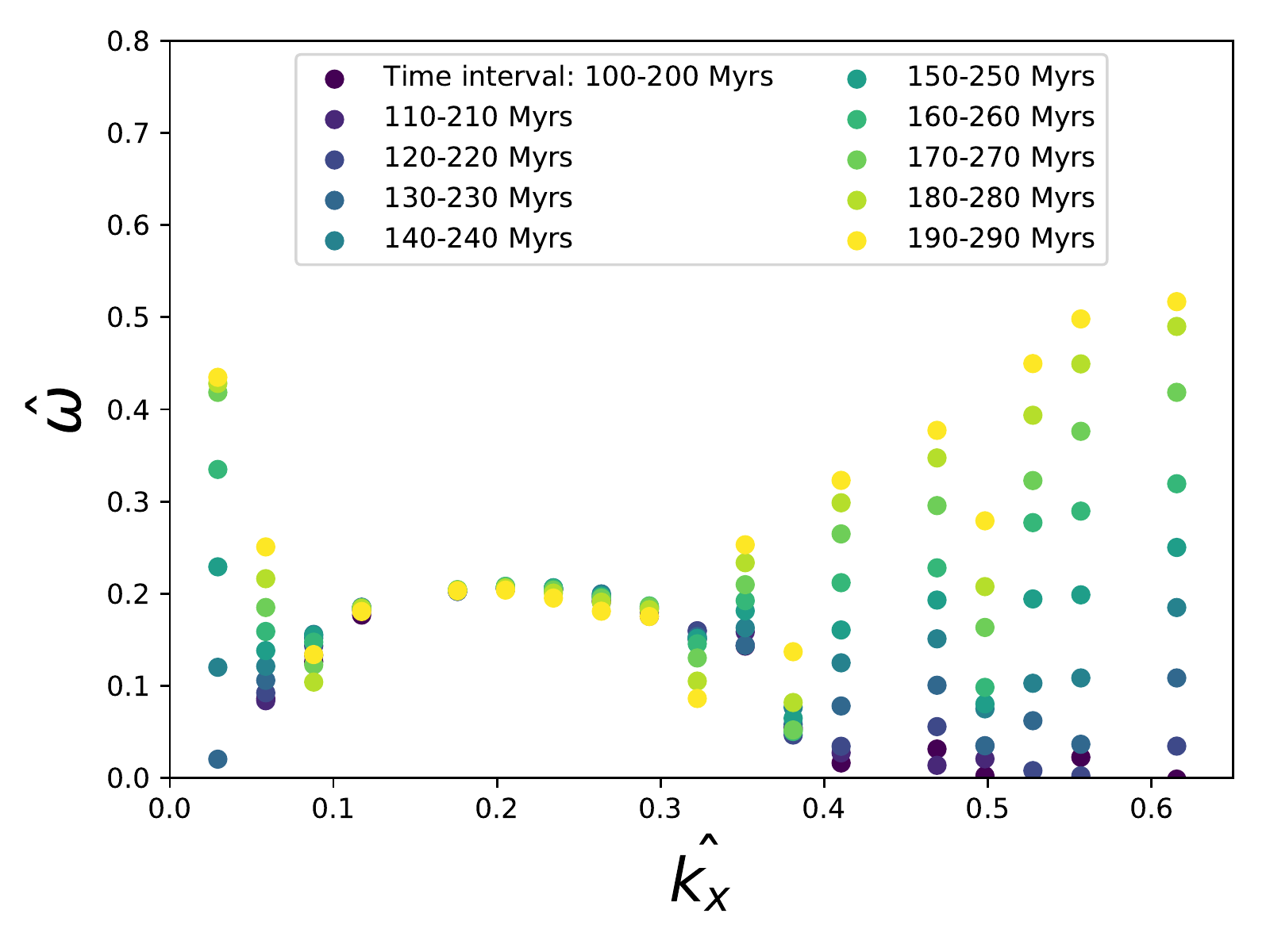}
\caption{Left panel: Fourier amplitudes of $B_{y}$ for a selection of excited modes in a $m = 2$, $c = 1$ unstable ISM using the Solar Neighborhood parameter set. Growth rates for this case are calculated by fitting a line to each curve between 100 and 200 Myrs (denoted by the vertical dashed lines) and calculating the slopes. Right panel: Dimensionless growth rate $\hat{\omega}$ vs $\hat{k_{x}}$ when different 100 Myr time intervals are chosen for line-fitting. This example nicely shows the nonlinear interactions that pump some of the short and long wavelength, initially very slowly growing modes (e.g. the $\lambda = 1/2$ and $\lambda = 8$ kpc modes) after 200 Myrs. This merging and transfer of energy between modes causes them to grow much faster than expected from linear theory, and this is prevalent in our simulations as the evolution surpasses the linear regime.
}
\label{fig:fourier_m3_c1.5}
\end{figure*}

In Figure \ref{fig:fourier_m3_c1.5}, we have plotted the Fourier amplitude of $B_y$ for different wavelengths in the simulation. We see for modes with wavelengths between 4/5 kpc and 4 kpc that in the range of 100-200 Myrs, the modes generally grow exponentially, as expected for unstable modes in the linear regime. However, other modes, most notably the 1/2 kpc and 8 kpc modes,  oscillate between different amplitudes but show no real growth in this range. From the stability analysis, we expect the 1/2 kpc mode and short wavelengths like it to be stable but that the 8 kpc mode should be unstable with a very small growth rate. Thus, we believe that for both of these modes and others of similar wavelengths, their oscillating amplitudes are due to them being non-linearly pumped by the faster growing modes in the simulation. As the simulation reaches 200-300 Myrs, the simulation leaves the linear regime and all of the modes begin to grow similarly. Thus, when calculating growth rates, we focus on the 100-200 Myrs regimes, while noting that the growth rates of modes that are stable or only slightly unstable will be more random due to this nonlinear pumping and may not follow the linear analysis.

Despite the resulting mode coupling (see e.g. Figure \ref{fig:fourier_m3_c1.5}), we consistently find a tight match between simulation growth rates and our linear stability analysis (see Section \ref{sec:simresults2D}); therefore, we will only present results using this ``horse race" method.

It is worth noting that almost all studies in the literature carry out a similar ``horse race" by perturbing \emph{cell-by-cell} (e.g. \citealt{RodriguesParker2015}), which means the perturbations change depending on grid layout and resolution, or utilize the ``leading the horse to water" method. However, we have found few studies that have made a direct comparison of simulation growth rates to linear theory (as an example of work that has, see \cite{KosinskiParkerCooling2007} where linear growth rates of Parker and thermal instabilities were compared with simulation growth rates).
\\ \\
\section{Simulation Results: 2D}\label{sec:simresults2D}
We first describe our transition to a smooth gravitational potential. Although a constant gravitational field is more tractable analytically, we found a smooth transition necessary (and more realistic) for our simulations. Using the constant field did give reasonable growth rates in some cases, but it was not reliable, partially because of the impossible strain placed on simulations to resolve an abrupt transition from positive to negative gravitational force at the midplane (see Appendix \ref{sec:constantgrav} for a short summary of our struggles). Therefore, we were driven to a smooth gravitational potential by computational necessity as well as scientific realism.


\begin{table}[]
    \centering
    \begin{tabular}{c c c}
        $\Sigma (M_{\odot} \rm pc^{-2})$ & $\Sigma_{g} (M_{\odot} \rm pc^{-2})$ & H (pc)  \\
        \hline
        Solar Neighborhood Setup \\
        \citep{McKeeSolar2015} \\
        47.1 & 13.7 & 250 \\
        \hline
        \cite{RodriguesParker2015} Setup \\
        100.0 & 10.0 & 500 \\
        \hline
    \end{tabular}
    \caption{Two parameter sets of total (stellar + gas) surface density ($\Sigma$), gas surface density ($\Sigma_{g}$), and scale height (H)}
    \label{tab:paramTable}
\end{table}

\subsection{Smooth Gravity and Growth Rates}\label{subsec:gravity}

\cite{GizParker1993} were the first to study the Parker instability with a smooth gravitational acceleration, in their case $g(y)\propto\tanh{y}$, which corresponds to a self-gravitating, isothermal mass layer. They pointed out that this setup, regardless of the exact function for a smooth gravitational acceleration, is analogous to a quantum harmonic oscillator and results in both continuum modes (as in Parker's analysis) \emph{and} discrete modes, where the modes are set by a parameter $\ell$. They found that the discrete modes are the most unstable and therefore, we compare to these modes in our simulations. In Parker's analysis (and that of many other authors, including HZ18), these modes are suppressed because the potential well has no width near $y = 0$. 

\cite{KimParker1998} subsequently compared three cases: constant gravity, linear gravity, and the $\tanh$ gravity profile. They found that the linear case was the most unstable while constant gravity led to the most stable system. Recently, \cite{RodriguesParker2015} ran simulations in 3D with the $tanh$ gravitational acceleration with cosmic ray diffusion. In most analyses of 2D simulations, it is found that the undular modes ($k_\parallel >> k_\perp$) are preferred by the instability over the interchange modes ($k_\perp >> k_\parallel$) due to the inability of gas to fall into magnetic valleys with the interchange modes. However, due to their 3D setup, \cite{RodriguesParker2015} found no preference for undular or interchange modes, an important point to note when comparing simulations to observations.

It is also worth noting some key differences between mode symmetries. Even in a constant gravity case, simulating both sides of the disk (with $g$ abruptly changing sign at the midplane) allows for midplane warping modes, whereas simulations that are cut at the midplane theoretically keep the midplane untouched. This is an important point appreciated in previous works, including \cite{MatsumotoParker1988} and \cite{BasuParker1997}, which showed that the midplane warping mode ('even' in $v_y$) is naturally selected over the symmetric mode ('odd') when random perturbations are applied to the stratified system. This even mode also leads to the greatest density enhancements in the midplane, which is most conducive to giant molecular cloud formation. These modes typically also have faster growth times than the odd modes due to the extra freedom for gas to cross the midplane and convert gravitational potential energy to kinetic energy. However, in their work with the realistic gravitational potential, \cite{GizParker1993}, \cite{KimParker1998}, and \cite{RodriguesParker2015} all found that 
the modes show no preference for a specific symmetry, odd or even.

For our simulations, we will use the $tanh$ gravity profile, following \cite{RodriguesParker2015}. Using $\hat{y}$ as the vertical direction, we write the density, gravitational acceleration, and temperature as functions of y: 

\begin{equation}
     \rho(y) = \frac{\Sigma_{g}}{2H} \rm sech^{2} \left(\frac{y}{H} \right)
\end{equation}

\begin{equation}
     g(y) = -2\pi G \Sigma \rm tanh \left(\frac{y}{H} \right)
\end{equation}
where $\Sigma$ is the total (gas + stars) surface density of the galaxy, and H is the scale height.

\begin{equation}
\label{eq:tempEqn}
     T = \frac{\pi G \Sigma m_{p} \mu H}{k_{B}(1+ \alpha + \beta)}
\end{equation}
where $m_{p}$ is the proton mass, we choose $\mu = 1$ as the mean molecular mass, 
and where $\Sigma_{g}$ is the disk total surface density. 
This definition of our temperature follows from magneto-hydrostatic equilibrium -- a balance between the gravitational force and the sum of the vertical pressure gradient (from both gas and cosmic rays) and the magnetic pressure gradient.

For our simulation setup, we mainly follow the parameter choices of \cite{RodriguesParker2015}, with $\Sigma = 100.0$ $\rm M_{\odot} pc^{-2}$, $\Sigma_{g} = 10.0$ $\rm M_{\odot} pc^{-2}$, and $H = 500$ pc. To make our conclusions more robust, we also run a few simulations with a second set of parameters appropriate for the Solar Neighborhood \citep{McKeeSolar2015}, with $\Sigma = 47.1$ $\rm M_{\odot} pc^{-2}$, $\Sigma_{g} = 13.7$  $\rm M_{\odot} pc^{-2}$, and $H = 250$ pc. This has the added benefit that we can comment directly on the ramifications for our Solar Neighborhood, which is a standard ISM model in many ISM ``patch" analyses (e.g. \cite{KimTIGRESS2018}). These two parameter sets are shown in Table \ref{tab:paramTable} for clarity. For each choice of surface densities and scale heights, we then vary the magnetic field and cosmic ray parameters $m$ and $c$ (or $\alpha$ and $\beta$). This, in turn, defines the temperature of the ISM. Throughout our work, we found no major differences between our results for the two parameter sets, especially in a qualitative sense. Quantitatively, they lead to different growth curves, which is to be expected.  

\begin{figure}[]
\centering
\includegraphics[width= 0.43\textwidth]{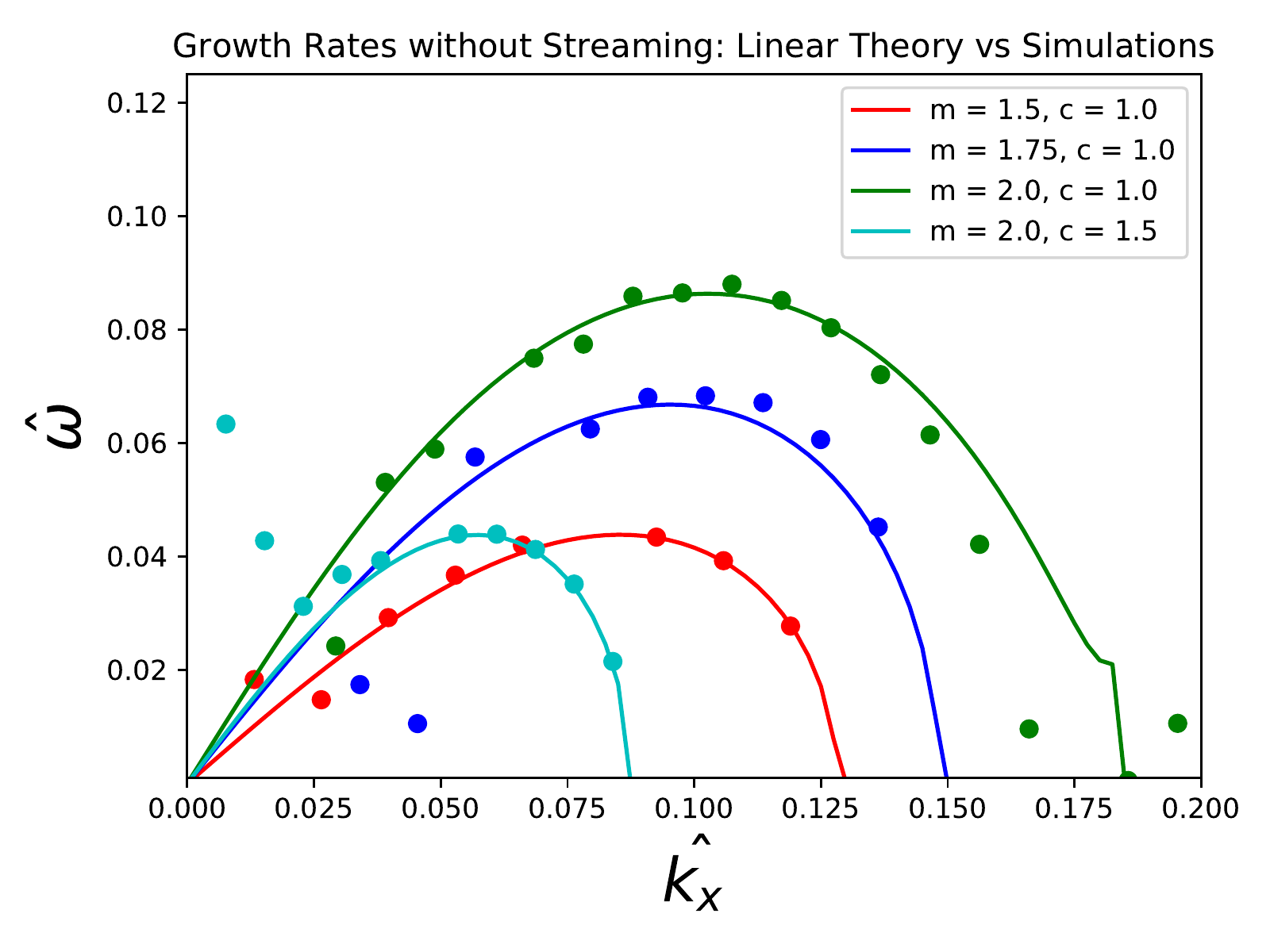}
\caption{Dimensionless growth rate comparison between the linear stability analysis and 2D FLASH simulations. This test was for the Modified Parker setup (no streaming or cooling) with the Solar Neighborhood galaxy parameter set. The solid lines represent the linear stability analysis while the dots represent the Fourier transform of wavelengths periodic in the box of the simulation.}
\label{fig:linear_vs_simulation_noStreaming}
\end{figure}

To validate our simulations, we begin with a growth rate comparison of our Modified Parker simulations to the growth rates calculated using a linear stability analysis, now accounting for a smooth gravity profile. To see how to solve these linearized equations, see \cite{GizParker1993} who provide a nice, detailed analysis of the Parker instability with a $tanh$ gravitational profile.\footnote{Note that solving the same set of equations with smooth gravity \emph{and} streaming is much more difficult. We leave this to future work.} Our linear analysis comparison for Modified Parker follows \cite{GizParker1993}. 

For many different values of $m$ and $c$, we ran out simulations in 2D that evolved through the linear phase. A comparison of the growth rates obtained from simulations against the growth rates from the linear theory are presented in Figure \ref{fig:linear_vs_simulation_noStreaming}. We can see that at the highest growth rates, simulated growth rates line up very well with the linear analysis. For modes far from the peak growth rate, the simulated mode growth diverges from the linear theory in some cases. This may be attributed either to slow growth, which is more difficult for the simulation to capture, or to pumping by faster growing modes which have reached nonlinear amplitude - an  effect not captured by the linear analysis.

\begin{figure}[]
\centering
\includegraphics[width = 0.49\textwidth]{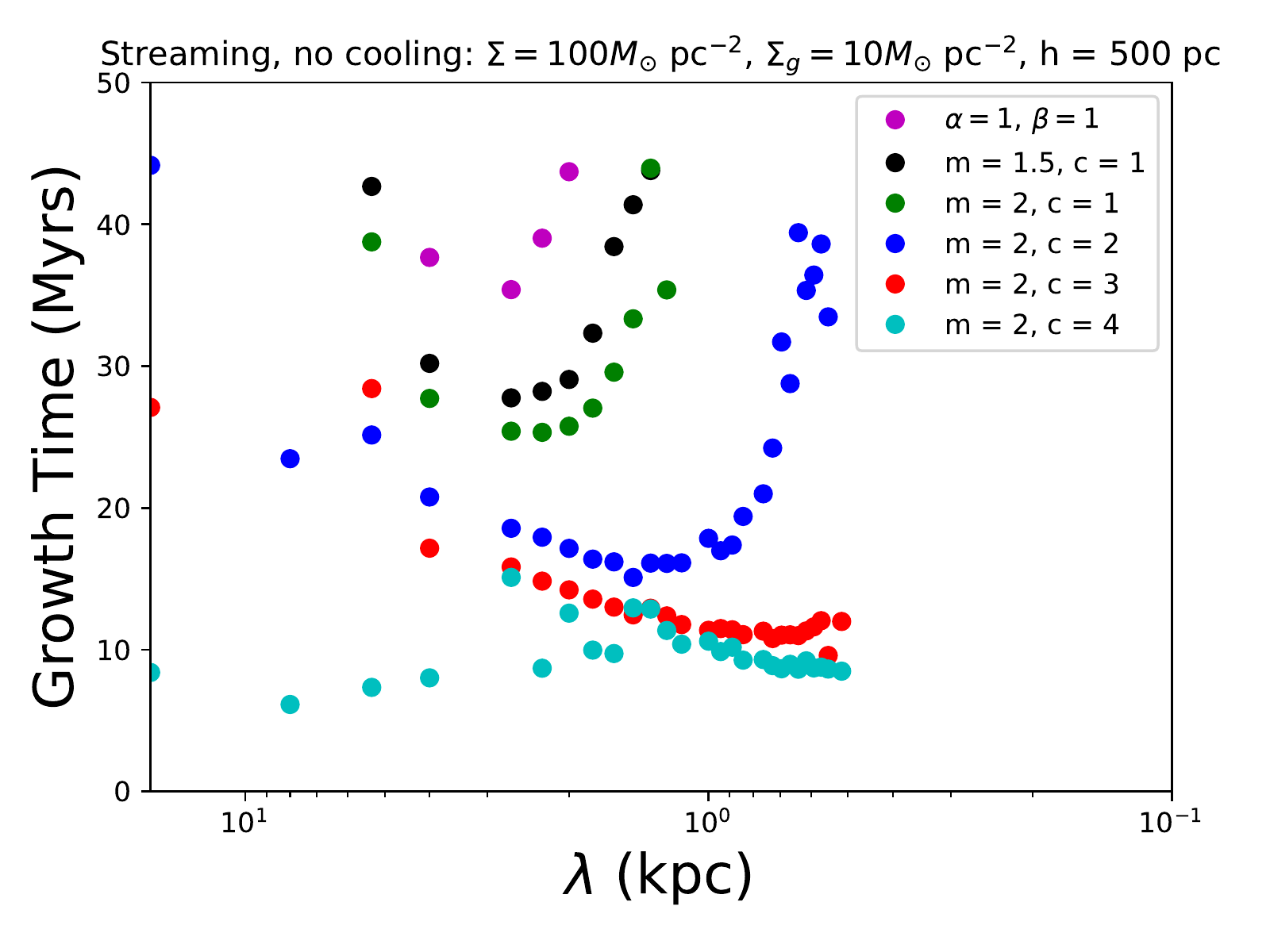}
\caption{Growth time curves for \cite{RodriguesParker2015} parameters and varying ISM compositions. As m and c increase, the instability becomes faster over a larger range of wavelengths, which is consistent with HZ18. } 
\label{fig:streaming_growthRates}
\end{figure}

In 2D simulations with streaming, although we do not have a linear stability analysis to compare to directly, we find a good qualitative match between our growth curves and the results of HZ18. Figure \ref{fig:streaming_growthRates} shows our derived growth times for simulations of varying (m,c) values for \cite{RodriguesParker2015} galaxy parameters. We see that increasing $m$ and $c$ leads to shorter growth times and an expanded range of unstable modes. 
The $m = 2$, $c = 4$ simulation even looks to grow faster and faster at shorter wavelengths, which are not shown on our plot because we don't trust them to be well-resolved at our resolution of 31.25 pc. We expect that, as we found in the constant gravity case, at higher $c$ these growth curves will turn back towards stability, but we do not have the numerical resolution to probe that full $(m,c)$ range in our simulations.

Overall, we are very pleased with the match between theory and simulation, and we proceed to an analysis of the nonlinear regime.

\subsection{Nonlinear Evolution}
\label{subsec:evol_nonlin}

\begin{figure*}[]
\centering
\includegraphics[width =0.32\textwidth]{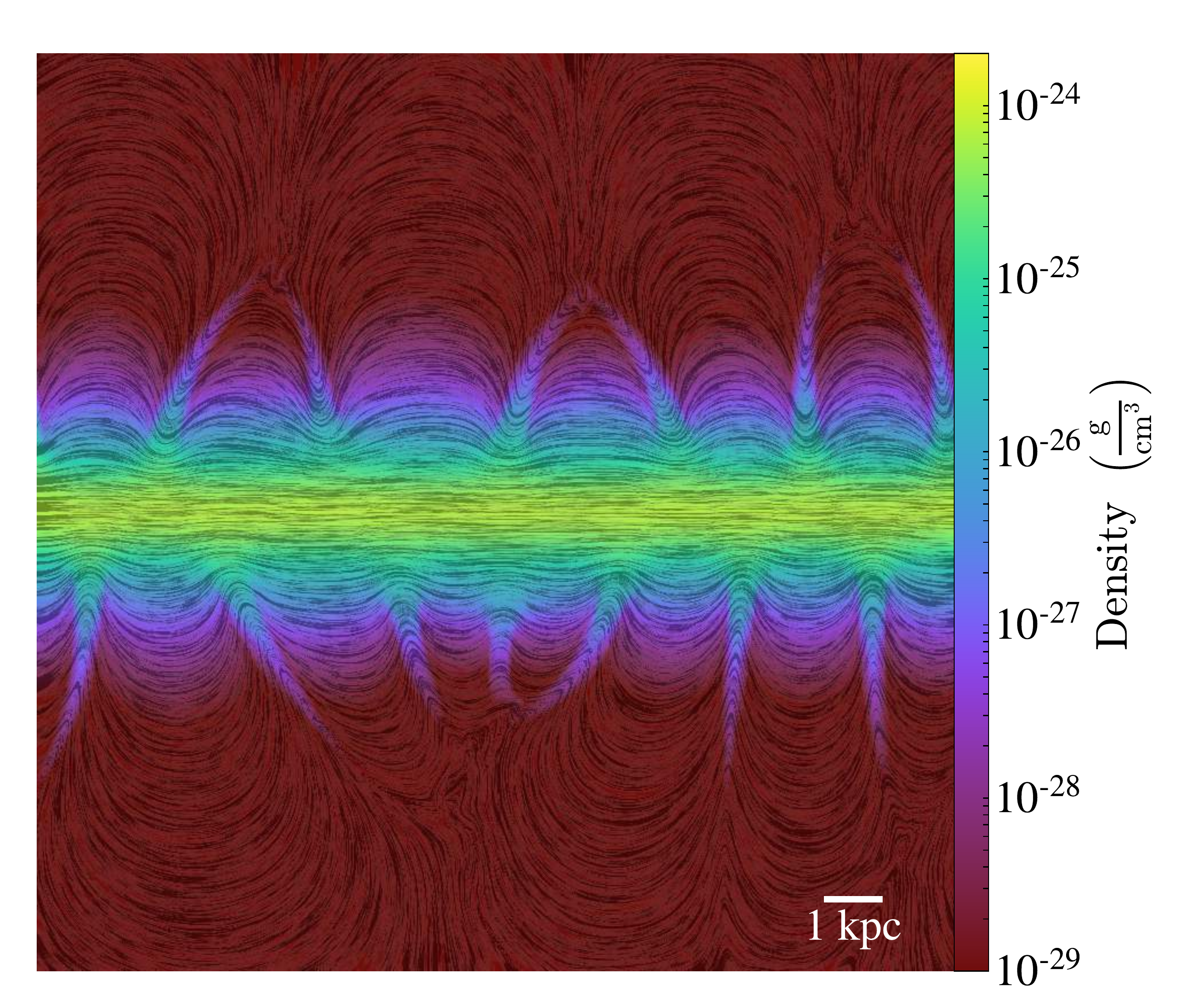}
\includegraphics[width=0.32\textwidth]{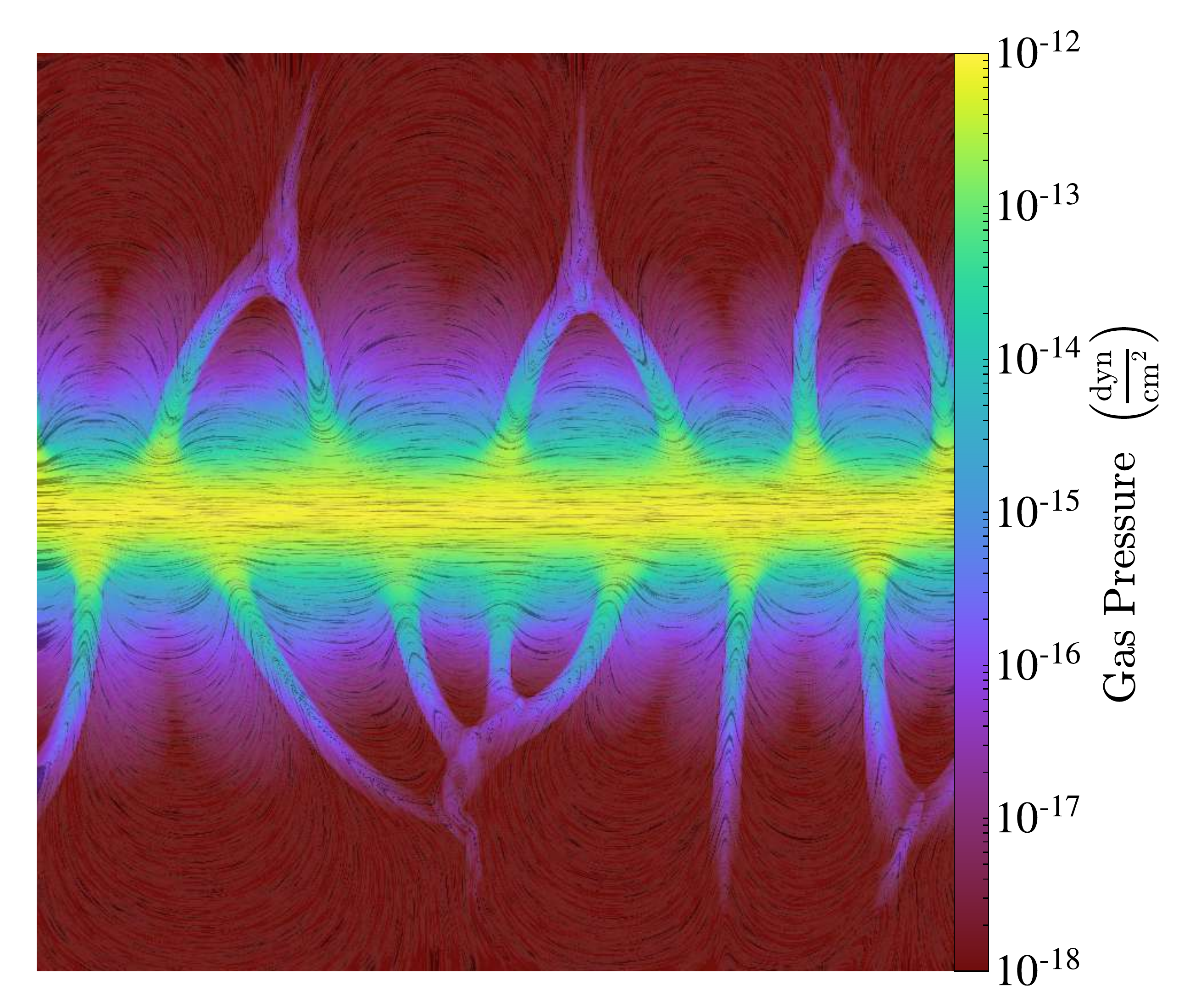}
\includegraphics[width=0.32\textwidth]{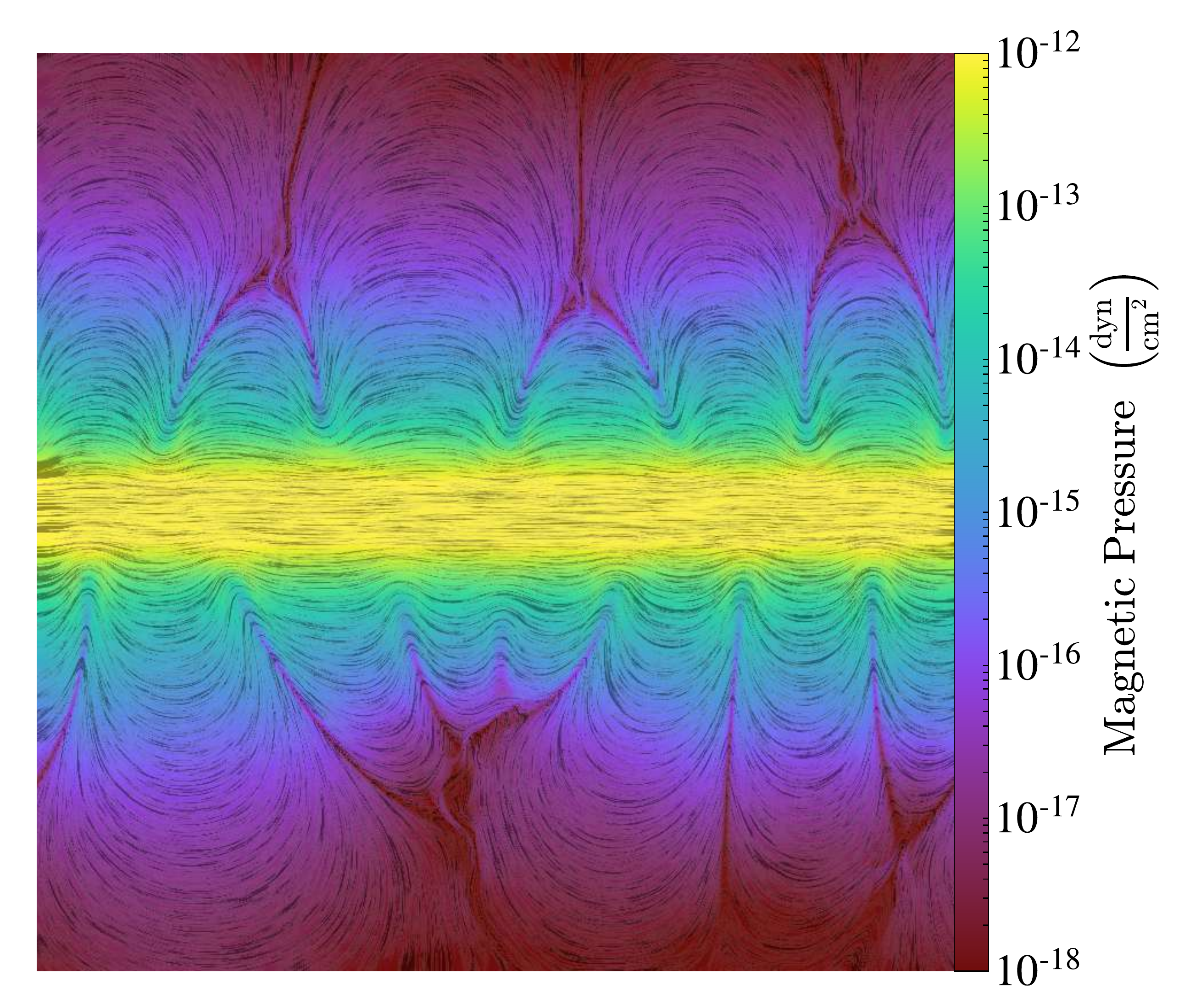}
\includegraphics[width=0.32\textwidth]{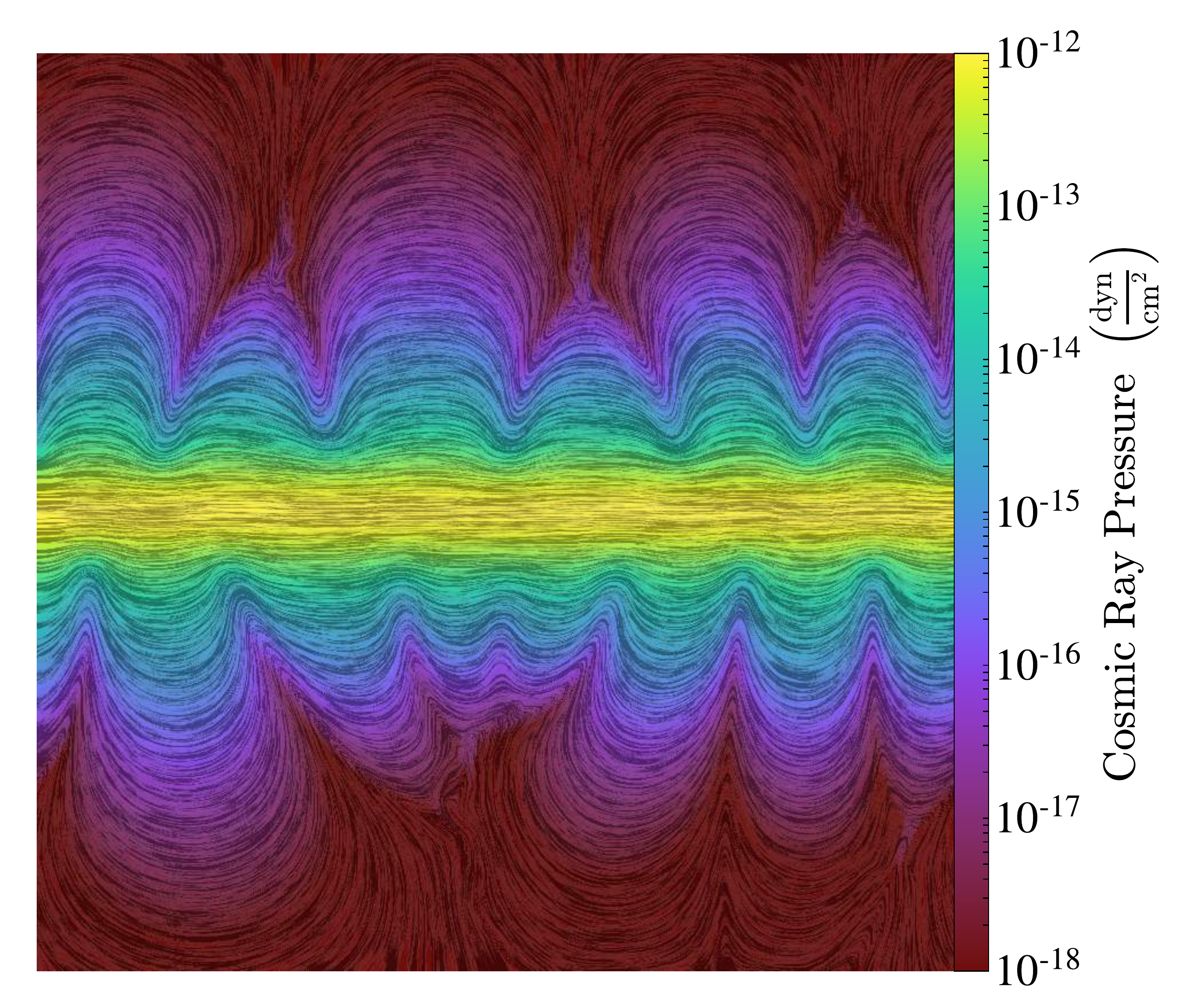}
\includegraphics[width=0.32\textwidth]{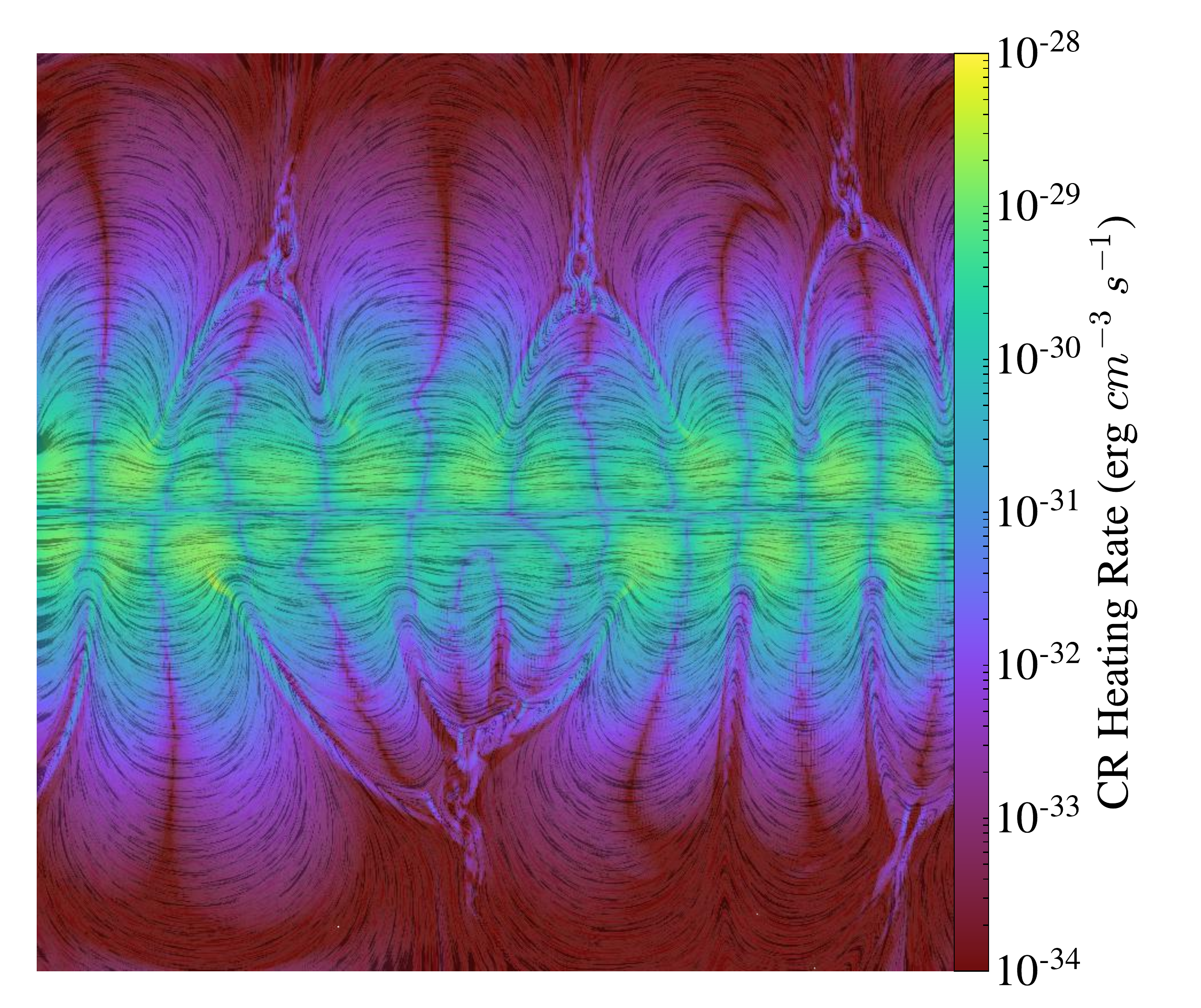}
\includegraphics[width=0.32\textwidth]{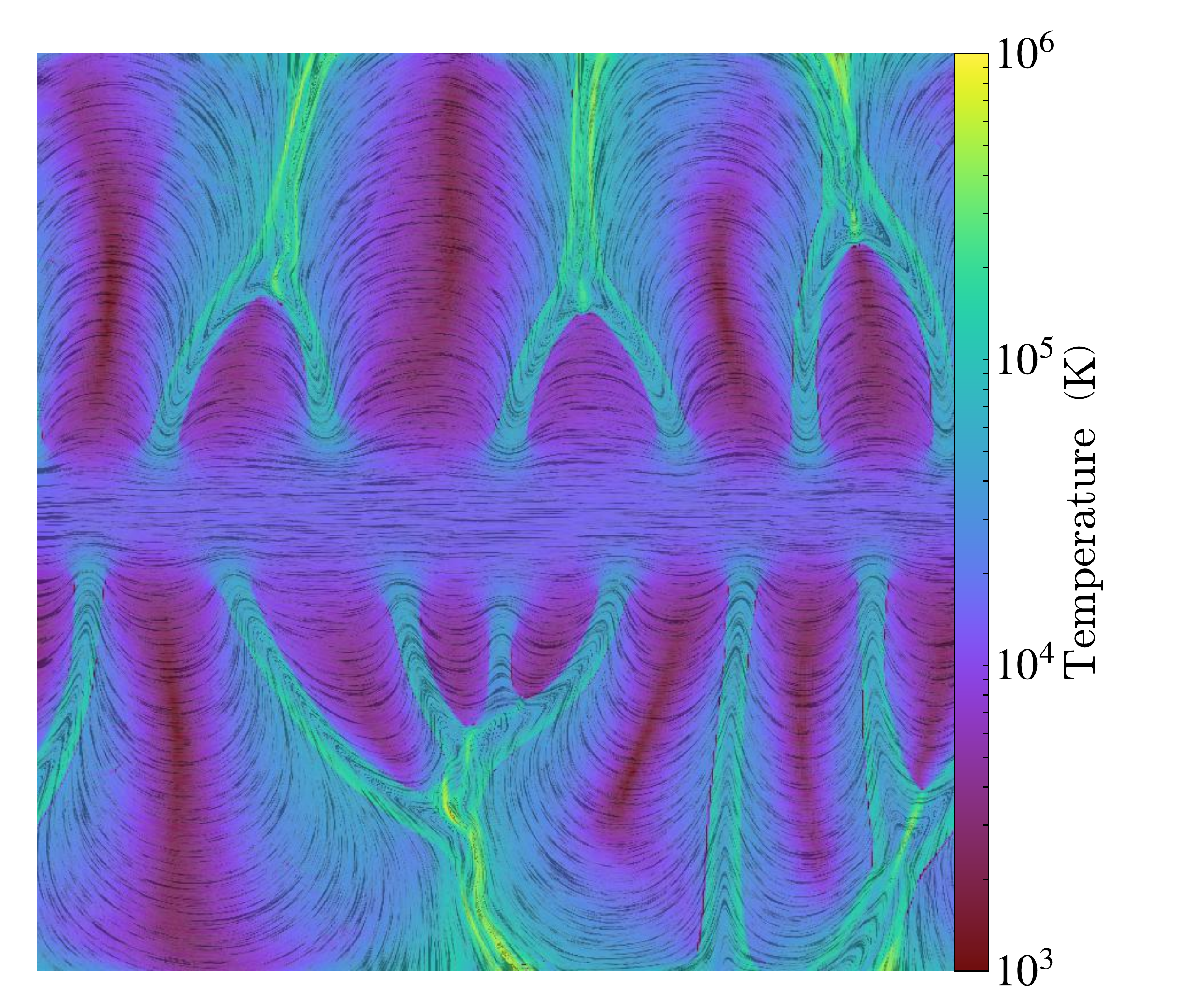}
\caption{An example of an evolved Parker instability with cosmic ray streaming and \cite{RodriguesParker2015} parameters. Here, we set $m = 2$ and $c = 1$ and don't include cooling. Line integral convolutions of the magnetic field are overplotted. The magnetic loops that are often associated with the Parker instability are seen. In the density plot, we also see the gas falling down in the valleys of the magnetic field as one would expect for the instability. Along the resulting Parker loops, compressive heating raises the temperature far above the equilibrium temperature of $\approx 10^{4}$ K, while cosmic ray heating increases the diffuse gas temperature to beyond $10^{5}$ K. This is especially true high above the disk where the heating \emph{per particle} is greatest.}
\label{fig:m2_c1_Rodrigues_slices}
\end{figure*}

The typical highly-evolved state of a Parker unstable system is shown in Figure \ref{fig:m2_c1_Rodrigues_slices} for \cite{RodriguesParker2015} parameters with cosmic ray streaming, no cooling, $m = 2$, and $c = 1$. Bearing in mind that the Parker instability is driven by the gravitational potential energy of gas supported above its thermal scale height by magnetic and cosmic ray pressure, it is unsurprising that the gas is more heavily concentrated in the magnetic valleys, forming the tendril-like structures of higher density gas seen in the plots. This leads to a density increase in the midplane with displaced, buoyant gas ballooning outward. The magnetic field is advected with these buoyant loops, resulting in a magnetic pressure increase of a few orders of magnitude in regions a few kpc above the disk. As the system becomes more and more nonlinear, these tendrils/arches of gas come together to form the bigger loops that are also observed in these slice plots. This behavior may be abetted by anomalous (numerical) magnetic diffusion, but we cannot quantify this effect with our current simulations.  Cosmic rays, streaming along the field lines, will begin to move away from the midplane, heating the gas as they do this. Importantly, cosmic rays seem to heat the gas most efficiently in the regions between these loops where the gas density is lower. We will focus on this behavior in Section \ref{subsec:heatingUnstable}.

\begin{figure}[]
\centering
\includegraphics[width 
=0.35\textwidth]{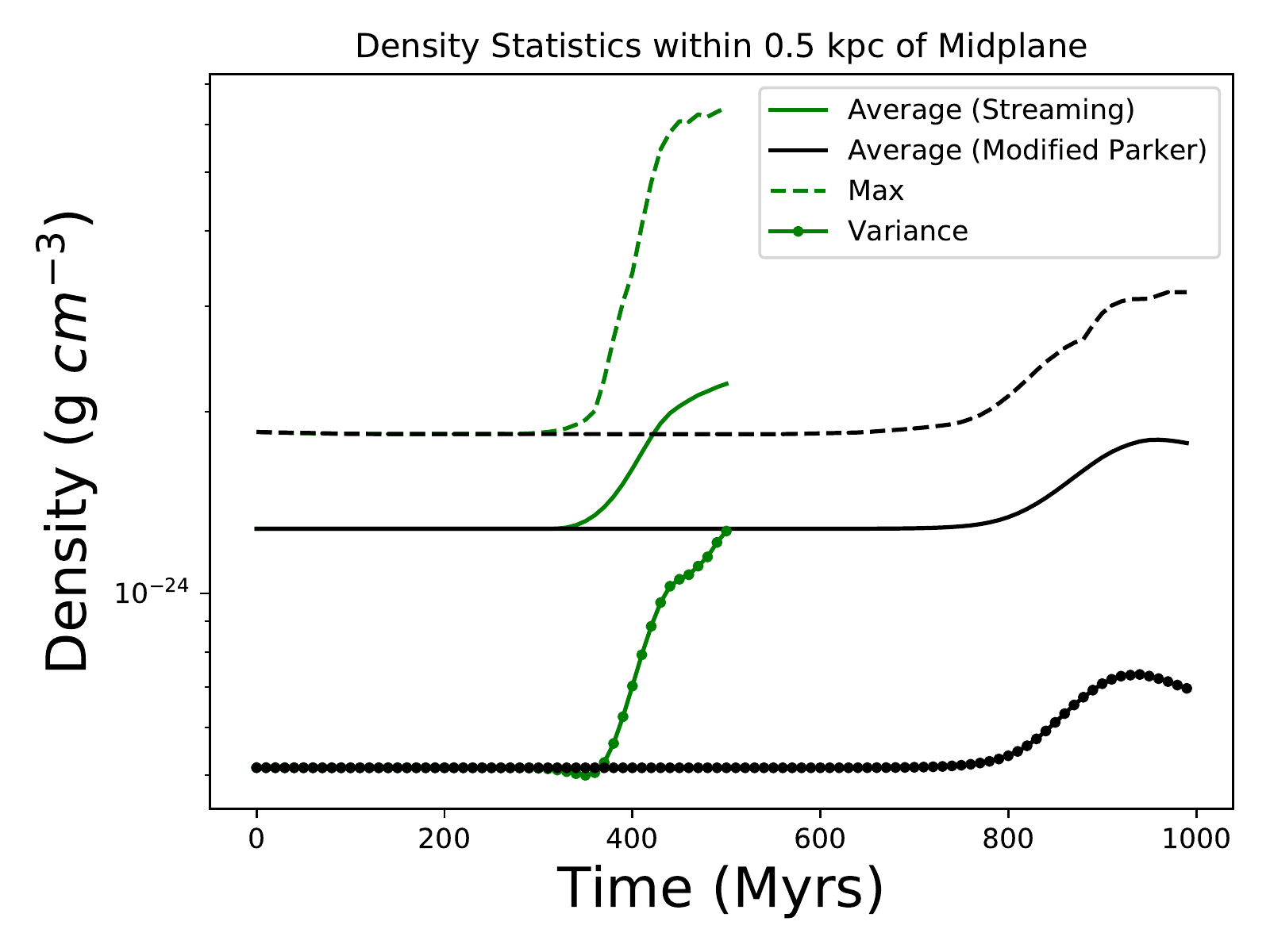}
\includegraphics[width =0.35\textwidth]{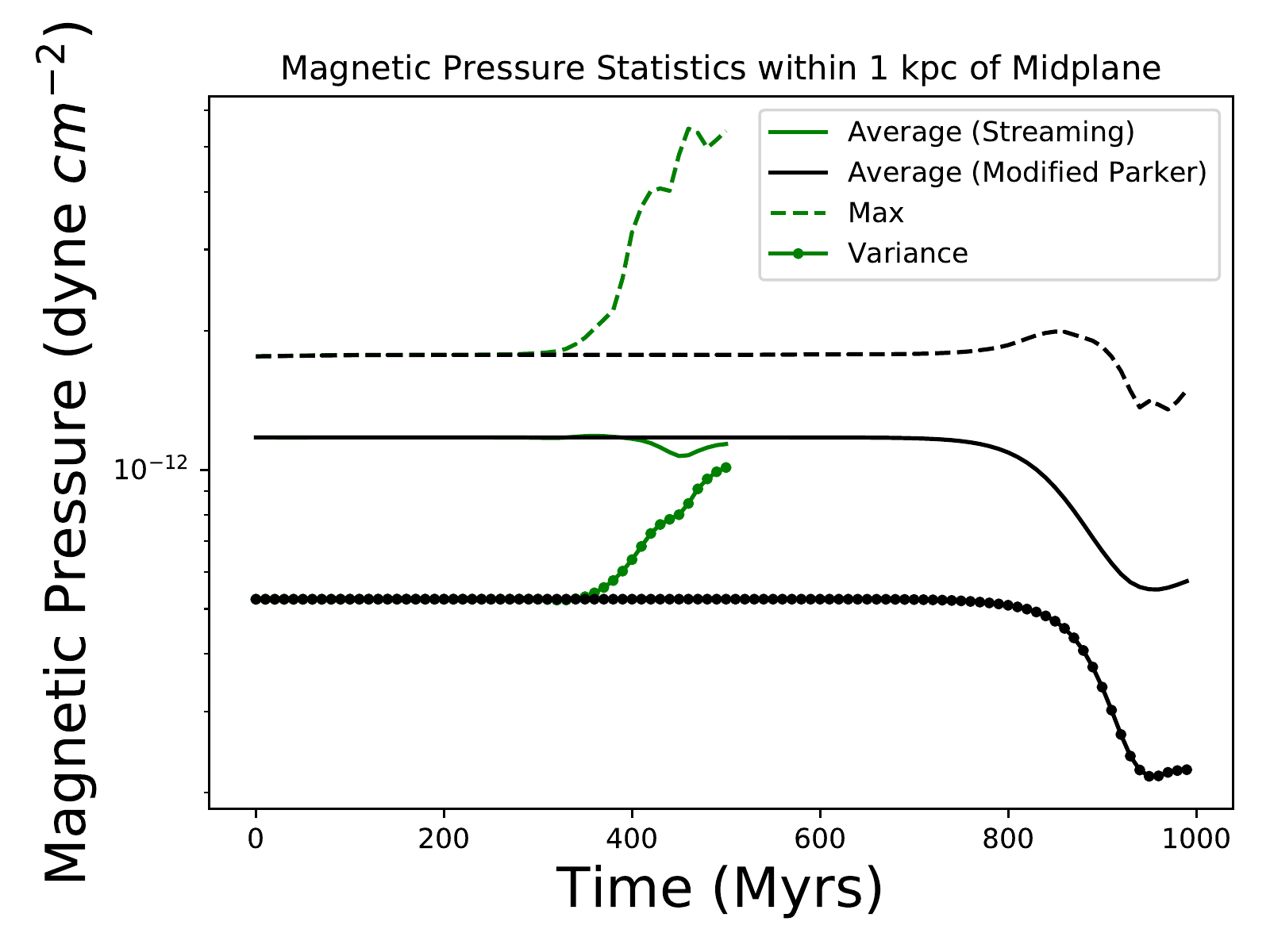}
\includegraphics[width =0.35\textwidth]{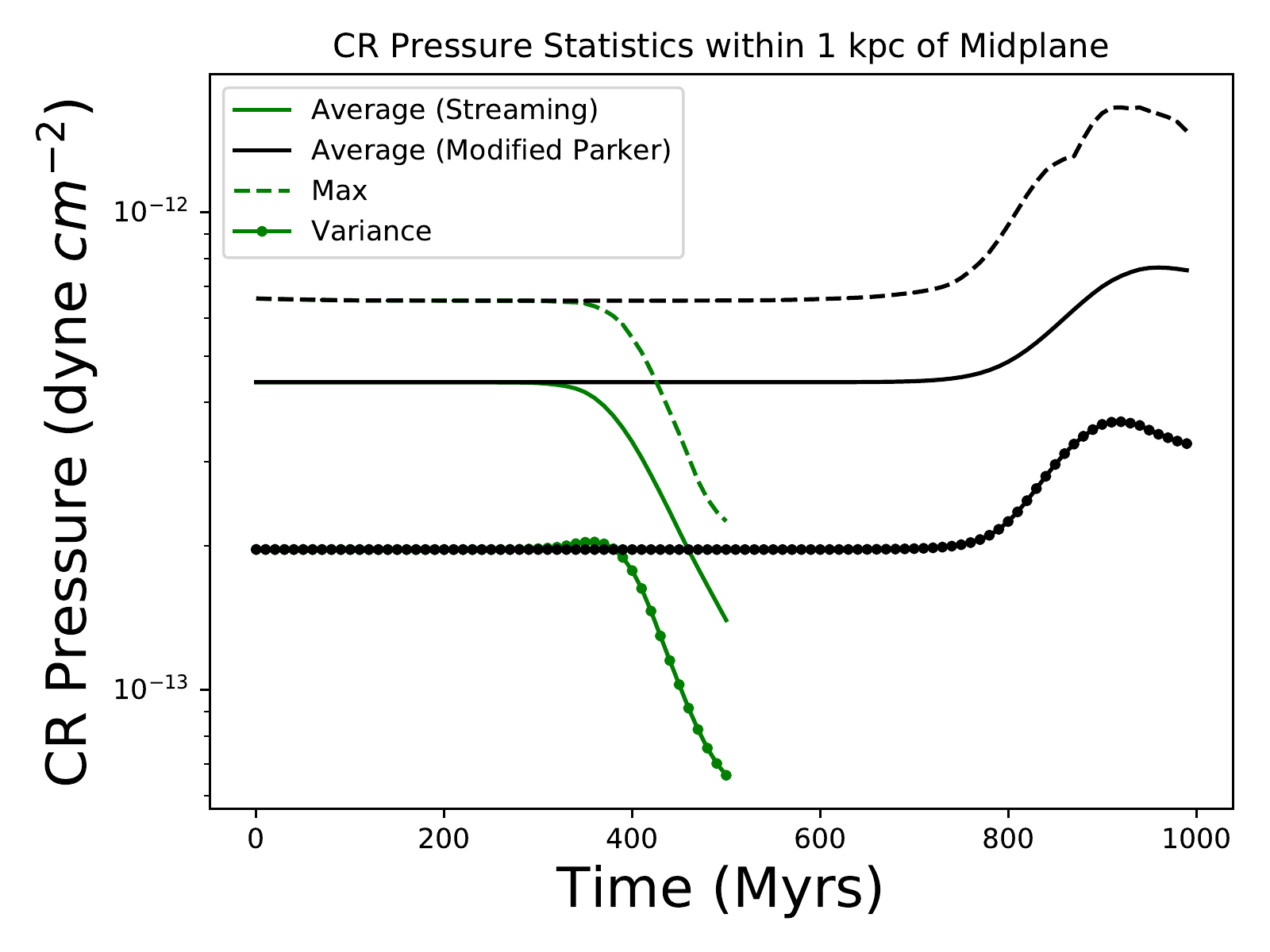}
\includegraphics[width =0.35\textwidth]{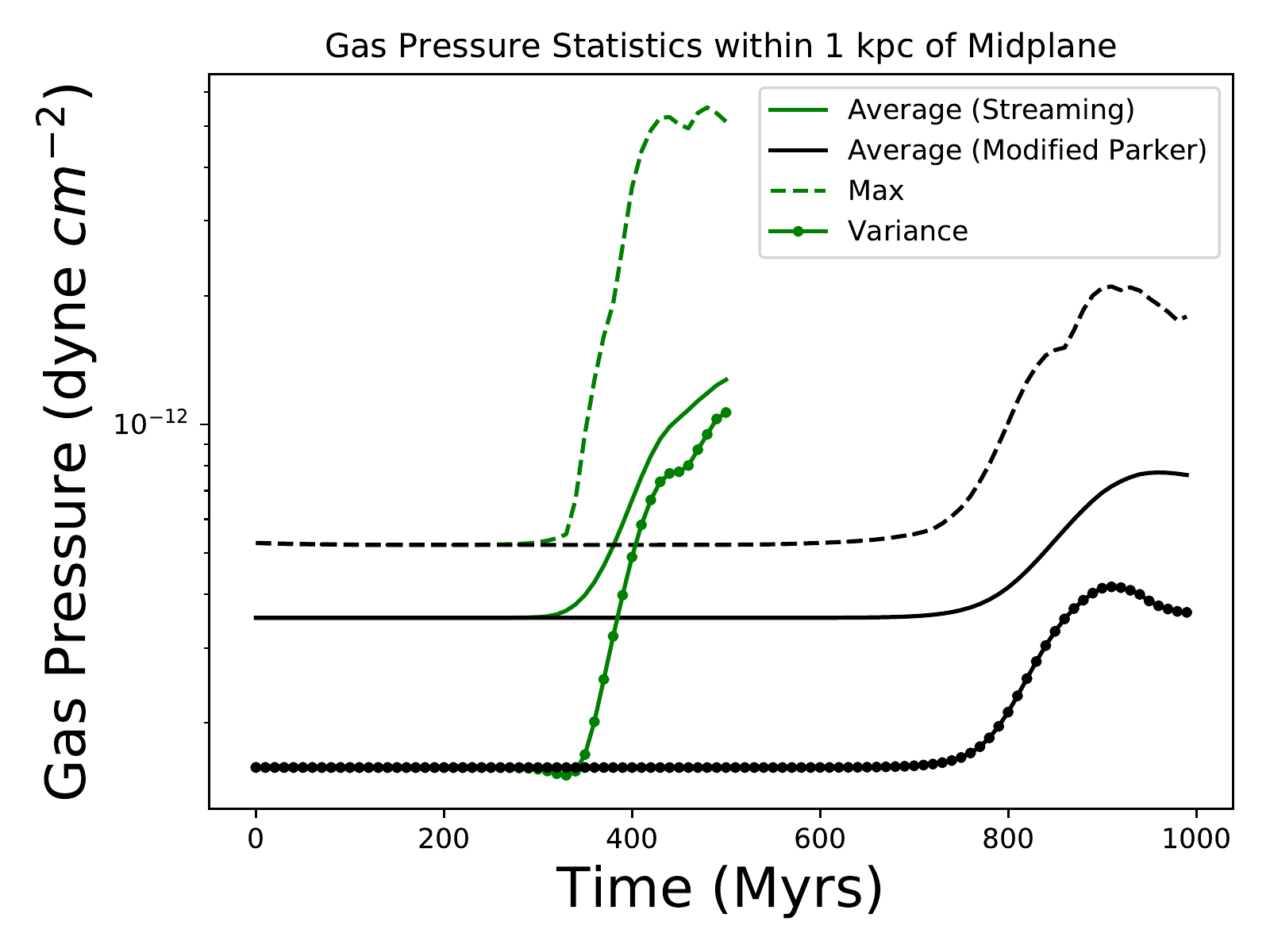}
\caption{Evolution of the mass-weighted density and pressures, within 1 kpc of the midplane, for Modified Parker simulations with and without cosmic ray streaming. Note that these simulations used Solar Neighborhood parameters but the qualitative results hold for \cite{RodriguesParker2015}. The density and gas pressures grow much more quickly and to higher values with streaming than they do without it. A huge difference is seen between the cosmic ray pressures as with streaming, they move out of the midplane giving more room for the gas to compress while when they just advect, they move with the gas and increase their own pressure in the midplane.}
\label{fig:solar_dens_Pres_evolution}
\end{figure}


As found by \cite{GizParker1993, KimParker1998} and \cite{RodriguesParker2015}, our system shows no preference for a particular type of symmetry. Instead, it appears to be a mixing of both of the 'even' and 'odd' modes. This lack of preferred symmetry is not due to our initial conditions. As outlined in \S\ref{sec:sim_method}, we apply a perturbation to velocity in the $y$ direction (the direction of gravity) that goes as a sine function. This initial perturbation therefore would seem to favor the antisymmetric system about the midplane, yet the symmetric modes still develop and mix with the antisymmetric ones, indicating that neither mode is dominant and both will arise in the system, regardless of the initial conditions.

In Figure \ref{fig:solar_dens_Pres_evolution}, we see the differences in ISM structure when streaming is added to our system. In general, we see that in Modified Parker with Streaming, the changes in all of our quantities happen at earlier times, due to the increased instability of this system compared with Modified Parker. The gas pressure and density, qualitatively, increase in both systems, but quantitatively, they increase much more sharply and to a larger value when streaming is added. This again reinforces that the system with cosmic ray streaming is more unstable as the gas compacts more easily into the valleys of the magnetic field as the cosmic rays stream away up the magnetic loops. An interesting comparison arises between the two systems with the magnetic pressure as well. In both systems, the average magnetic pressure slightly decreases as it is replaced by the gas pressure. However, we see a stark difference in the maximum magnetic pressure in the two systems. In the more stable system of Modified Parker, the maximum magnetic pressure roughly follows the trajectory of the average magnetic pressure within 1 kpc. However, with streaming, as the instability creates more dense, heavier pockets of gas, the magnetic field lines near the midplane that cannot lift away from the disk are compressed closer together and so the magnitude of the field sharply rises.

These two models differ most strongly in the cosmic ray pressure statistics. In Modified Parker, the cosmic rays only advect with the gas and, therefore, the cosmic ray pressure increases proportionally to the gas pressure. However, in Modified Parker with Streaming, we see that the cosmic ray pressure actually decreases near the midplane of the disk. Since the cosmic rays are no longer locked to the gas, they are able to move out of the way of the falling gas and follow the magnetic loops up away from the disk. This gives more room in which the gas can be compressed, possibly providing another reason for why Modified Parker with Streaming is so much more unstable than Modified Parker in the linear theory (in addition to cosmic ray heating). 

\subsection{Diffusion vs Streaming}
\label{subsec:diff_vs_streaming}

In an attempt to test our intuitions from HZ18 and further understand why cosmic ray streaming makes the system more unstable, we also decided to run simulations with cosmic ray diffusion and compare the two different models of transport. The linear theory of the Parker instability with cosmic ray diffusion can be found in the work of \cite{ryuparker2003}, \cite{kuwabaraparker2004}, and \cite{kuwabaraparker2006} but is not included in this work. In general, they all found that higher parallel diffusion coefficients lead to a more unstable system than systems with lower diffusion coefficients. For reference, Classic Parker is equivalent to $\kappa_\parallel \rightarrow \infty$ and Modified Parker is equivalent to $\kappa_\parallel = 0$ (Here ``parallel" means with respect to the magnetic field), so the results of our linear stability analysis reinforce their findings. 
In our system, we add diffusion to our Modified Parker setup with $\gamma_g = 5/3$ and $\gamma_c=4/3$ and assume only a parallel diffusion coefficient, $\kappa_\parallel$. 

\begin{figure}[]
\centering
\includegraphics[width = 0.49\textwidth]{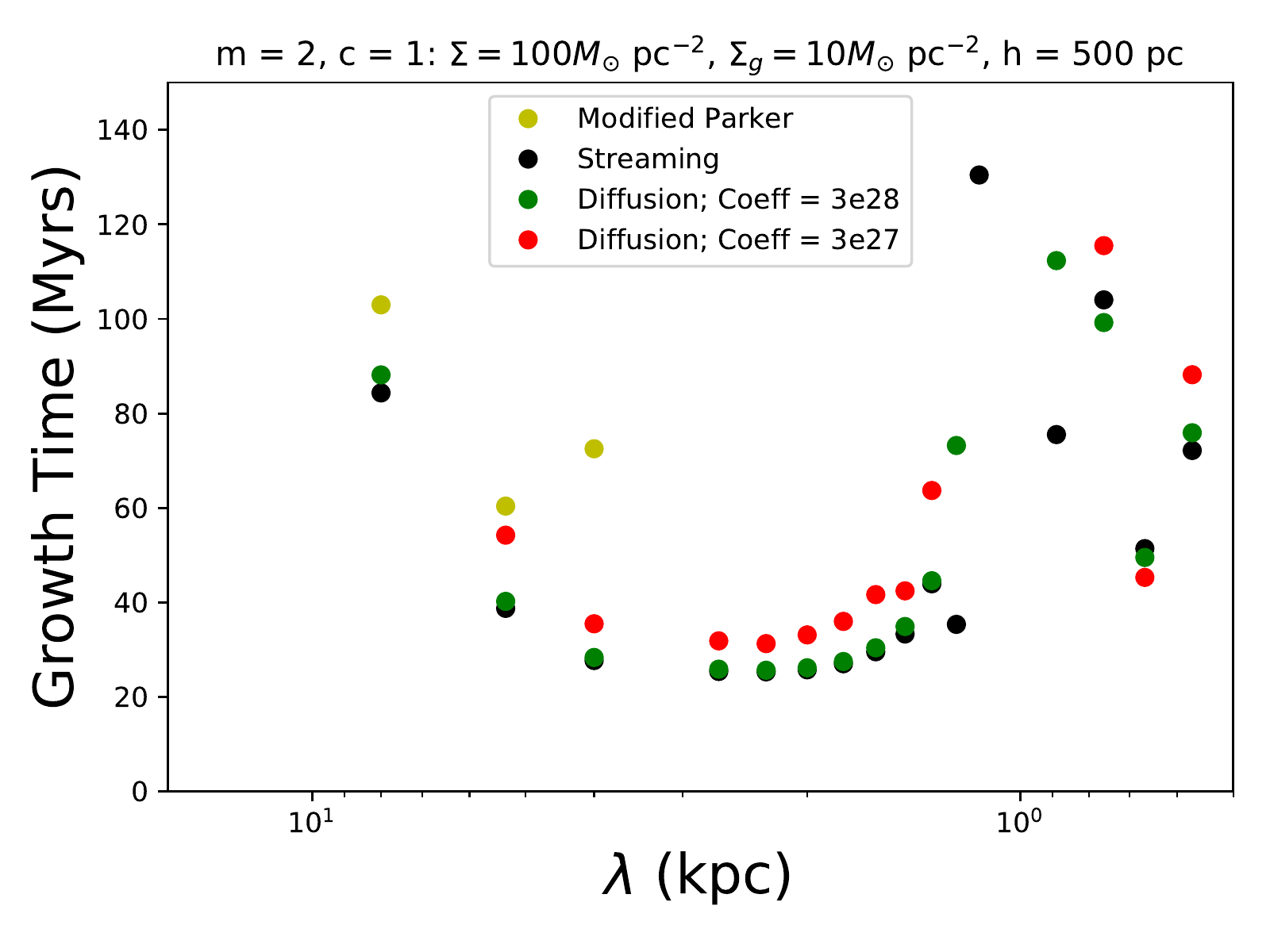}
\caption{m = 2, c = 1 growth time vs wavelength for the galaxy parameters from \cite{RodriguesParker2015} using different transport models. Here, we find the somewhat surprising result that diffusion growth times, even for a relatively small diffusion coefficient of $3 \times 10^{27}$ $\rm cm^{2}$ $\rm s^{-1}$ about $0.1$ of the canonical Milky Way value), are comparable to the streaming growth times.}
\label{fig:Rodrigues_diffusion_streaming}
\end{figure}

In Figure \ref{fig:Rodrigues_diffusion_streaming}, we plot the simulated growth rates of the diffusion and the streaming cases for $m = 2$, $c = 1$ and \cite{RodriguesParker2015} parameters.

We find with diffusion coefficients of $\rm 3 \times 10^{27}$ $\rm cm^{2}$ $\rm s^{-1}$ and $\rm 3 \times 10^{28}$ $\rm cm^{2}$ $\rm s^{-1}$ that the growth rates qualitatively match the results from \cite{ryuparker2003} and \cite{kuwabaraparker2004}. Lower diffusion coefficients allow stronger cosmic ray - -  gas coupling, therefore stabilizing the system. This reaches the highest level of stability in Modified Parker, where $\kappa_\parallel = 0$. Conversely, for our larger diffusion coefficient of $\rm 3 \times 10^{28}$ $\rm cm^{2}$ $\rm s^{-1}$, we find the instability almost matches the large growth rates found with cosmic ray streaming.

Along with the cosmic ray pressure differences shown in Figure \ref{fig:solar_dens_Pres_evolution}, this seems to indicate that the cosmic ray heating itself is not solely responsible for the increased instability we see in Modified Parker with Streaming. Since diffusion appears to be just as unstable as streaming for reasonable values of the diffusion coefficient, we must also conclude that even just the simple transport of the cosmic rays along the field lines and out of the magnetic valleys also helps destabilize the system. 

\begin{figure}[]
\centering
\includegraphics[width = 0.38\textwidth]{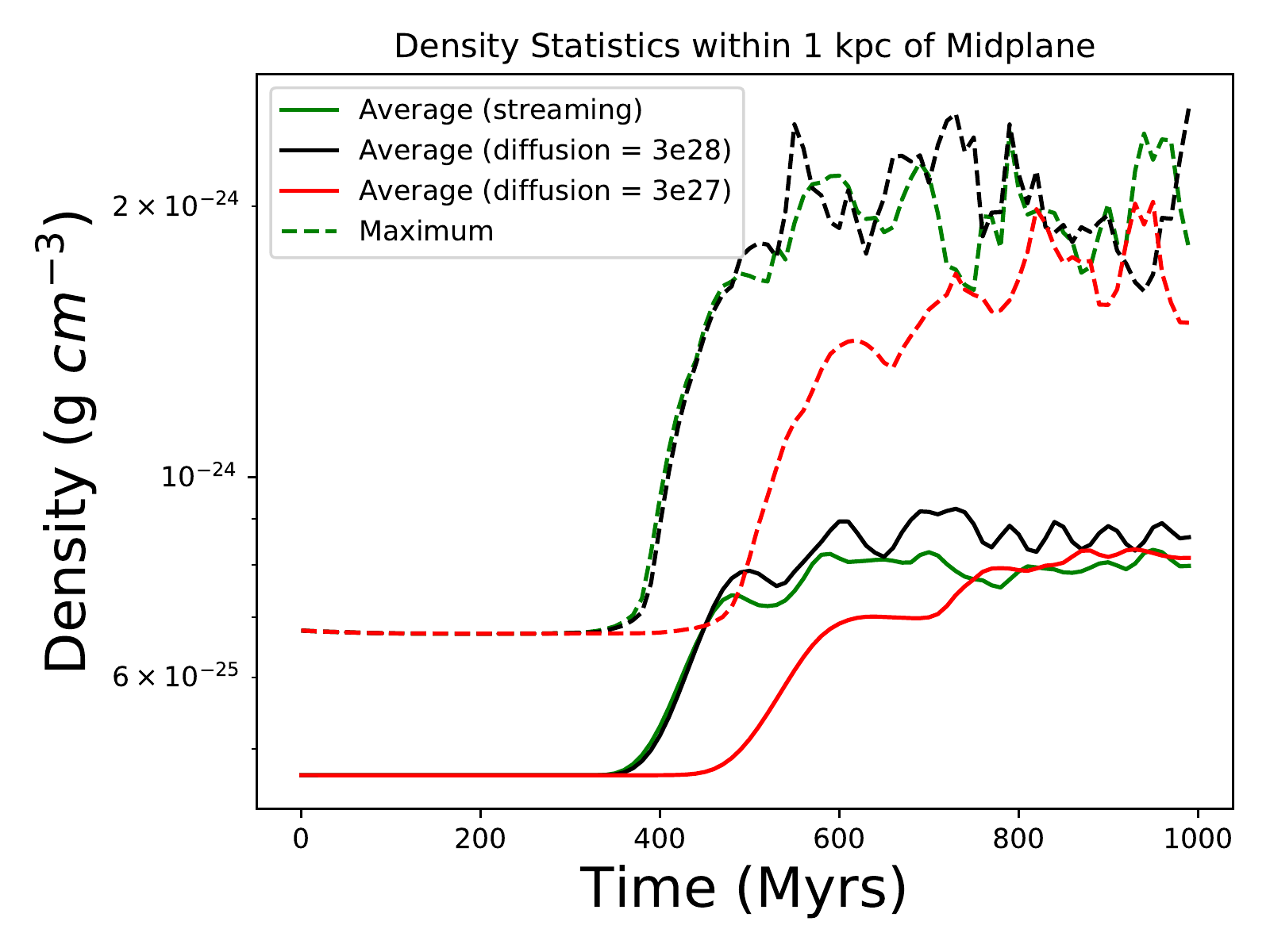}
\includegraphics[width = 0.38\textwidth]{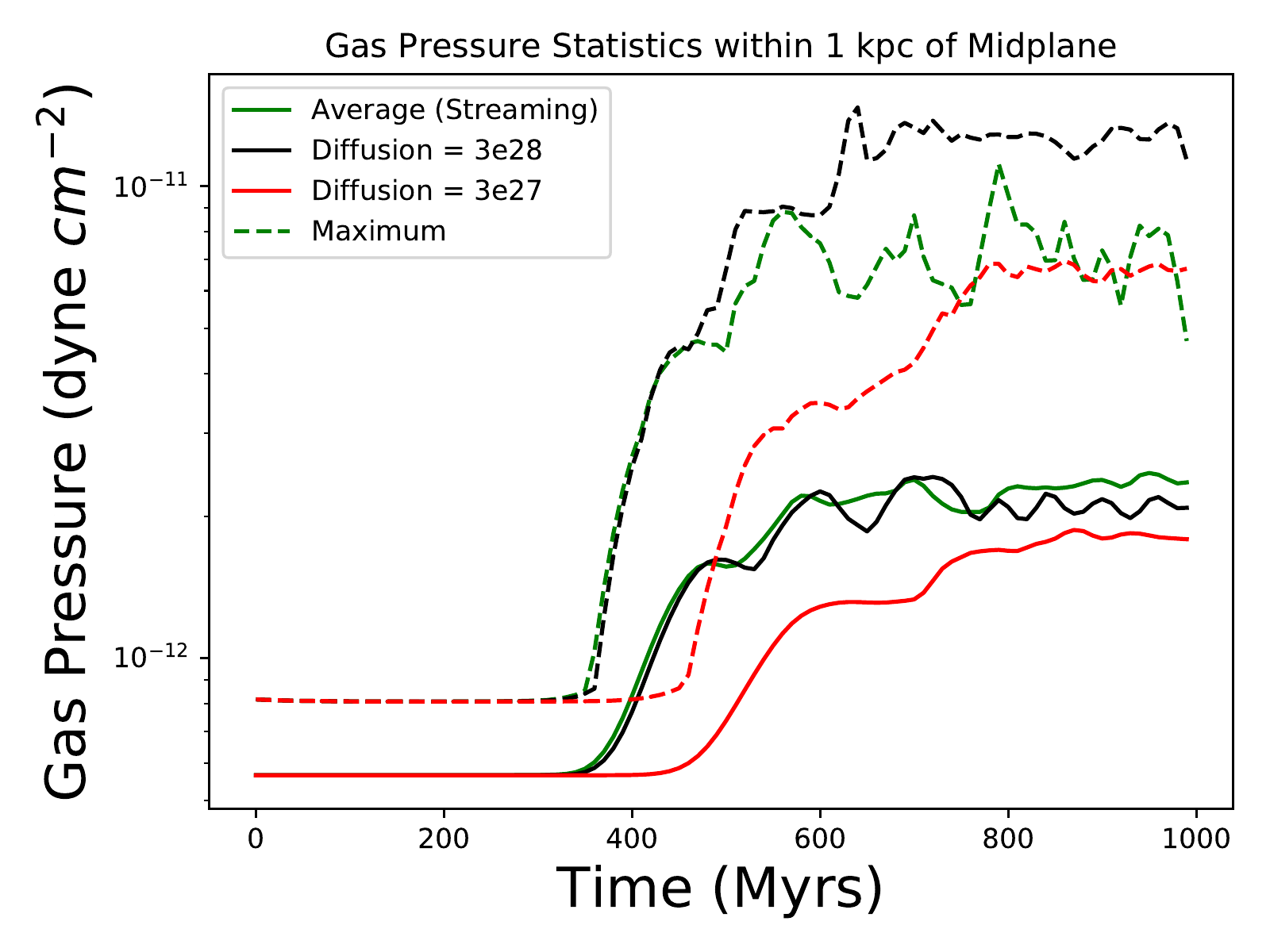}
\includegraphics[width = 0.38\textwidth]{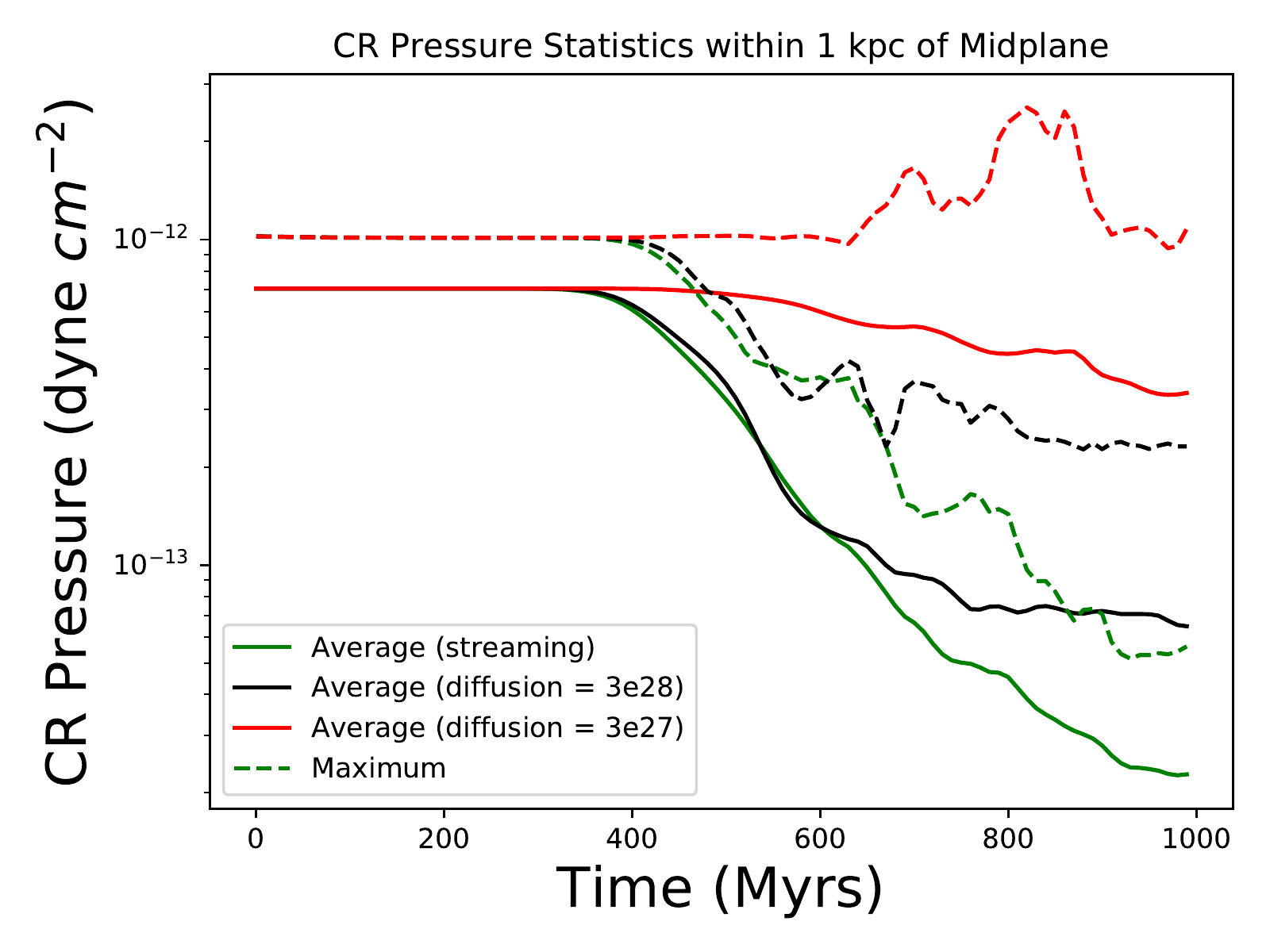}
\includegraphics[width = 0.38\textwidth]{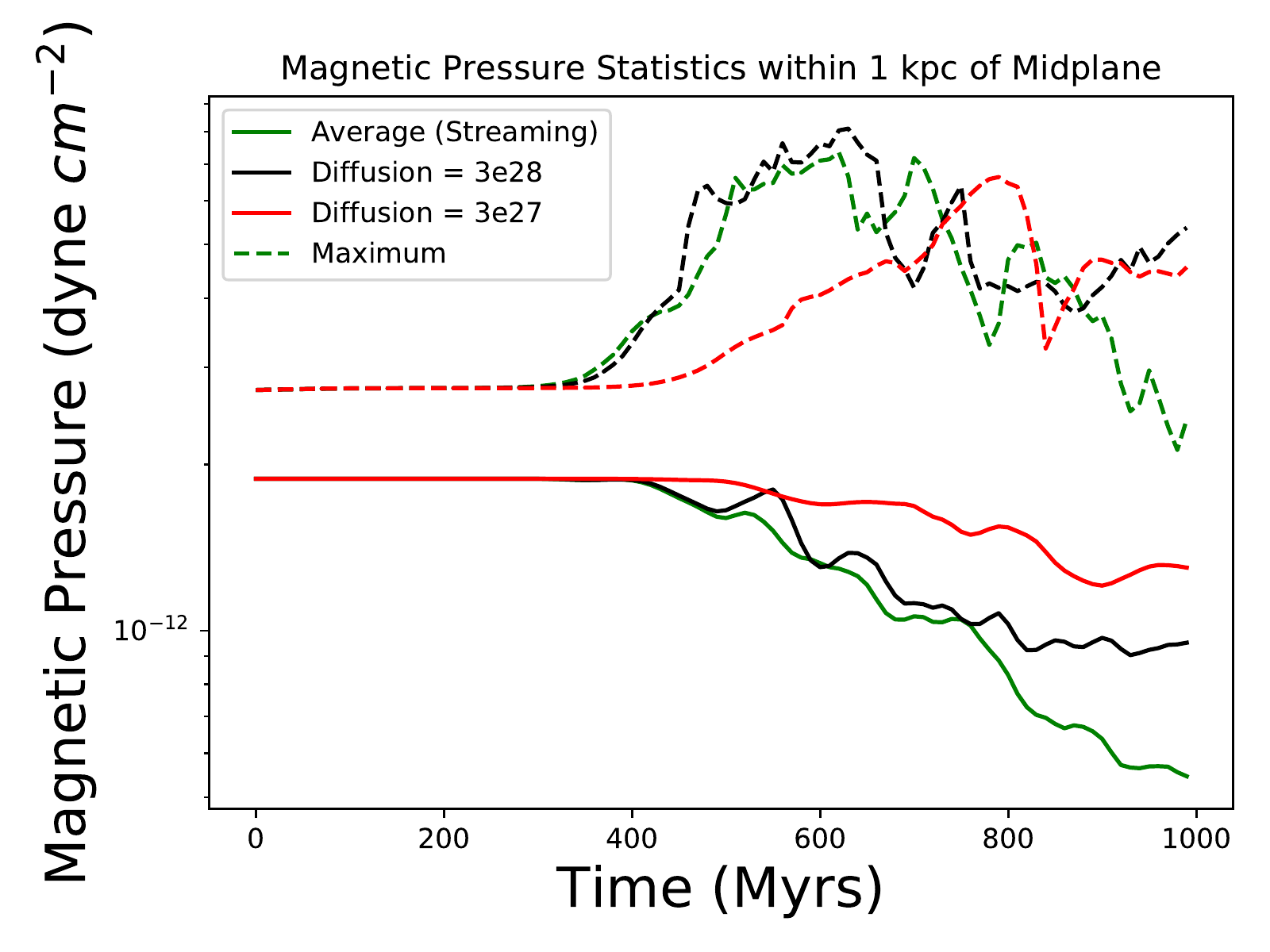}
\caption{Comparison of streaming vs diffusion (coefficients of $3 \times 10^{27}$ and $3 \times 10^{28}$ $\rm cm^{2}$ $\rm s^{-1}$) for the m = 2, c = 1 simulation with \citealt{RodriguesParker2015} parameters, showing the mass-weighted average and maximum of density, gas pressure, CR pressure, and magnetic pressure within the first kpc above and below the disk. Note the change in axis limits between the three pressure plots. }
\label{fig:Rodrigues_diffusion_streaming_comp}
\end{figure}

For further comparison between the two models, we plot the evolution of the density and pressures in Figure \ref{fig:Rodrigues_diffusion_streaming_comp} (these plots are similar to Figure \ref{fig:solar_dens_Pres_evolution} shown in \S\ref{subsec:evol_nonlin}) and again look at the time average of these quantities within 1 kpc of the midplane. We see that all three transport models follow the same trend for all four quantities. The densities and gas pressures increase for all models but more sharply increase with streaming and the larger diffusion coefficient. The magnetic pressure is again similar between the three as the overall average decreases as the loops move away from the midplane but the maximum increases for all three as the increase in gas density and pressure pushes together the field lines at the midplane and thus increase the field. 

The only difference between the three cases is shown in the cosmic ray pressure  statistics. For the lowest diffusion coefficient of $3 \times 10^{27}$ cm$^2$/s, we see that the cosmic ray pressure barely changes throughout the simulation, only slightly dipping near 600 Myrs. We also find that the larger diffusion coefficient and streaming look similar, both falling by about an order of magnitude at 600 Myrs. 

However, as the system becomes nonlinear past this point, the two lines diverge for the first time as the larger diffusion coefficient begins to asymptote at its lower pressure while streaming continues to fall by approximately another order of magnitude. 

However, around this point in the simulation (usually around 800-1000 Myrs), we observe reconnection events driven by numerical resistivity, causing the previously connected magnetic loops to tear into magnetic islands. Because FLASH conserves total energy, the resulting dissipation in magnetic energy is compensated by an increase in thermal energy (i.e. reconnection heating). This has been observed in past work, as well (for instance, see Hanasz et al. 2002). While the change in magnetic topology affects the subsequent evolution of the system, especially cosmic ray streaming and diffusion parallel to $\mathbf{B}$, we find that reconnection heating is very localized (see Appendix B) and plays a sub-dominant role compared to cosmic ray heating. Comparing the streaming and diffusion $= 3 \times 10^{28} cm^{2}/s$ simulations, we observe this reconnection to happen in almost identical places at identical times. Because the subsequent change in magnetic topology is consistent between these transport models, reconnection does not confuse our interpretation: cosmic ray streaming appears to more efficiently transport cosmic rays away from the midplane than other transport models included in this paper. This seems to explain why the streaming model has slightly higher growth rates than the diffusion models but also appears to make cosmic ray streaming the fastest process by which cosmic rays may escape from the galaxy into the CGM and IGM.


\begin{figure}
\centering
\includegraphics[width = 0.49\textwidth]{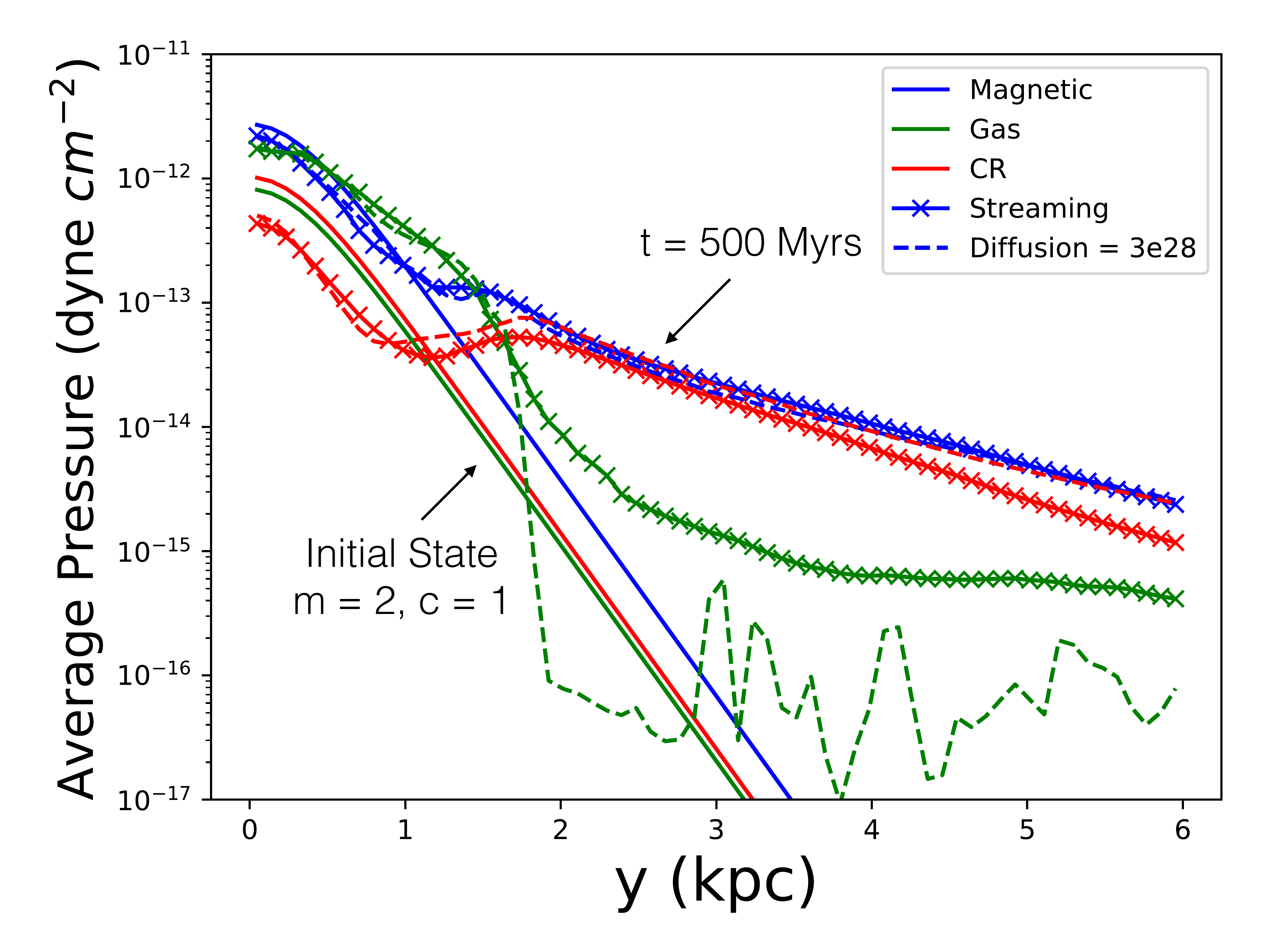}
\caption{Average magnetic, gas, and cosmic ray pressure as a function of height above the midplane in our m = 2, c = 1 2D simulations. Near the midplane, gas pressure increases at the expense of cosmic ray pressure, which diffuses or streams to greater heights. For cosmic ray and magnetic pressure at 500 Myrs, the diffusion and streaming cases line up well; however, cosmic ray heating results in a large increase in gas pressure beyond 2 kpc compared to the diffusion run. }
\label{fig:pressure_profiles}
\end{figure}

\begin{figure}
\centering
\includegraphics[width = 0.49\textwidth]{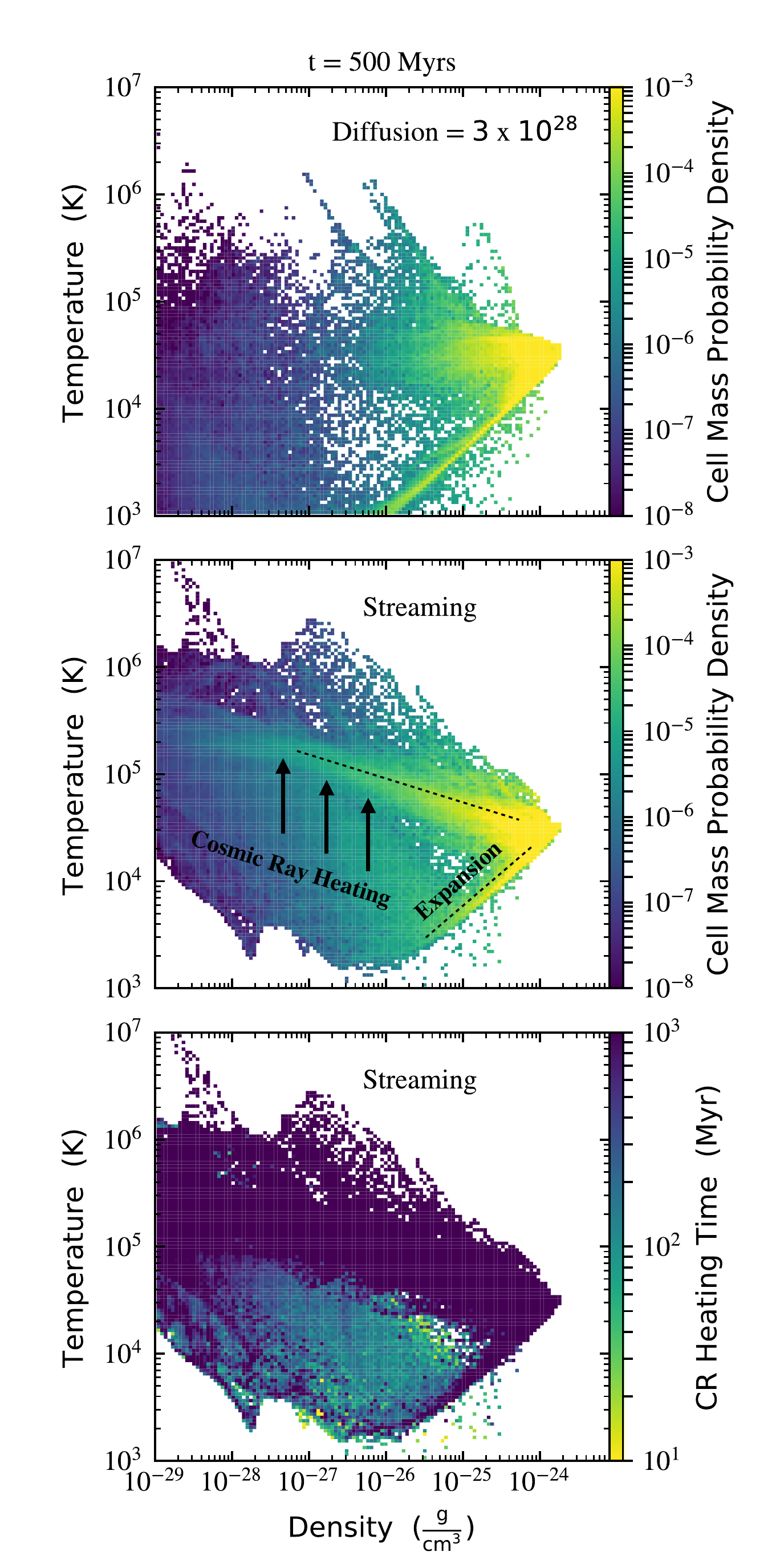}
\caption{Phase plots of diffusion and streaming simulations (without cooling) at $t = 500$ Myrs. We see that diffusion and streaming, despite almost identical growth rates, develop different phase structure, especially for the adiabatically expanding low-density gas present a few kpc above the midplane. The bottom panel shows the cosmic ray heating time, defined as $E_{\rm gas}/Q_{\rm CR}$. This heating time is shortest in the diffuse, low-temperature gas, causing it to shift up to higher temperatures. The diffusion simulation, on the other hand, maintains a reservoir of cold, low-density gas.}
\label{fig:phase_plots_diff_vs_streaming}
\end{figure}

\subsection{The Role of Cosmic Ray Heating}
\label{subsec:heatingUnstable}


Given the minute differences presented so far between the streaming and diffusion 
cases, it is reasonable to ask whether the cosmic ray heating term actually plays a role in the instability. In the linear regime, our simulations show only very small cosmic ray pressure perturbations. The cosmic ray heating term $\propto v_{A} \cdot \nabla P_{cr}$ is negligible then; however, the self-confinement picture tells us that cosmic rays spread out along magnetic field lines at the local Alfven speed, regardless of how small $\nabla P_{cr}$ is. Because this Alfvenic streaming acts similarly to diffusion, the two transport models lead to very similar linear growth stages.

In the nonlinear stage, however, as $\nabla P_{cr}$ becomes significant the heating term becomes very important. In Figure \ref{fig:pressure_profiles}, we plot the mass-weighted average pressures for our streaming and diffusion ($\kappa = 3 \times 10^{28}$) simulations at late times. The initial state is shown for comparison. Cosmic ray transport, in both cases, leads to a much larger cosmic ray scale height at $t = 500$ Myrs, while the average gas pressure increases near the midplane. The magnetic pressure, similar to cosmic ray pressure, increases dramatically at heights of a few kpc above the disk due to advection with the rising gas loops. It has been noted previously \citep{ParkerDynamo1992, HanaszParkerDynamo1993,HanaszParkerDynamo1997,hanaszreconnection2002} that this advection, combined with the Coriolis force, can stretch and twist magnetic fields, driving a fast magnetic dynamo. Including streaming, which expedites the instability, reinforces this possibility.


While the cosmic ray and magnetic pressure profiles overlap pretty closely between the diffusion and streaming runs, the gas pressure shows stark differences. In the streaming simulation, the gas pressure is a factor of 5-10 higher than the diffusion case at heights beyond $\approx 2$ kpc. We attribute this directly to cosmic ray heating in this cosmic ray dominated, extraplanar diffuse ionized gas (eDIG) medium. 


The effects of this heating are even more apparent when looking at the phase of the eDIG gas, which shows big differences between diffusion and streaming transport models. In Figure \ref{fig:phase_plots_diff_vs_streaming}, we show a color map of the mass fraction (cell mass probability) with axes of temperature and density. At time $t = 0$ (not shown), this plot is simply a horizontal line at constant temperature. In the nonlinear regime, however, the gas has formed multiple phases. Two sharp lines forming a ``v"-shape are present in each plot. The line extending to low densities and temperatures is formed by adiabatic expansion of the rising bubbles, while the line to low densities and high temperatures shows the pile-up of compressed (heated) gas extending from the magnetic valleys to a few scale heights above the disk. We see for diffusion that the distribution function is pretty evenly distributed around the peak at about $10^{-24}$ g/cm$^3$ and a few $\times 10^4$ K. It does slightly favor the lower temperature side of the plot, especially in the very low density limit. However, with streaming, we find that there is essentially no portion of the mass at low ($\rho$, T), while much more gas now exists above $10^4$ K. The bottom panel of Figure \ref{fig:phase_plots_diff_vs_streaming}, shows that the cosmic ray heating time, defined as $E_{\rm gas}/Q_{\rm CR}$ (where $Q_{CR}$ is the cosmic ray heating rate), is lowest precisely in the low ($\rho$, T) region corresponding to the expanding gas within Parker loops. Note that the cosmic ray heating is calculated in the same way as in the MHD equations where $Q_{CR} = \vert v_A \cdot \del P_{CR}\vert$.

\begin{figure}[]
\centering
\includegraphics[width=0.42\textwidth]{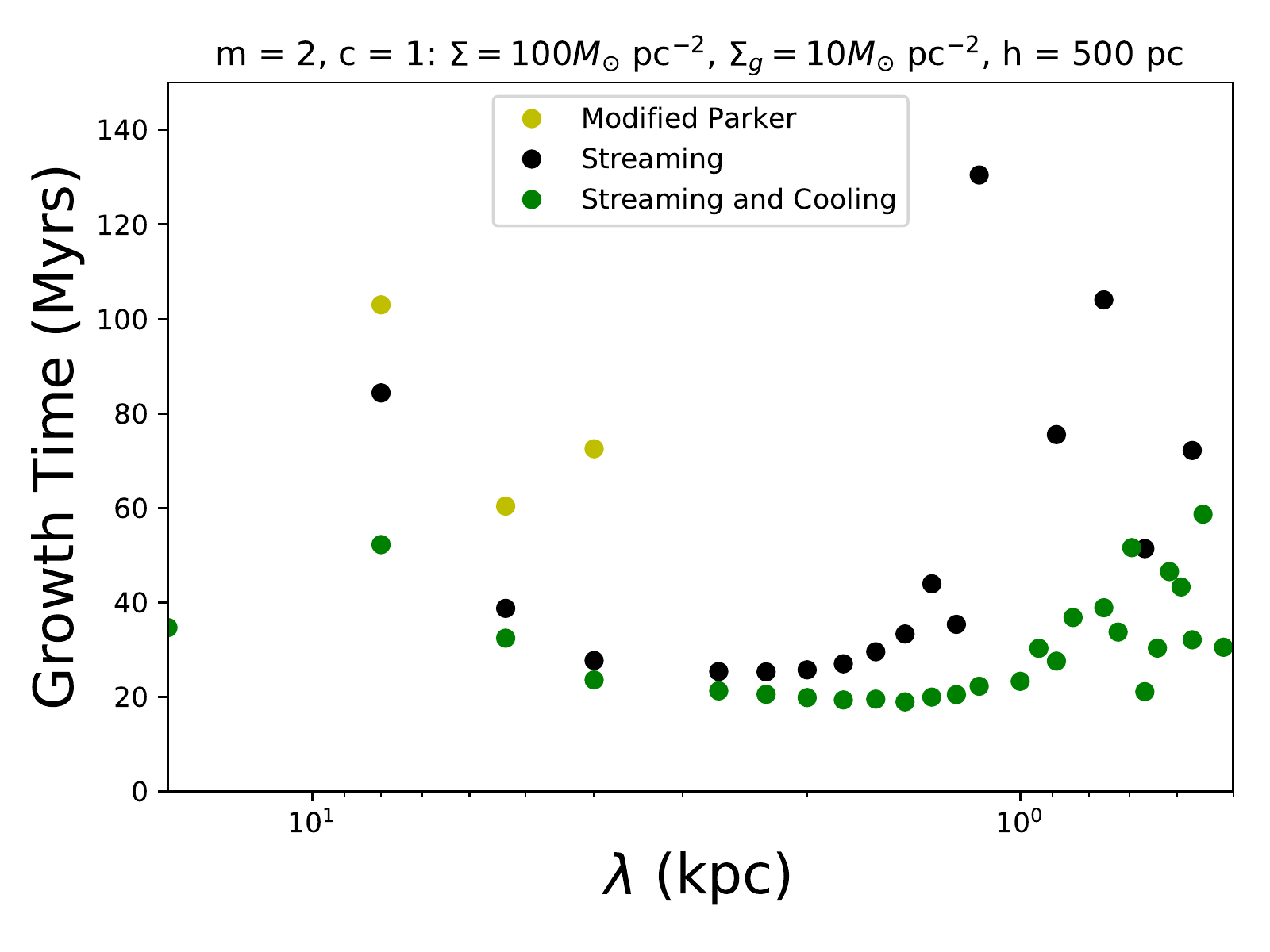}
\caption{A plot of the growth times of our 2D simulations without streaming and cooling, with streaming, and with streaming and cooling. We see that contrary to the linear stability analysis, the streaming and cooling system seems more unstable than if cooling was not included, despite the initial temperature being on the high side of $13000K$. However, as expected, implementing cosmic ray streaming is always more unstable than with other models of cosmic ray transport.}
\label{fig:streaming_and_cooling_growthTimes}
\end{figure}

\begin{figure*}[]
\centering
\includegraphics[width =0.32\textwidth]{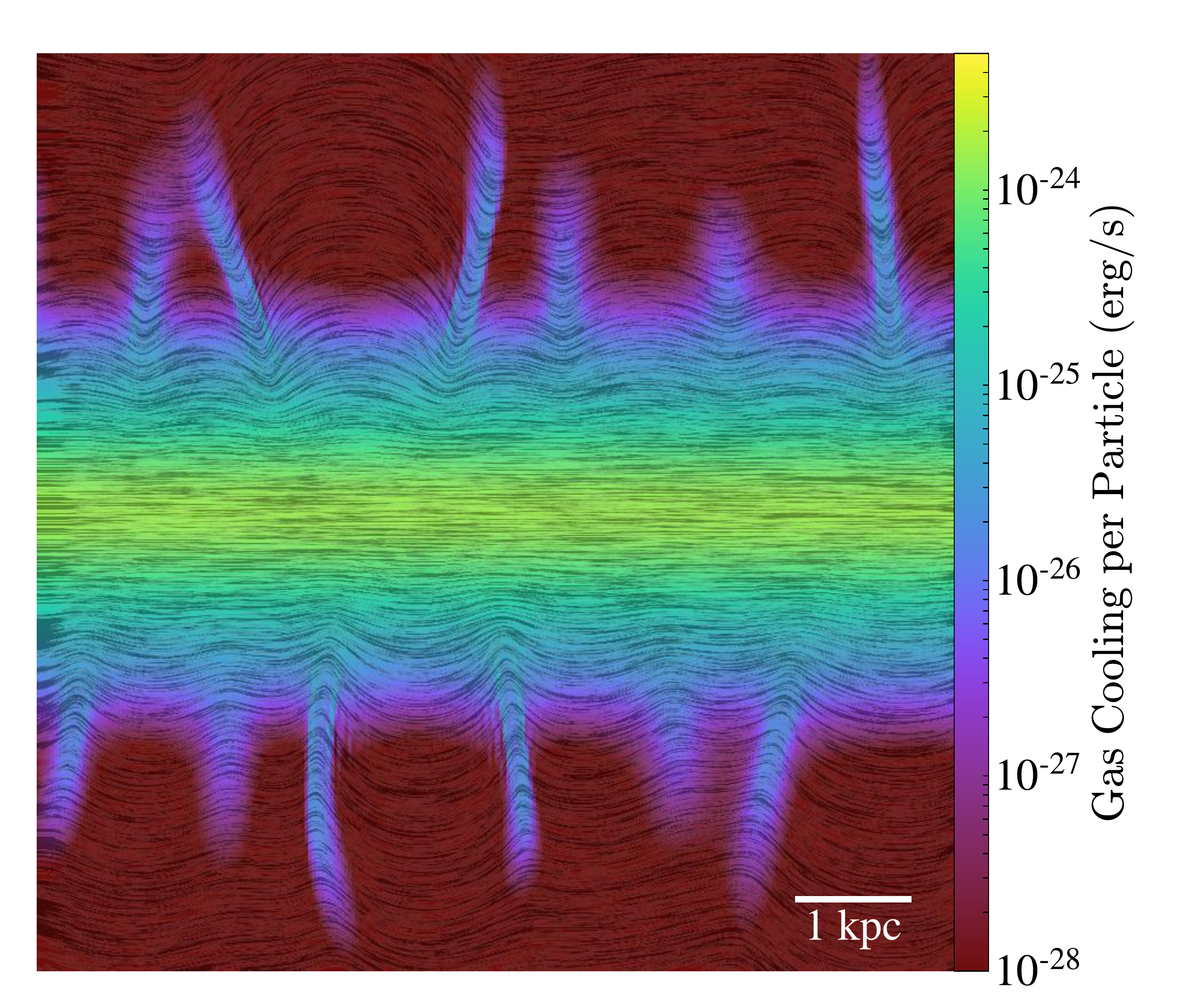}
\includegraphics[width =0.32\textwidth]{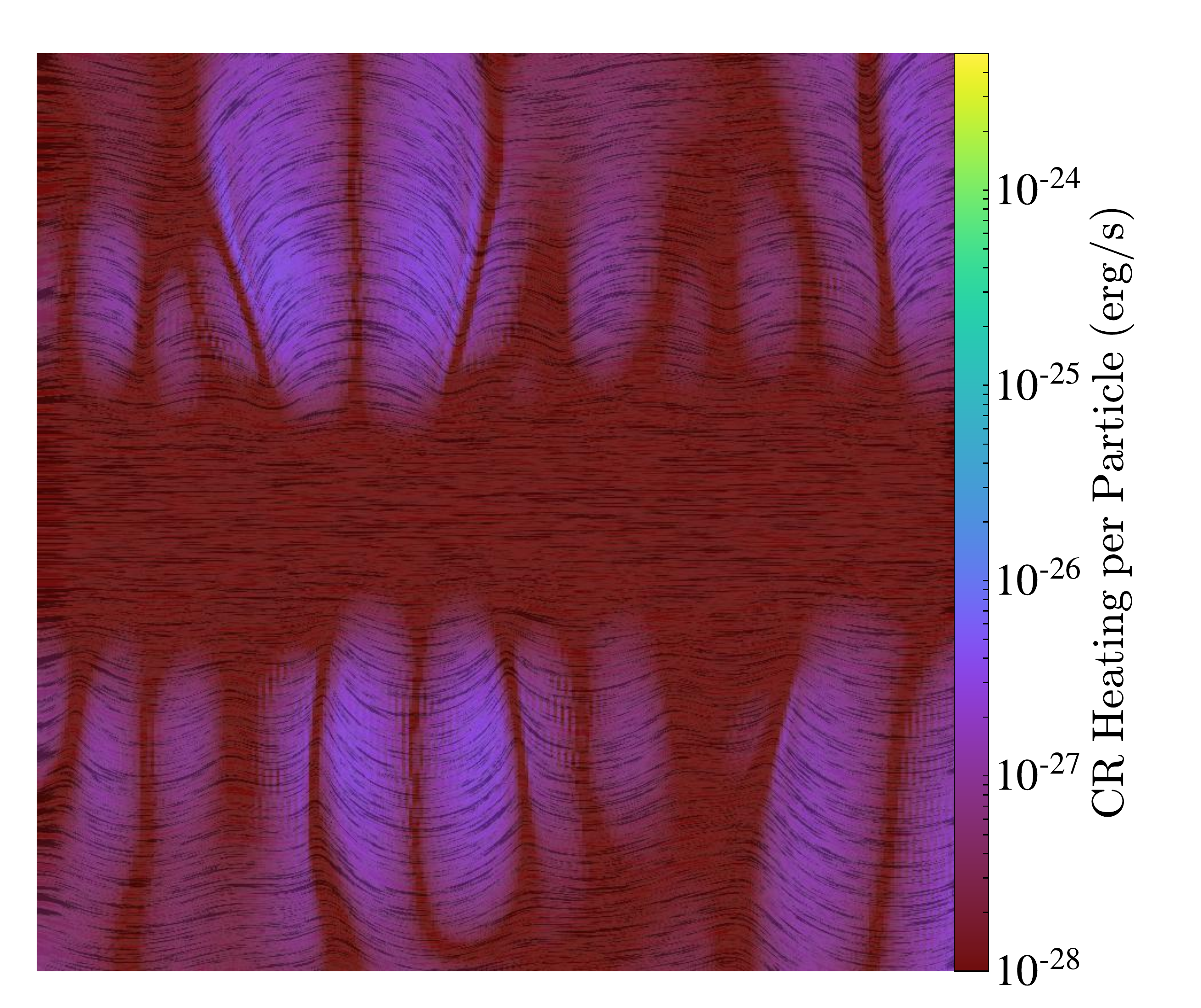}
\includegraphics[width =0.32\textwidth]{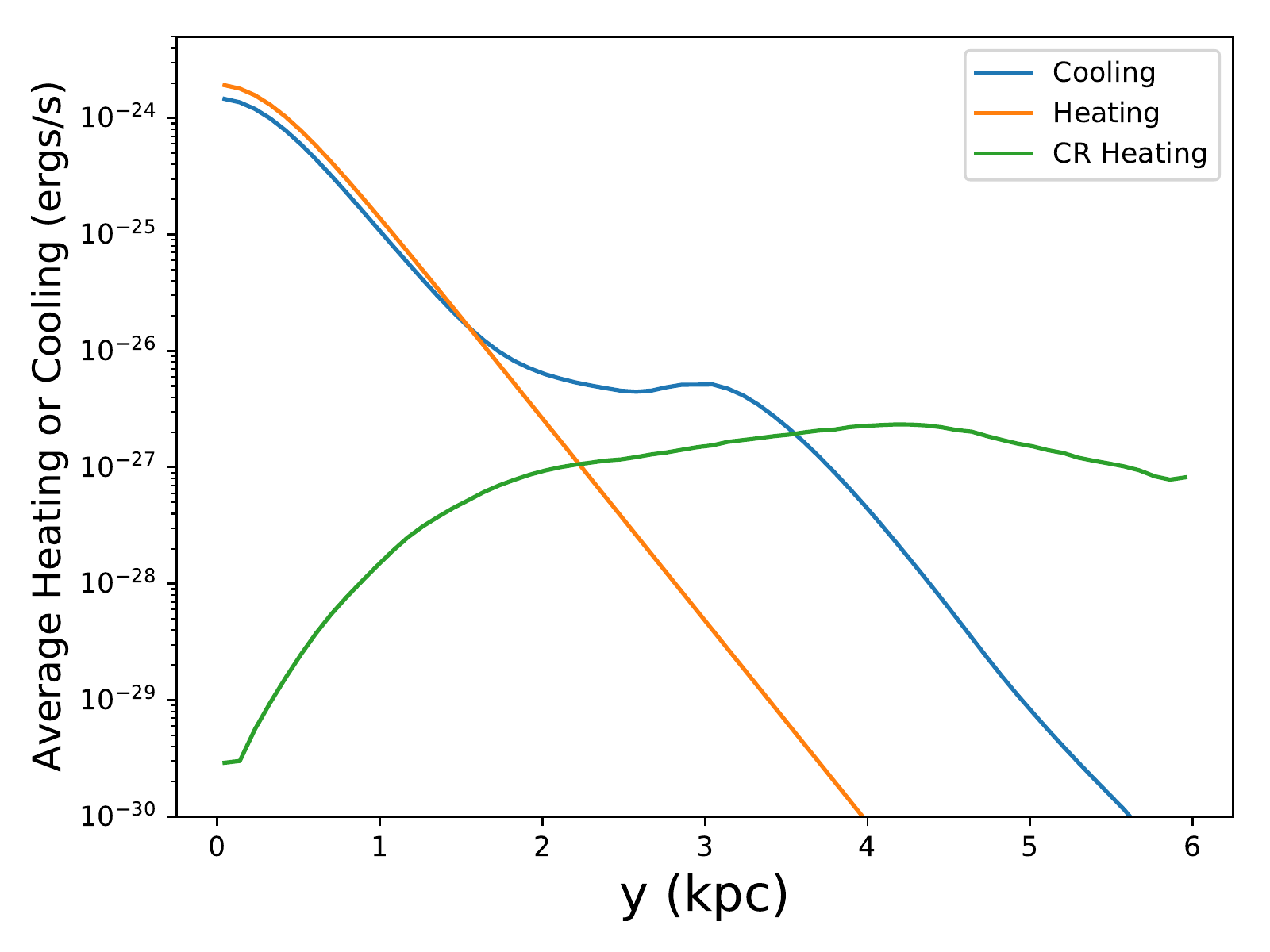}
\caption{Per particle heating and cooling at $t = 250$ Myrs, with streaming and cooling. The spatial off-set between gas cooling and cosmic ray heating is apparent. As seen in the right panel, above 3-4 kpc, the mass-weighted average cosmic ray heating term dominates the cooling term. The orange line shows the additional heating term necessary to balance cooling in hydrostatic equilibrium. }
\label{fig:perParticleHeatingCooling}
\end{figure*}

Physically, it is extremely important that the lower density gas is the population most affected by the cosmic ray heating. If this gas is heated, it is much easier for the instability to proceed as it creates more room for the magnetic field loops to grow from their own buoyancy. As these loops get steeper, more and more gas will fall into the valleys, further destabilizing the system. On the other hand, if the high density gas is heated, one would expect it to stabilize the system as the gas is then more difficult to compress in these valleys. As we see in Figure \ref{fig:phase_plots_diff_vs_streaming}, however, the high density gas in the streaming model appears to peak at roughly the same temperature as the diffusion case. Therefore, even if the cosmic ray heating only plays a small role in the actual instability when compared against a diffusion model, we see that the heating plays an important role in shaping properties of the thermal gas, especially a few kpc above the disk. Whether this heating can compete with cooling, which threatens to negate the heating term, is the subject of the next section. 

\section{Simulations with Cooling}
\label{sec:coolingSection}

\begin{figure}[]
\centering
\includegraphics[width=0.48\textwidth]{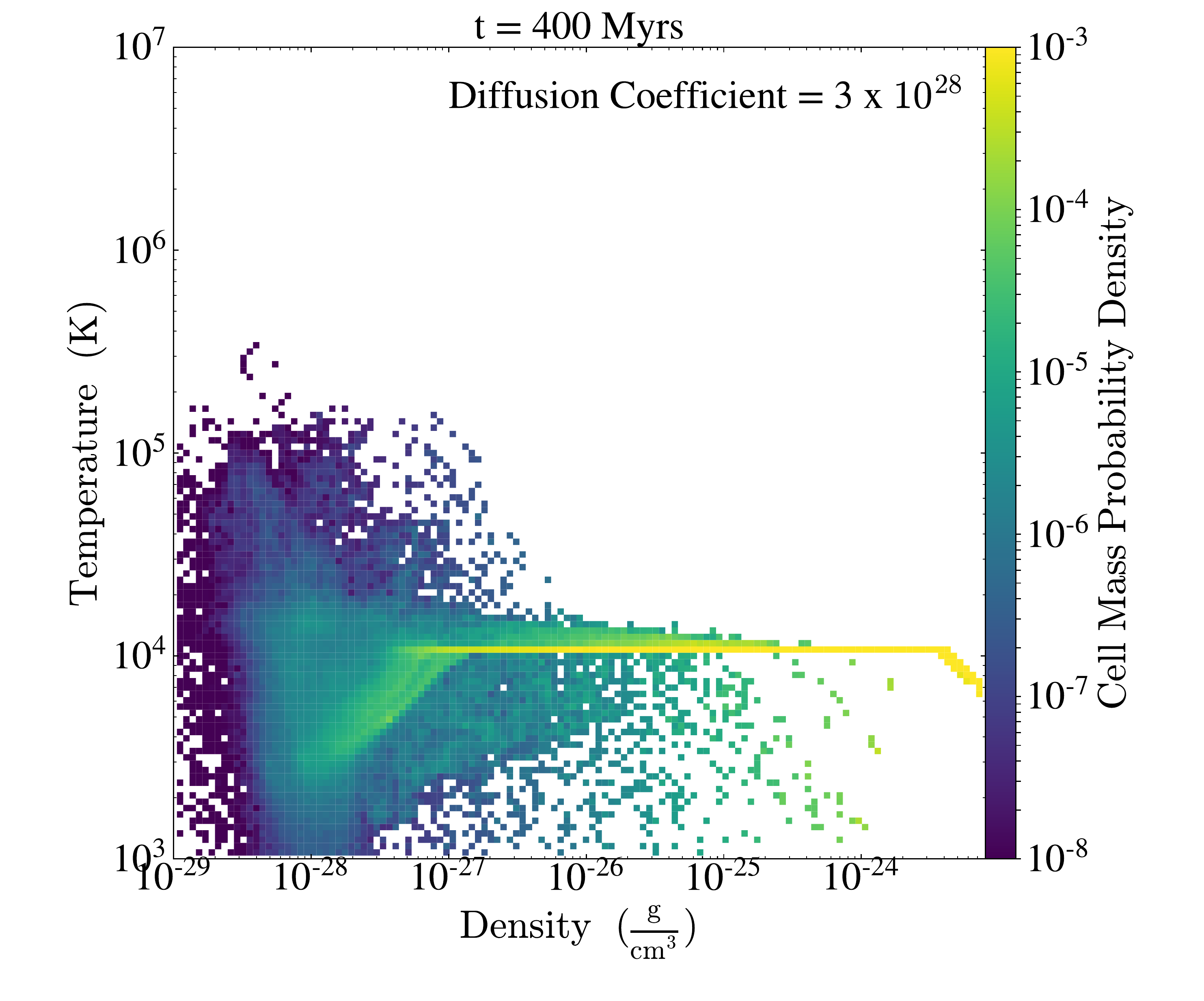}
\includegraphics[width=0.48\textwidth]{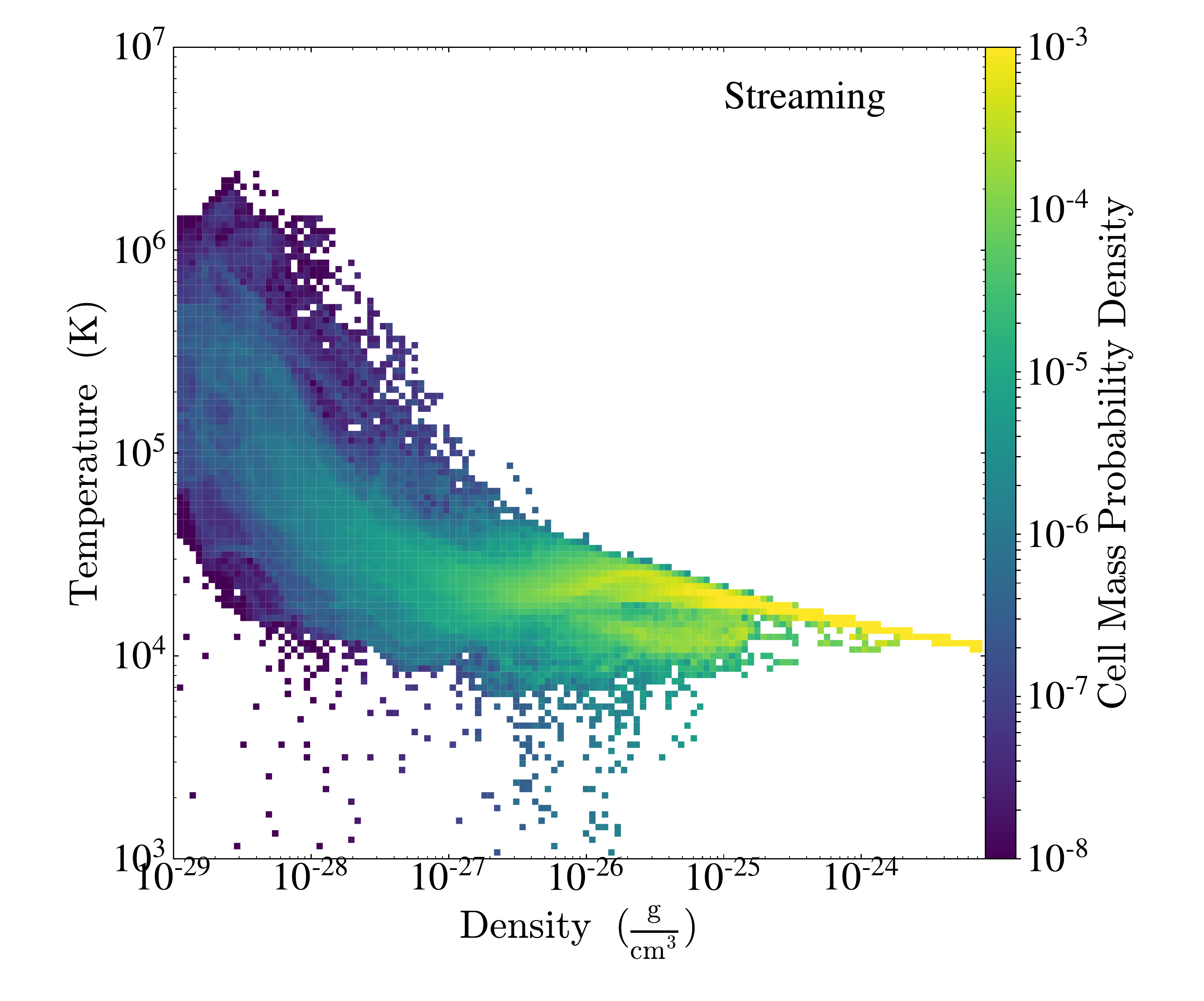}
\caption{Phase plots at t = 400 Myrs for m = 2, c = 1 simulations with cooling. In the streaming figure, note the lack of gas at low ($\rho$, T), as well as the upward rising slope (instead of flat slope) at higher densities. Both are attributed to the cosmic ray heating that is not present with only diffusion.}
\label{fig:phasePlots_withCooling}
\end{figure}

To address the interplay between radiative cooling and cosmic ray heating, we ran a few simulations with cooling included. As we found through trial and error, we must be careful not to confuse the Parker instability and thermal instability. Our choices of scale height, surface densities, and composition ($m$ and $c$ values) determine the temperature of our medium, which may be in a thermally unstable regime. Simulations with initial temperatures in the cold unstable phase tend to show instability right away, with pockets of dense gas forming in the midplane on the scale of individual cells. This behavior was also noted by \cite{MouschoviasParkerCooling2009}. Although the Parker loops form on much larger scales than thermal instability, we find in the nonlinear regime that large pockets of gas develop near the midplane and sit at the imposed temperature floor of 300 K, which changes the large-scale structure of the ISM.

As noted in \S\ref{subsec:cooling}, the cooling curve defines a warm stable phase on the rapidly rising part of the cooling curve where $\partial \Lambda/\partial T > \Lambda/T$. This analysis does not include our height-dependent heating rate put in to perfectly offset cooling and maintain an equilibrium state. Despite this, we find this thermal stability analysis to hold quite well. For the set of simulations (with cooling included) shown in this work, we always choose a temperature on the steep portion of the cooling curve, which shows good stability and maintains the equilibrium state; we can then watch the Parker instability evolve without clearly developing thermal instability in the cold unstable phase. However, as noted in \S\ref{subsec:cooling}, the \cite{InoueCooling2006} analytic fit deviates from the real cooling curve at high temperatures, notably overestimating cooling at $T \approx 10^{5.5} K$ where the O
VI peak should trigger thermal instability in a warm unstable phase. Therefore, we present results from here on using a more accurate, tabulated cooling curve \citep{SutherlandDopita1993} at high temperatures combined with a fit at $T < 8000$ K \citep{Slyz2005}. We also ran the same simulations with the \cite{InoueCooling2006} cooling curve and found no significant changes to our conclusions.

Figure \ref{fig:streaming_and_cooling_growthTimes} shows an intriguing result, which seems to hold for both higher and lower values of $(m,c)$: that cosmic ray streaming destabilizes the system even in the presence of cooling. In fact, streaming and cooling combined show faster growth times than streaming individually. This seems to contradict the conclusion of HZ18, which implicated the cosmic ray heating term as the destabilizing factor. As we've seen from our analysis in previous sections, though, streaming and diffusion give very comparable growth curves, suggesting instead that the transport itself is most important in the linear regime. Our analysis with cooling supports this idea, as cosmic ray transport and gas cooling don't seem to counteract each other. However, there are other factors that may be playing a role in this discrepancy, including the fact that the linear theory assumes a constant gravitational acceleration. As \cite{GizParker1993} and \cite{KimParker1998} note, smooth gravity generally promotes instability more than constant gravity. It is also possible, therefore, that the overall effects of cooling are different in these two systems as well.   

Analysis of the nonlinear regime shows us why this is the case: cooling and cosmic ray heating have only a small spatial overlap. Figure \ref{fig:perParticleHeatingCooling} shows the per-particle heating and cooling rates in at an evolved time for our m = 2, c = 1 simulation. While cooling dominates in the midplane region and in the compressed filaments weighing down magnetic field lines, cosmic ray heating acts mainly in the expanding bubbles a few kpc above the midplane. This can be seen also in the right panel of Figure \ref{fig:perParticleHeatingCooling}, which shows the mass-weighted average of the cosmic ray heating and cooling as functions of height above the disk. While cooling (and the heating term implemented to offset it) dominate within the first few scale heights, the cosmic ray heating term becomes important in the extraplanar gas, exceeding cooling by a few orders of magnitude at 5 kpc above the disk.  

We note that there \emph{is} some overlap between cosmic ray heating and radiative cooling in the dense pockets, as well, but this may in fact be another reason why the instability acts faster with streaming and cooling rather than with streaming alone. Without cooling, the heating in that region would stabilize the medium by increasing pressure. With cooling, however, that energy transferred between cosmic rays and thermal gas is immediately radiated away from the system, resulting in a net loss of pressure support. Instead of the usual Parker instability picture where cosmic ray pressure is displaced by gas pressure, in this case, some of the cosmic ray pressure simply disappears from the system as it is transferred to thermal gas and quickly radiated away. 

Regardless, Figure \ref{fig:phasePlots_withCooling} shows similar behavior compared to our non-cooling simulations, but the differences between diffusion and streaming are now more enhanced.  There is a complete lack of low ($\rho$,T) gas in the streaming case, while in the diffusion case, the combination of adiabatic and radiative cooling has pushed a significant mass fraction to that regime. Compared to Figure \ref{fig:phase_plots_diff_vs_streaming}, which shows a clear upward-trending line denoting the compressionally heated gas, both phase plots with cooling now lack that line, as gas near the peak of the cooling curve is now unstable. Compressed gas is now kept much tighter near the initial temperature $\approx 10^{4}$ K, especially in the diffusion simulation, which maintains a flat line in ($\rho$, T) space down to $\rho \approx 10^{-27} \rm g cm^{-3}$. The streaming simulation shows more of a slope due to cosmic ray heating preferentially of the lower density gas, for which cooling is less efficient, but even of the dense, warm gas too. We attribute this entirely different phase structure, again, to cosmic ray heating. Some implications for this are discussed in Section \ref{sec:summary}. 

\section{Nonlinear Simulations: 3D}
\label{sec:nonlinear3D}
For a small set of parameters, we also ran 3D simulations where perturbations are applied in the $\hat{y}$ direction again, but with random phases and wavelengths in both the $\hat{x}$ and $\hat{z}$ directions. Our perturbation amplitude is now $A = 10^{-6}$ (see Equation \ref{seedPerturbEqn}) to give the system a similar kick to the 2D simulations. 

\begin{figure}[]
\centering
\includegraphics[width=0.43\textwidth]{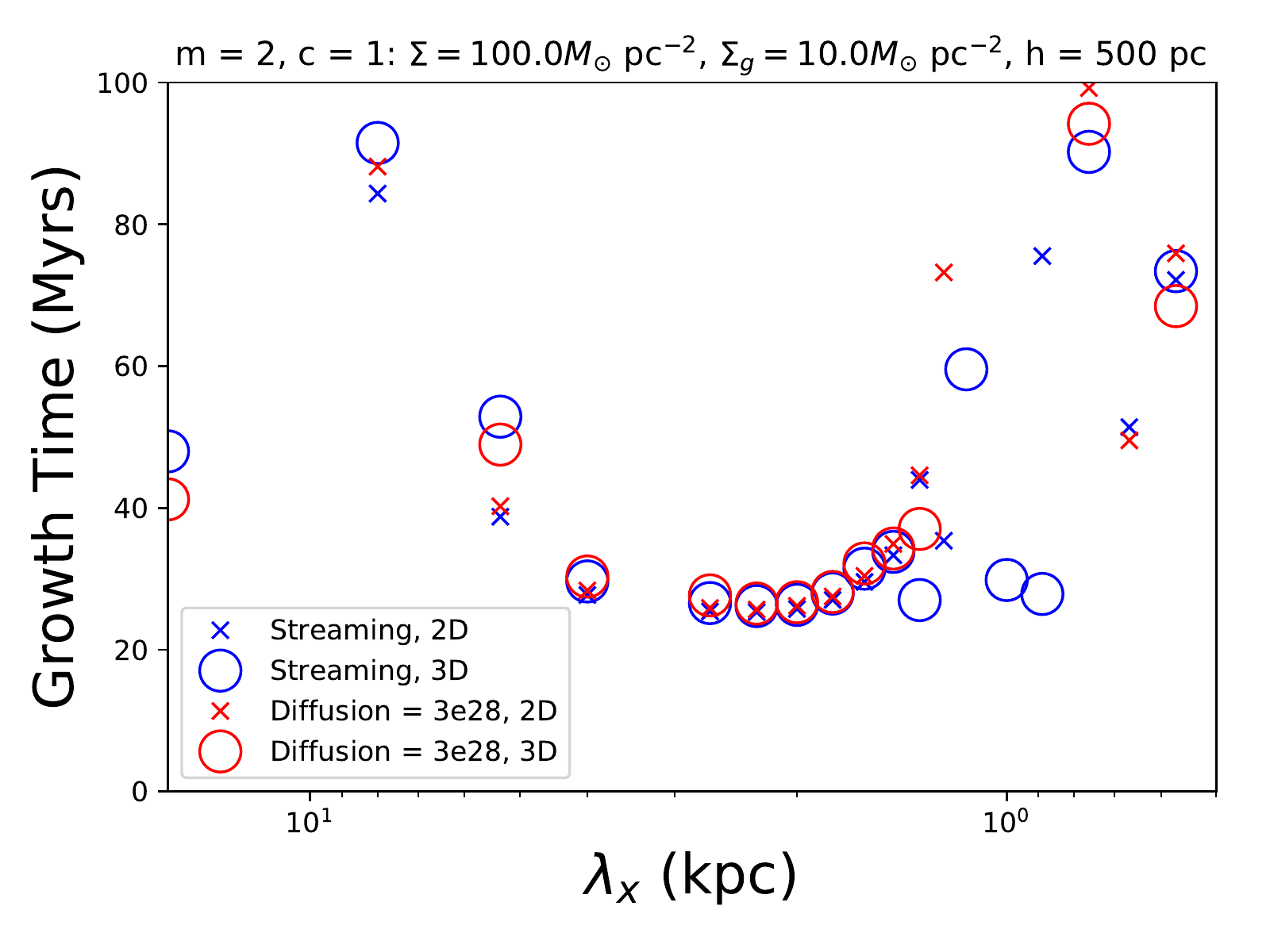}
\caption{2D vs 3D growth times with \cite{RodriguesParker2015} galaxy parameters. We see that contrary to the linear stability analysis from \cite{HeintzParker2018} where the 2D case with streaming had a larger growth rate peak, the growth rates for the 2D cases and 3D cases are essentially the same, with maybe the 2D case only being slightly more unstable.}
\label{fig:3DGrowthTimes}
\end{figure}

Figure \ref{fig:3DGrowthTimes} shows the growth rates of the 3D simulations compared to 2D simulations for diffusion = $3 \times 10^{28}$ and streaming. We see a tight overlap between 3D and 2D growth curves. Although we have not done a linear stability analysis of the smooth gravity setup with streaming or diffusion, we expect from our constant gravity analysis that the 3D growth rates would be slightly lower than 2D, while the 3D system should be more unstable at shorter wavelengths. We find that the differences between the two systems in our simulations are quite small, especially compared to the linear analysis. At long wavelengths, we possibly see a sign of the 2D system having larger growth rates than the 3D case with the 3D case beginning to overtake the 2D case at shorter wavelengths as we would expect. It is possible, however, that we are just in an $(m,c)$ regime where the differences should not be very large once smooth gravity is accounted for. We intend to explore this in future work once we have a smooth gravity and cooling linear stability analysis to compare to. 

As with our 2D simulations with a realistic gravity profile, we find there to be no symmetry favored by the instability in 3D. In fact, the instability qualitatively looks extremely similar (other than in the $k_z$ direction, obviously) to the 2D case in terms of the magnetic loops, formation of dense pockets of gas, and symmetry of the modes. We find many of the conclusions made about our 2D system still hold for our 3D simulations as well.

\begin{figure}
\centering
\includegraphics[width = 0.4\textwidth]{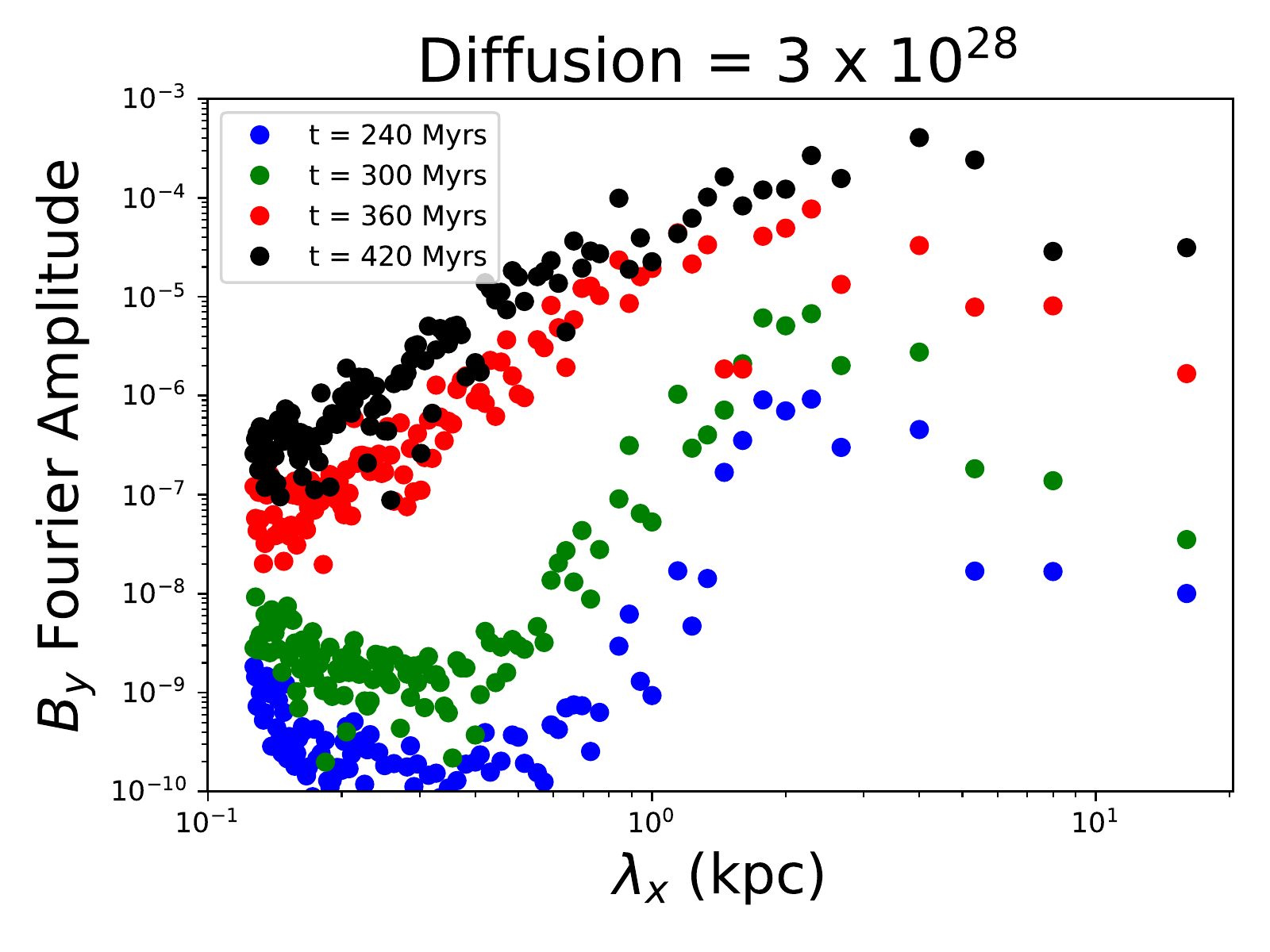}
\includegraphics[width = 0.4\textwidth]{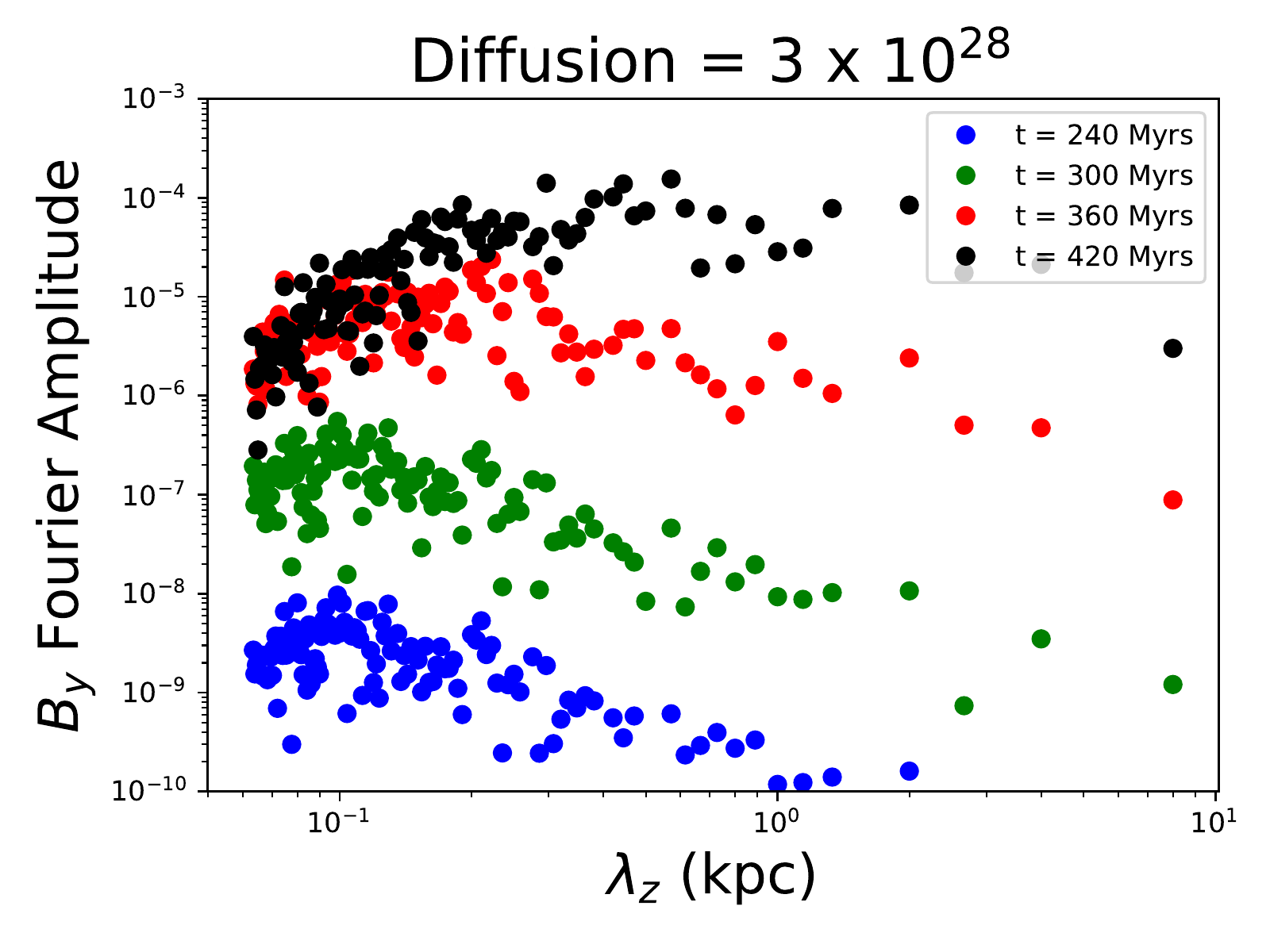}
\caption{Fourier amplitudes as a function of $\lambda_{z}$ for every perturbed mode in our 3D simulation box. During the linear regime (earlier times), most of the power is in modes with very short $\hat{z}$ wavelengths, but at late times, the power slowly shifts to larger $\lambda_{z}$. This affects observational signatures of the instability, such as Faraday rotation measure (Figure \ref{fig:faradayRM_diff3e28}).}
\label{fig:modePower}
\end{figure}

In our 3D simulations, due to the existence of both a $k_x$ and $k_z$ wavevector, we find that neither the undular or interchange mode is favored in a 3D setting as \cite{RodriguesParker2015} found in their work. In Figure \ref{fig:modePower}, we plot the Fourier amplitude of $B_{y}$ for the perturbed $\hat{x}$ and $\hat{z}$ wavelengths in our system. The dominant modes during the linear growth regime follow our intuition from previous linear stability analyses (HZ18): $k_{z} \rightarrow \infty$ gives the fastest growing mode, and this is true for each of our Modified Parker, diffusion, and streaming simulations. In the nonlinear regime, at late times, the mode growth shifts to larger wavelengths, most noticeably in the $\hat{z}$ direction, where modes nearly 1 kpc in size now have considerable power. This is naturally explained by small Parker loops connecting into bigger structures, but given the very small wavelengths (of order our resolution) in the $\hat{z}$ direction, we can't rule out some smoothing of magnetic structures due to numerical magnetic field diffusion. 

As further analysis of our 3D simulations, we also create some mock observations that may be useful to disentangle the Parker instability from other processes. One would naively expect that the characteristic Parker loops would be visible in edge-on synchrotron intensity maps; however, \cite{RodriguesParker2015} find their mock intensity maps to be dominated by disk stratification, with only negligible variation at a given height due to Parker loops. A more clear Parker instability signature may be present in face-on Faraday rotation measure maps, which probe a convolution of the electron number density with the vertical magnetic field. The rotation measure can be computed as:

\begin{equation}
    \phi = (0.812 \rm rad m^{-2}) \int \frac{n_{e}(y)}{1 \rm cm^{-3}} \frac{B_{y}}{1 \mu G} \frac{dy}{1 \rm pc}
\end{equation}
where $\hat{y}$ is the line-of-sight, assuming a ``top-down" view of our simulation box. The electron number density, $n_{e}(y)$, is a tabulated function of ($\rho$, T) assuming photoionization equilibrium with the extragalactic UV background \cite{wiersma2009}.

\begin{figure*}[]
\centering
\includegraphics[width = 0.49\textwidth]{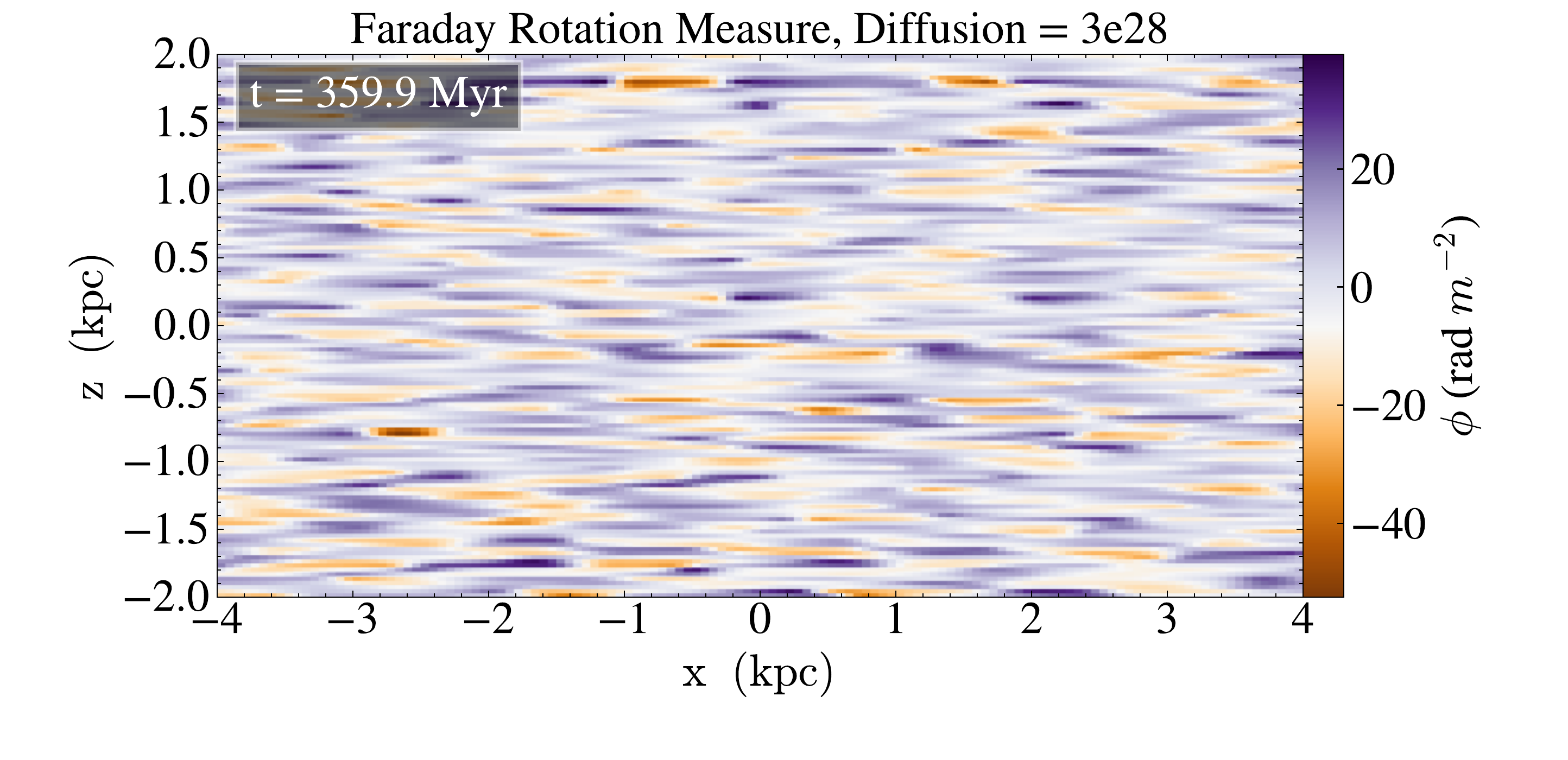}
\includegraphics[width = 0.49\textwidth]{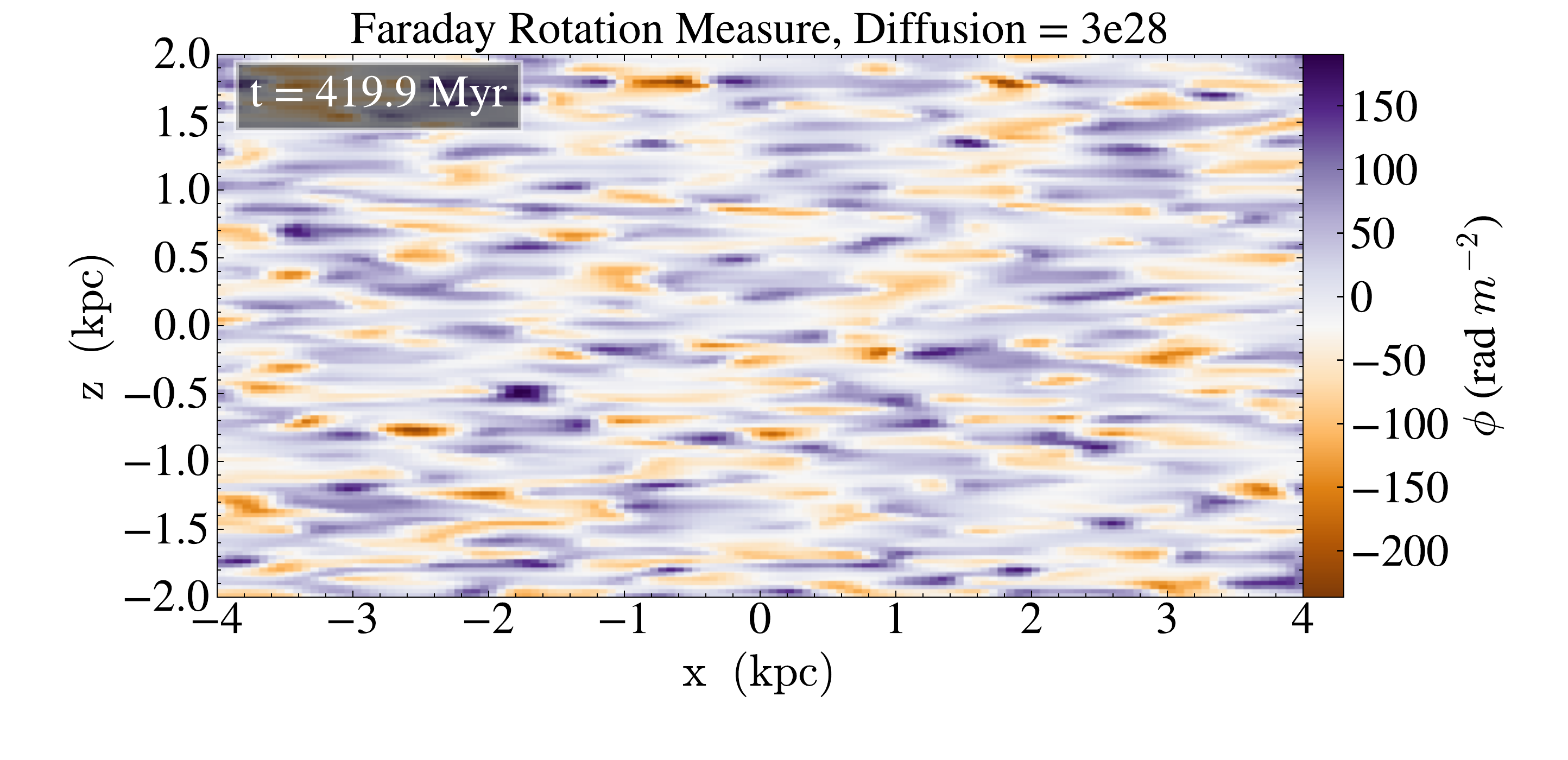}
\caption{Face-on ($\hat{y}$ out of the page) Faraday rotation measure maps of our 3D simulation at different times. The shift in mode power from very short to larger $\lambda_{z}$ is evident. However, with the small wavelengths, this shift may be due to numerical magnetic diffusion. Also note the significant differences in colorbar scale. Depending on when an observer ``catches" the Parker instability, observational signatures, such as these distinct Faraday rotation measure maps, may look quite different.}
\label{fig:faradayRM_diff3e28}
\end{figure*}

Figure \ref{fig:faradayRM_diff3e28} shows mock Faraday rotation maps made from our 3D $\rm \kappa_\parallel = 3 \times 10^{28}$ cm$^2$/s simulations. We see in the direction parallel to the background magnetic field, the wavelength of our Faraday rotations is on order of 1 kpc. However, we again see that in the horizontally perpendicular direction to the magnetic field, the shortest possible wavelengths for our simulation resolution are favored by the instability. As we saw in Figure \ref{fig:modePower}, at later times, in the nonlinear regime, these short wavelengths begin to merge together into larger wavelengths (that are still shorter than in the parallel direction). 

The Parker instability has been notably difficult to observe despite it likely being pervasive in galaxies. In early observations, the Parker instability was often mistaken for what were eventually determined to be superbubbles due to their similar magnetic loop structure. However, the Parker instability differs from these superbubbles in that it does not clear the surrounding medium out of the midplane; instead, it further allows the gas to collapse, another potential signature of the Parker instability. While one cannot easily distinguish these differences in an edge-on synchrotron polarization map, these differences should appear in face-on Faraday rotation measure maps and their associated structure functions, which may present one way to see the signature of the instability \citep{RodriguesParker2015}. 

Going beyond this analysis, we note that the very short wavelengths present in our Faraday rotation measure maps may provide an indicator of the Parker instability in its early nonlinear phase, while larger modes transverse to the background magnetic field would indicate a later stage in the instability evolution. These important visuals may provide useful measures to compare against observations when searching for the Parker instability. Given that \cite{RodriguesParker2015} show a correlation between the Faraday rotation measure structure function and the cosmic ray content ($c$ value) of the galaxy, we wonder whether one could additionally infer the cosmic ray \emph{transport} method at-play. However, given the almost identical growth curves and Faraday rotation measure maps generated by our diffusion and streaming simulations, depending on the system, prospects for using Parker instability observations to distinguish between diffusive and streaming transport seem bleak. 

\section{Summary \& Conclusions}
\label{sec:summary}
In this paper, we carried out a linear stability analysis of the Parker instability under three different treatments of cosmic ray transport, extending our previous analysis (HZ18) to include radiative cooling. For Classic Parker and Modified Parker, radiative cooling further destabilizes the system as the thermal gas is now easier to compress into the valleys of the magnetic field. With streaming, the system becomes more stable due to the cooling counteracting the cosmic ray heating. At temperatures that are known to be thermally unstable to the condensation mode,like $5000K$, we find for all three transport models that the Parker modes evolve into thermal instability at shorter wavelengths.

We then extended our analysis by running MHD simulations of the Parker instability for various cosmic ray and magnetic field strengths and for various cosmic ray transport models. 
After modifying our linear stability analysis to include a smooth gravitational potential \citep{GizParker1993, KimParker1998, RodriguesParker2015}, which is more realistic and more tractable for simulations than a constant gravity, we ran a set of 2D simulations iterating over $m$ and $c$ values for two different ISM parameter choices \citep{McKeeSolar2015,RodriguesParker2015}. 

Our simulations with streaming display the clear trends of HZ18, namely that streaming extends the range of instability to shorter wavelengths and promotes faster growth compared to Modified Parker (see Figure \ref{fig:streaming_growthRates}). The difference in disk evolution between Modified Parker with and without Streaming (Figure \ref{fig:solar_dens_Pres_evolution}) has implications for the formation of molecular, star-forming clouds since the average compression increases with streaming, \emph{and} these compressions occur in more locations since the dominant Parker wavelength is now shorter with Streaming. As discussed in HZ18, this may provide some insight to why simulations of cosmic ray driven winds can simultaneously sustain both star formation and large-scale outflows while the same simulations with a Modified Parker cosmic ray treatment quench star formation and develop puffy, seemingly stable disks. 

One of the most interesting trends that we find, however, is that the growth rates for streaming and diffusion are nearly identical using a typical diffusion coefficient of $\kappa_{\parallel} = 3 \times 10^{28} \rm cm^{2}$ $\rm s^{-1}$ (see Figure \ref{fig:Rodrigues_diffusion_streaming_comp}). This suggests that, instead of cosmic ray heating acting as the dominant destabilizing mechanism, there is a large role played by cosmic rays, whether by diffusion or streaming, moving away from the compressing pockets and to the lower density regions of gas supported by the magnetic loops.

Although the growth curves for diffusion and streaming are very similar, we do find important differences in the nonlinear regime. 
\begin{itemize}
    \item Average gas pressure beyond a few kpc from the midplane is increased by a factor of 5 - 10 for streaming when compared to diffusion ($\kappa_{\parallel} = 3 \times 10^{28}$), despite the magnetic and cosmic ray pressure profiles looking nearly identical at a given time (Figure \ref{fig:pressure_profiles}.)
    \item Phase diagrams (Figures \ref{fig:phase_plots_diff_vs_streaming} and \ref{fig:phasePlots_withCooling}) show very different evolutionary tracks. While both show a clear signature of adiabatic expansion to low T, $\rho$, much of this low-density gas lies at higher temperatures in the streaming case. 
\end{itemize}
    
Figure \ref{fig:perParticleHeatingCooling} shows the mechanism at work: cosmic ray heating is most important in these diffuse, expanding bubbles, while cooling is most efficient in the dense, compressing pockets. Because these heating and cooling are spatially off-set, they generally do not cancel each other out, and in some cases, they may even lead to slightly increased instability over the streaming-only case, similar to the results found in the linear stability analysis when the temperature was lowered to $5000K$. This heating of the extraplanar diffuse gas region gives credence to theories of a cosmic ray heated warm ionized medium \citep{WienerWIM}, which may provide an important supplemental heating mechanism in addition to photoionization, turbulent dissipation, and magnetic reconnection \citep{ReynoldsWIM}.





In all cases, we find a saturated state where further mode growth is suppressed due to magnetic tension; however, the buoyant Parker loops we create always leave the top of our simulation domain, which is a full 32 scale heights above/below the midplane. These loops help advect cosmic rays and the magnetic field to great heights, while cosmic ray transport further shifts the cosmic ray population and forms a cosmic ray dominated eDIG layer. Galactic wind simulations with cosmic ray transport generally result in a similar picture, with thermal gas dominating in the disk and cosmic ray pressure dominating in the halo. These cosmic ray dominated halos have numerous implications for galaxy evolution and interpretation of recent outflow and CGM observations, which find a coexistence of low- and high-ions that can be better explained by a cosmic ray pressure supported medium than a thermally supported one \citep{SalemCGM2016, butskyWinds2018}. Even in our idealized simulations, we find that cosmic ray transport begins to develop a cosmic ray dominated halo, and it instigates this shift by enhancing the Parker instability. 




In addition to our 2D simulations, we also ran 3D simulations of the Parker instability in order to make comparisons with our 2D results and to provide a few mock observations that may prove useful in helping detect the Parker instability. We find in 3D no preference for the undular or interchange modes and instead get a mix of the two, with the wavelength in the horizontal direction perpendicular to the field becoming as short as allowed by our simulation resolution. This changes over time, though, as modes coalesce to form larger wavelengths in both the parallel and transverse directions. This leads to very different Faraday rotation measure maps, which would propagate to differences in the structure function of the medium if we had calculated it. \cite{RodriguesParker2015} show that such structure functions may be used to infer the cosmic ray content of galaxies, which is correlated with magnetic structure in their simulations. Given that our diffusion and streaming simulations generate almost identical growth curves and structures, it would be a challenge to go further than this and try to infer the dominant cosmic ray transport method in the galaxy from its structure function. Overall, the results of our 3D simulations seem closely correlated to our 2D simulation results, even in regards to the growth rates of the instability. 

Given the wide range of phenomena pinned partially on the Parker instability (star formation, galactic dynamo, etc.), this modern treatment of the instability has far-reaching implications if the instability can act on timescales comparable to turbulence and star formation in the disk. Compared to Modified Parker ($\gamma_{c} = 4/3$), diffusion, streaming, and cooling are all generally destabilizing and push the most unstable mode down to shorter wavelengths, causing a fast shift in the composition of the ISM. 

By scaling the growth times of our linear stability analysis with realistic galactic parameters, we can determine if the instability is acting on relevant timescales and how that may change based on the surface density of the galaxy as well as the ratio of the three pressures (thermal gas, magnetic, cosmic ray). For our dimensionless variable definitions, our growth rate is scaled by:
\begin{equation}
\begin{split}
    \omega = \hat{\omega}&(0.531 \rm{Myrs^{-1}})(\frac{(1+\alpha+\beta)}{\gamma_g})^{1/2} \\ &(\frac{\Sigma}{50 \rm M_\odot pc^{-2}})^{1/2}(\frac{500 \rm pc}{H})^{1/2} 
\end{split}
\end{equation}
which can be derived by modifying the definition of $\hat{\omega}$. We plugged in the formulas for $H_0$ in eq. \ref{eq:dimensionless} and the sound speed in terms of temperature ($a_g \equiv \sqrt{\gamma_g k_B T/ \Bar{m}}$) and then used eq. \ref{eq:tempEqn} to plug in for $T$. The coefficient is then attained from combining all of the constants in the formula ($\Bar{m}$, $k_B$, and the normalization constants). From \cite{SalemCROutflows2014} and \cite{ruszkowskiwinds2017}, we can model our galaxy by assuming a stratified, isothermal, and self-gravitating disk, similar to what we have done throughout this analysis. In that system, $\Sigma \approx 400 \hspace{1pt} \rm e^{-R/R_0}$ $\rm M_\odot pc^{-2}$ where it is assumed $R_0 = 3.5$ kpc and the vertical scale height, $H=350$ pc. The setup of the disk is comparable to a Milky Way type galaxy \citep{KlypinMWModel2002,BovyMWSurface2013}. We also assume that the three pressure exist in equipartition as they do in the Milky Way so $\alpha = \beta = 1$.

As one can see based on our definition above, our growth rates are scaled by the square root of both the sum of the pressures and the surface density of the galaxy so minor changes in these values will not have a large effect on the growth times of the instability. However, for the parameter space prescribed above, at a distance of $R=2$ kpc (which gives $\Sigma \approx 220$ $\rm M_\odot pc^{-2}$), we obtain a growth rate for the Parker instability of 53 Myrs. If we move farther away from the center of the disk, to a distance more similar to our solar neighborhood ($R = 8$ kpc, $\Sigma \approx 40$ $\rm M_\odot pc^{-2}$, close to \cite{McKeeSolar2015}), the growth time grows to be $127$ Myrs., a little over twice that of the growth time at $r=2$ kpc. Therefore, the Parker instability has a better chance of being comparable to star formation and turbulence timescales when it is closer to the center of the disk for a particular galaxy. A radial distance of 2 kpc still works for the instability since we found often in our simulations that the wavelength of the instability was of order 1 kpc. 

For starburst galaxies and higher density galaxies than the Milky Way, we expect the Parker instability to grow even more quickly, perhaps growing on the scale of 1 Myr. While it is possible that these higher density environments also have shorter timescales for turbulence and star formation, it is possible that the Parker instability could arise from an equilibrium created by these two processes. Furthermore, the Parker instability acts on wavelengths that are larger than the typical sizes of both turbulence and star formation so the different length scales may also allow both to occur simultaneously (as noted by \cite{zweibelparker1975}).

A scaling of $\alpha$ and $\beta$ shows the tendency for higher $\alpha$ and $\beta$ to increase the growth rate but the minor changes away from one that would match the realistic parameters of most galaxies is too small to have a large impact on the growth rate. For example, for our Milky Way system at 2 kpc, the instability only grows 10 Myrs faster when both $\alpha$ and $\beta$ are increased to two. Also note, of course, that our scaling relation here is not a stability criterion and much of the instability physics is contained within $\hat{\omega}$.




As with most work, it is important to note that many assumptions have been made in this work. Our systems assume no differential rotation or turbulence, which have been shown to be stabilizing, as well no self-gravity which is destabilizing. Furthermore, additional effects like phase transitions caused by the radiative cooling are not included (see \cite{MouschoviasParkerCooling2009} to see the effects these transitions can have). Lastly, in our setup, cosmic rays are included as a preexisting fluid along with magnetic fields and thermal gas. The injection of cosmic rays at local sources would affect the outcome, but we expect our main conclusions to hold in future simulations accounting for additional ISM processes. A study of how the Parker instability behaves in a more complicated environment, without some of these simplifications, is left to future work.  

\acknowledgments
The authors would like to thank the referee for insights and suggestions that improved this paper. We also thank Thomas Berlok, Adrian Fraser, Matthew Kunz, Naomi McClure-Griffiths, Peng Oh, Eliot Quataert and Josh Wiener for helpful discussions and technical support throughout this project. We would also like to especially thank Mateusz Ruszkowski and Karen Yang for providing us with the cosmic ray module they developed for FLASH. The software used in this work was in part developed by the DOE NNSA-ASC OASCR FLASH Center at the University of Chicago. The analysis and visualization package \emph{yt} \citep{ytPaper} was crucial to this work, and we thank the developers for making it open-source. This work used the Extreme Science and Engineering Discovery Environment (XSEDE), which is supported by National Science Foundation grant number ACI-1548562 \citep{xsede}. Specifically, our computing resources stemmed from allocations TG-AST170033, TG-AST190019, and PHY180037 on the Stampede2 and Comet supercomputers. Simulations were also run on the University of Wisconsin - Madison HPC cluster. C.B. is supported by the National Science Foundation Graduate Research Fellowship Program under Grant No. DGE-1256259. We would also like to thank the Vila Trust and the WARF Foundation at the University of Wisconsin as well as NSF Grant AST-1616037. 

\appendix
\section{Implementing Radiative Cooling}
\label{sec:coolingderivation}
To implement radiative cooling into our linearized perturbation equations \ref{eq:contin} - \ref{eq:thermenergy}, we start from the First Law of Thermodynamics, $dU = dQ - P_gdV$ where $U$ is the total energy, $Q$ is the heating rate, $P_g$ is the gas pressure and $V$ is the volume. Taking a time derivative and assuming that:
\begin{equation}
\begin{split}
    \frac{dU}{dt} = \frac{1}{\gamma_g - 1}k_B \frac{dT}{dt}, \qquad \frac{dQ}{dt} &= \frac{\Bar{m}}{\rho}(n\Gamma - n^2\Lambda(T)), \\ 
    \frac{dV}{dt} = -\frac{\Bar{m}}{\rho^2}&\frac{d\rho}{dt}
\end{split}
\end{equation}
where $k_B$ is Boltzmann's constant, $T$ is the thermal gas temperature, $\rho$ is the gas density, and $\Bar{m}$ is the average mass of the gas, we find that the first law becomes:
\begin{equation}
    \frac{\rho}{\Bar{m}(\gamma_g - 1)}k_B\frac{dT}{dt} = \Big(n\Gamma - n^2\Lambda(T)\Big) + \frac{P_g}{\rho}\frac{d\rho}{dt} - \mathbf{u_A}\cdot\nabla P_c
\end{equation}
Perturbing this system and making substitutions for $\rho$ and $\Bar{m}$, we find:
\begin{equation}
    \frac{P_g}{T(\gamma_g - 1)}\frac{d\delta T}{dt} = (\delta n\Gamma -2n\delta n \Lambda(T) - n^2 \frac{d\Lambda(T)}{dT}\delta T) + \frac{a_g^2}{\gamma_g}\frac{d\delta\rho}{dt} - (\mathbf{u}_A \cdot \del\delta P_c + \mathbf{\delta u}_A \cdot \del P_c)
\end{equation}
Finally, we note that since we begin in thermal equilibrium, $n\Gamma = n^2 \Lambda(T)$ and we use for $\delta T$ and $\delta n$:
\begin{equation}
    \frac{\delta T}{T} = \frac{\delta P_g}{P_g} - \frac{\delta \rho}{\rho} \qquad \frac{\delta n}{n} = \frac{\delta \rho}{\rho} 
\end{equation}
Making these substitutions, using the mass continuity equation, and rearranging terms, we arrive at our equation of state for the thermal gas (eq. \ref{eq:thermenergy}):
\begin{equation}
\begin{split}
    \Big(\frac{\partial}{\partial t} + (\gamma_g - 1)\frac{n^2T}{P_g}\frac{d\Lambda(T)}{dT}\Big)\delta P_g &= - \mathbf{\delta u}\cdot\mathbf{\del}P_g 
    - \gamma_g P_g \mathbf{\del}\cdot\mathbf{\delta u} \\
    -\Big(\gamma_g - 1\Big)\Big(\frac{n^2 T}{\rho} \frac{d\Lambda(T)}{dT} - \frac{n^2\Lambda(T)}{\rho}\Big)\delta\rho
      &- (\gamma_g - 1)(\mathbf{u}_A \cdot \del\delta P_c + \mathbf{\delta u}_A \cdot \del P_c)
\end{split}\label{eq:appendixthermenergycooling}
\end{equation}
\section{Convergence Study}
\label{sec:convergence}
\begin{figure}
\centering
\includegraphics[width = 0.4\textwidth]{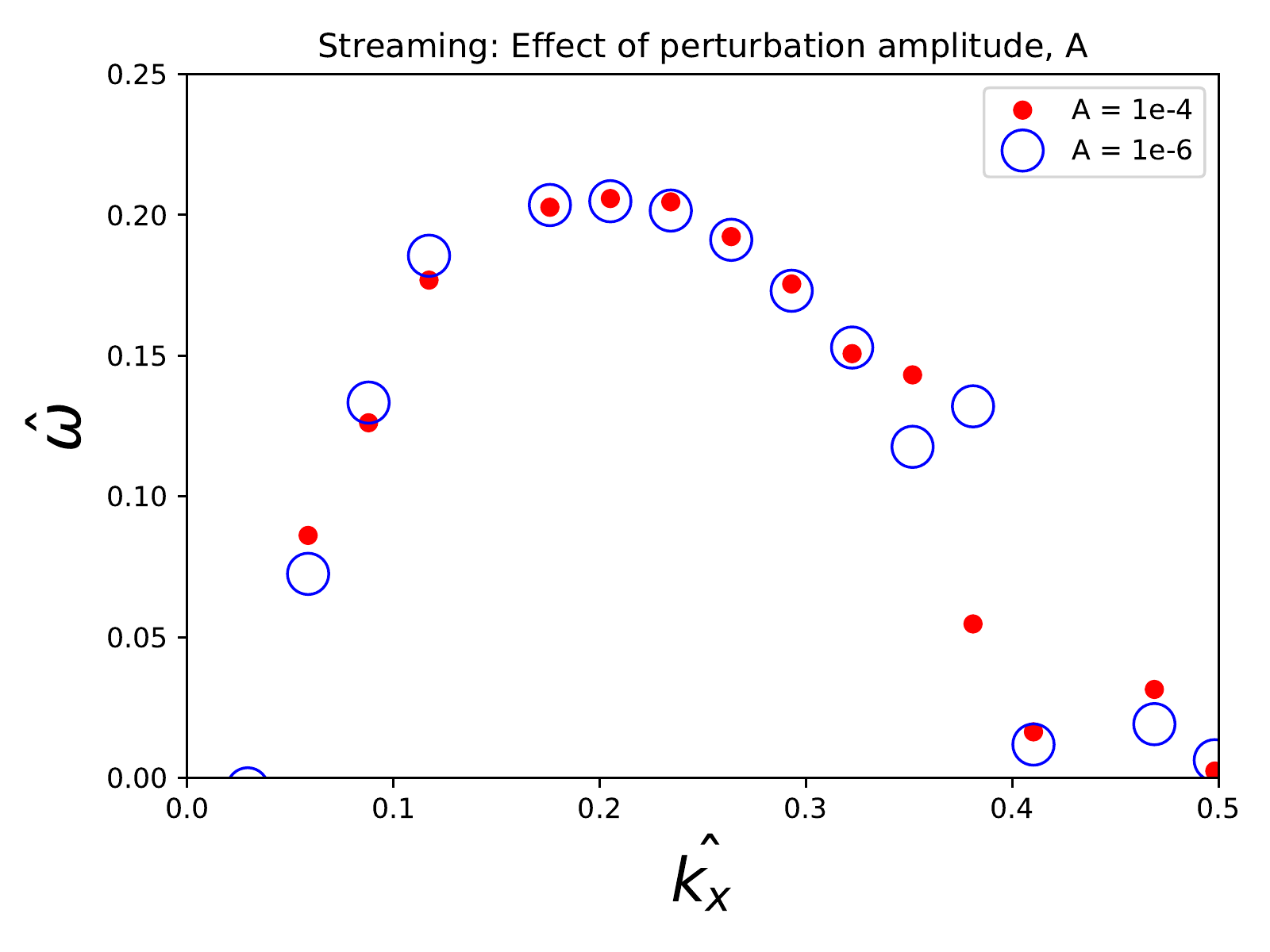}
\includegraphics[width = 0.4\textwidth]{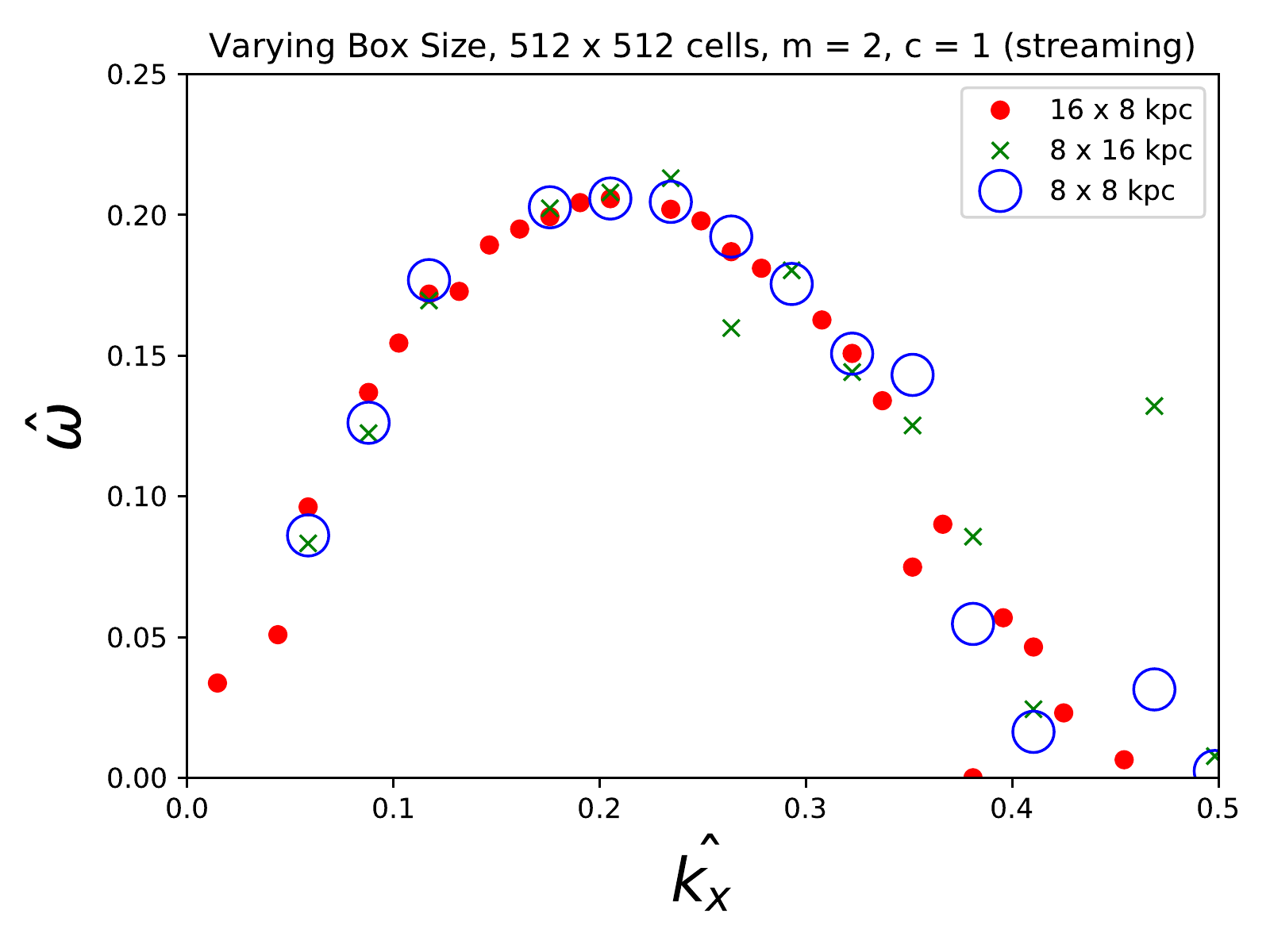}
\includegraphics[width = 0.4\textwidth]{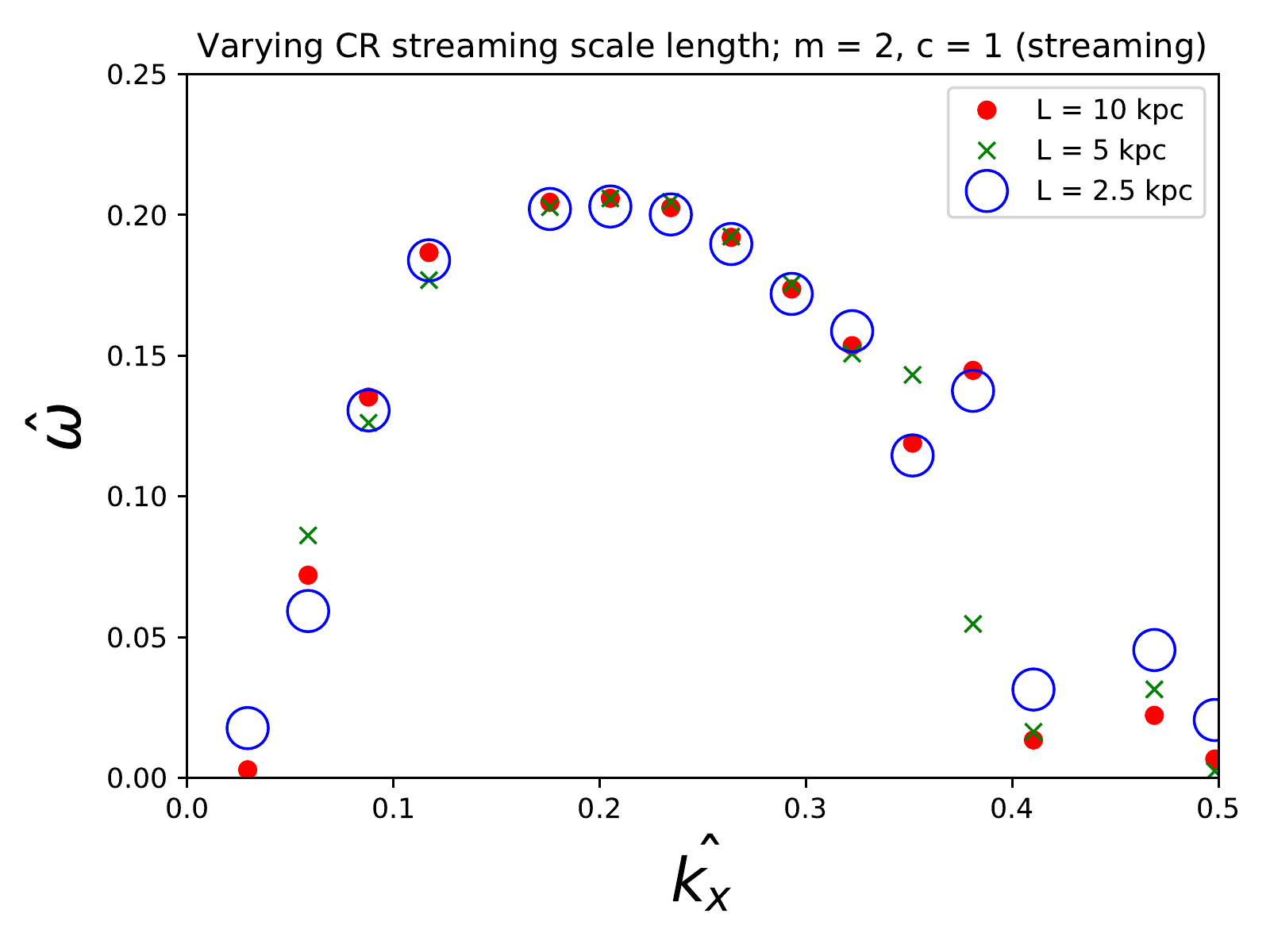}
\includegraphics[width = 0.4\textwidth]{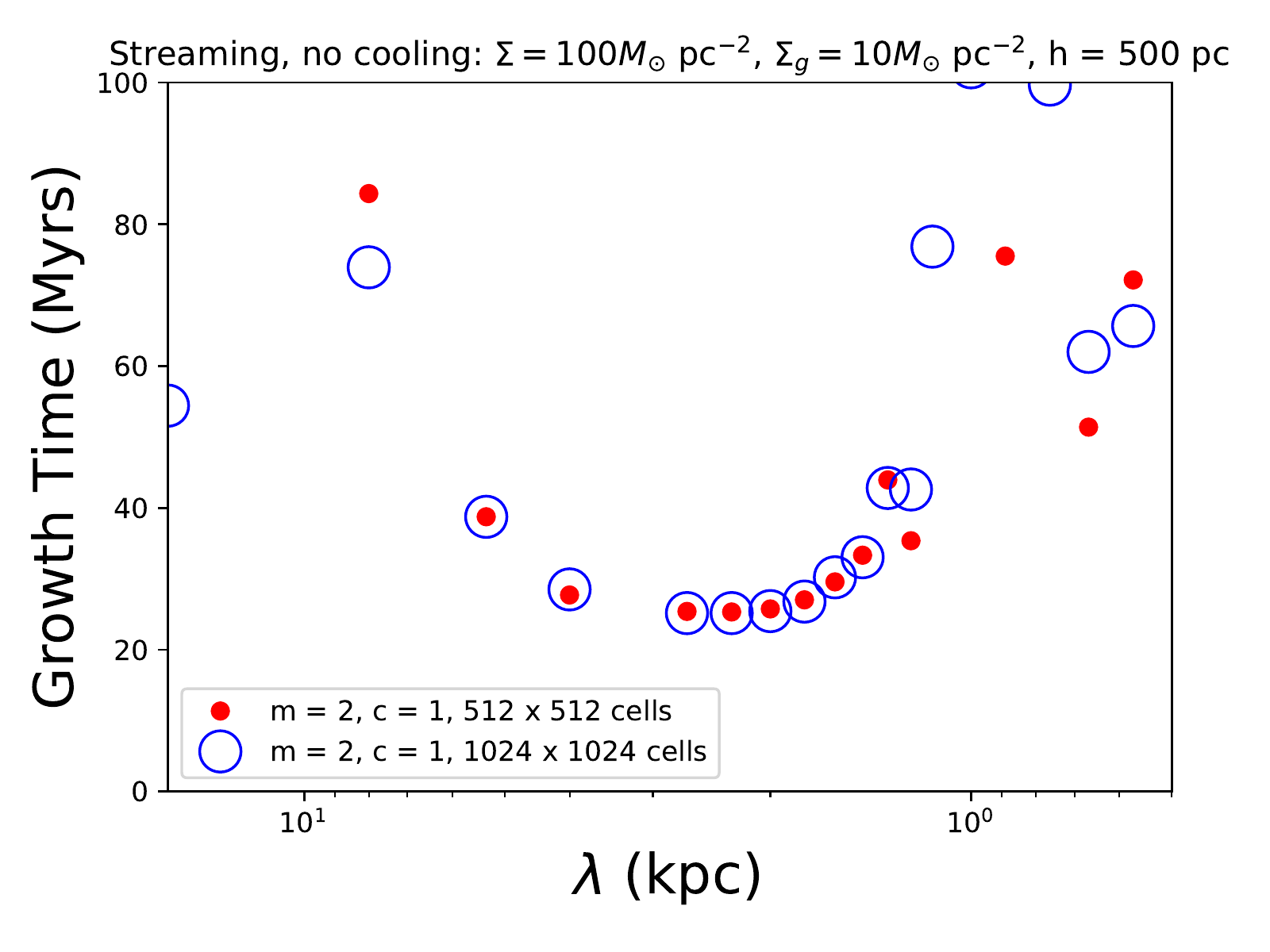}
\caption{Streaming, no cooling, m = 2, c = 1. \emph{Upper Left}: Solar Neighborhood parameters \citep{McKeeSolar2015}. Growth times for two different perturbation amplitudes, A (Eq. \ref{seedPerturbEqn}), showing no significant differences. \emph{Upper Right}: Solar Neighborhood parameters. Effect of box size and dimension, each with 512 x 512 cells. The taller box shows more spurious growth rates (likely due to decreased resolution), but all follow a similar curve, suggesting good convergence as long as the box height is many scale heights, which is appropriate for comparison to the linear stability analyses that assume $k_{y} \rightarrow 0$. \emph{Lower Left}: Solar Neighborhood parameters. Convergence with respect to the characteristic cosmic ray scale length, L, used in the regularization approximation \citep{SharmaRegularization2009}. 
\emph{Lower Right}: \cite{RodriguesParker2015} parameters. Growth rates for m = 2, c = 1 at our fiducial resolution of 31.25 pc (512 x 512 grid cells); there is very little difference with a simulation at double the resolution.}
\label{fig:convergence_plots}
\end{figure}

In this section, we show convergence studies with respect to resolution, box size, and the characteristic cosmic ray scale length, L, that is used in the regularization method for cosmic ray streaming. Figure \ref{fig:convergence_plots} shows a subset of the convergence checks we did to make sure the growth rates with streaming are precise. We did these checks for both Solar Neighborhood \citep{McKeeSolar2015} parameters and \cite{RodriguesParker2015} parameters. Our fiducial choices of scale length $L = 5$ kpc, grid size (512 x 512 cells for 2D), perturbation amplitude (A = $10^{-4}$ in 2D), and box size (8 kpc x 8 kpc for Solar Neighborhood parameters, 16 kpc x 16 kpc for \cite{RodriguesParker2015} parameters) are all well-motivated.

We additionally re-ran a subset of our 2D simulations with lower timesteps to tease out the effects of reconnective heating, which occurs when oppositely-directed magnetic field lines are dragged together in the Parker loops. We found through trial and error that using a CFL number $\gtrapprox 0.6$ resulted in numerical instability, as one would expect, where numerical reconnection converted magnetic energy to thermal energy. This was especially prevalent near the top of two merging Parker loops, where the resulting outburst of hot gas heated much of the surroundings and was not efficiently counteracted by cooling.

\begin{figure}[t]
\centering
\includegraphics[width = 0.32\textwidth]{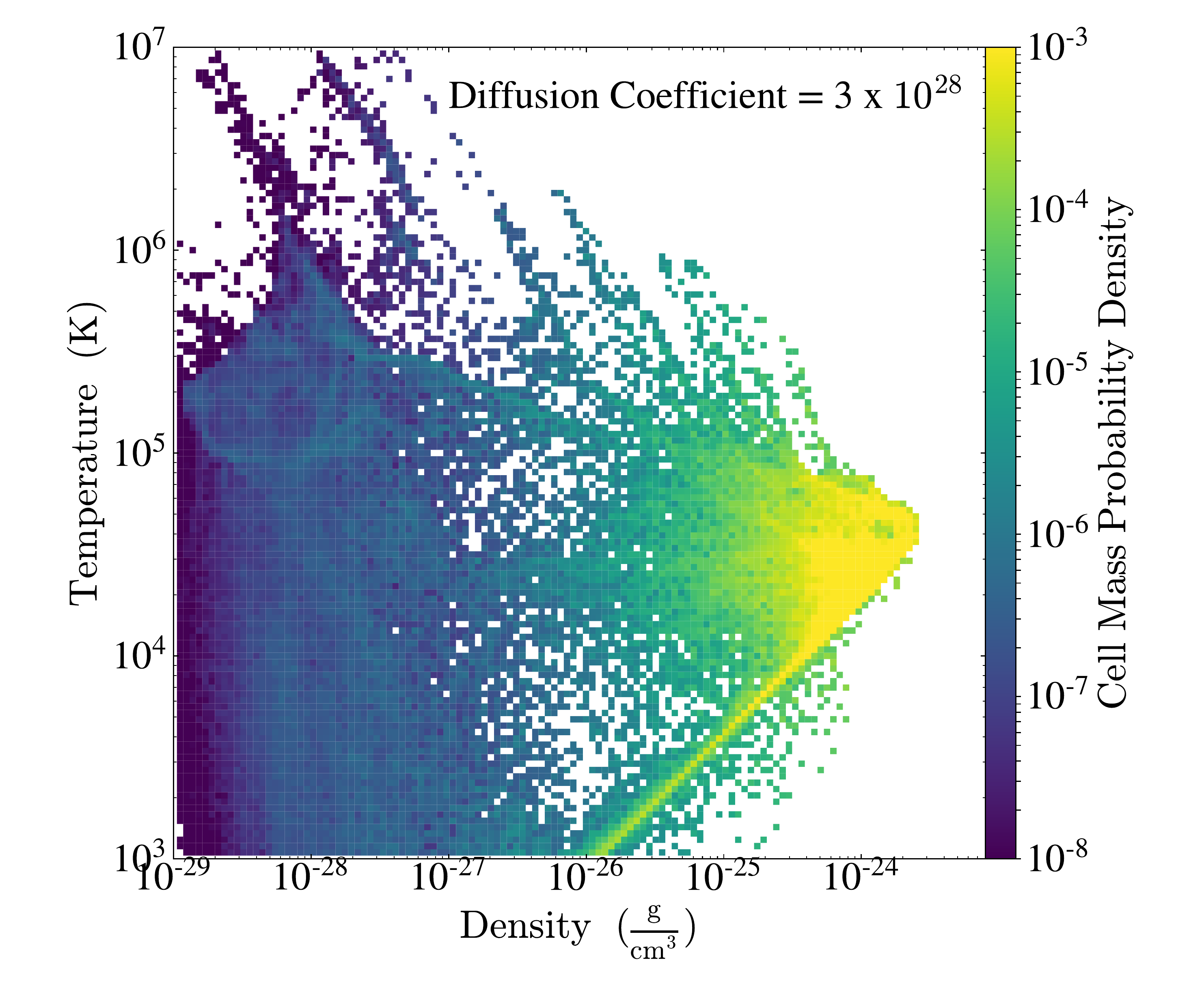}
\includegraphics[width = 0.32\textwidth]{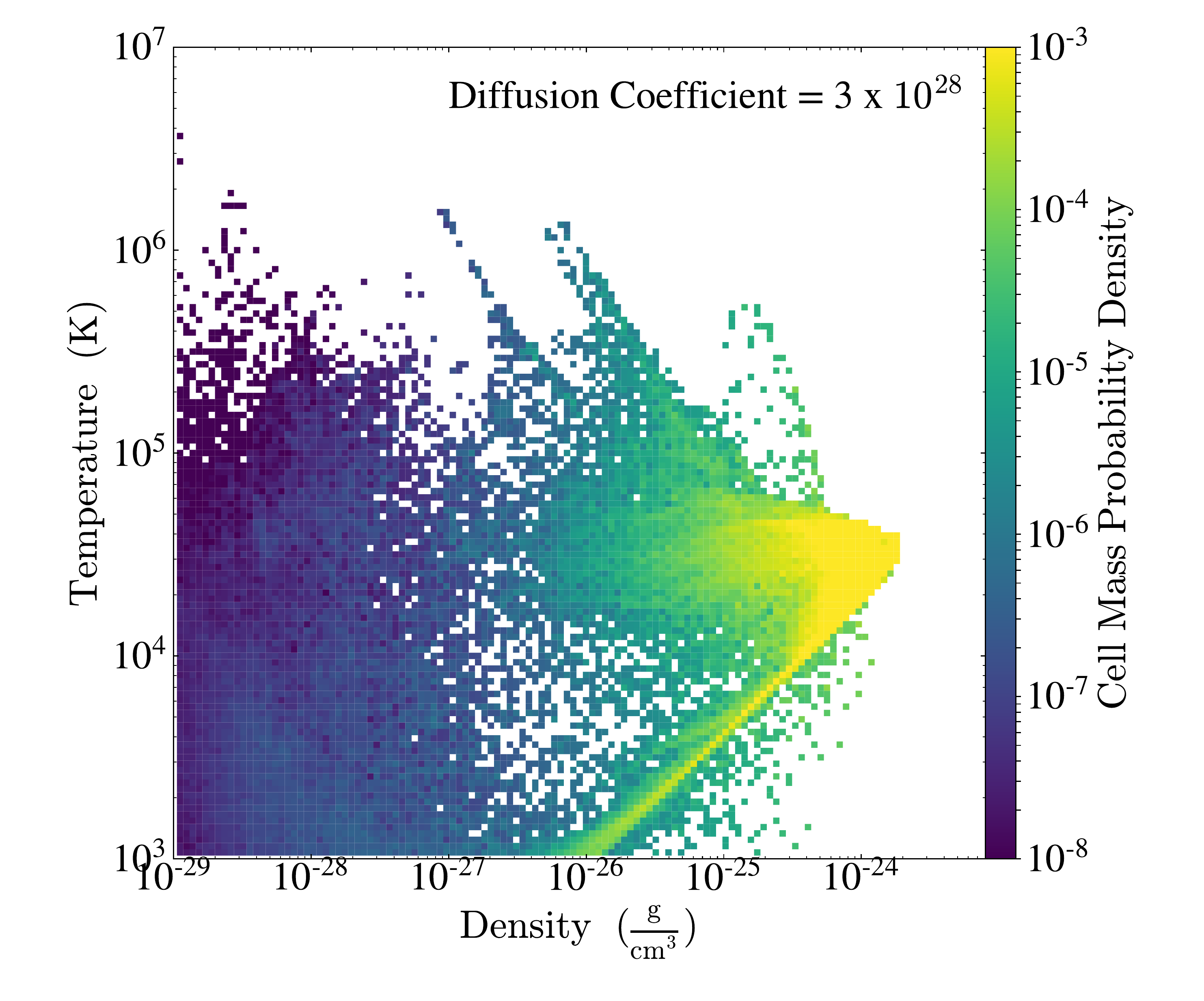}
\includegraphics[width = 0.32\textwidth]{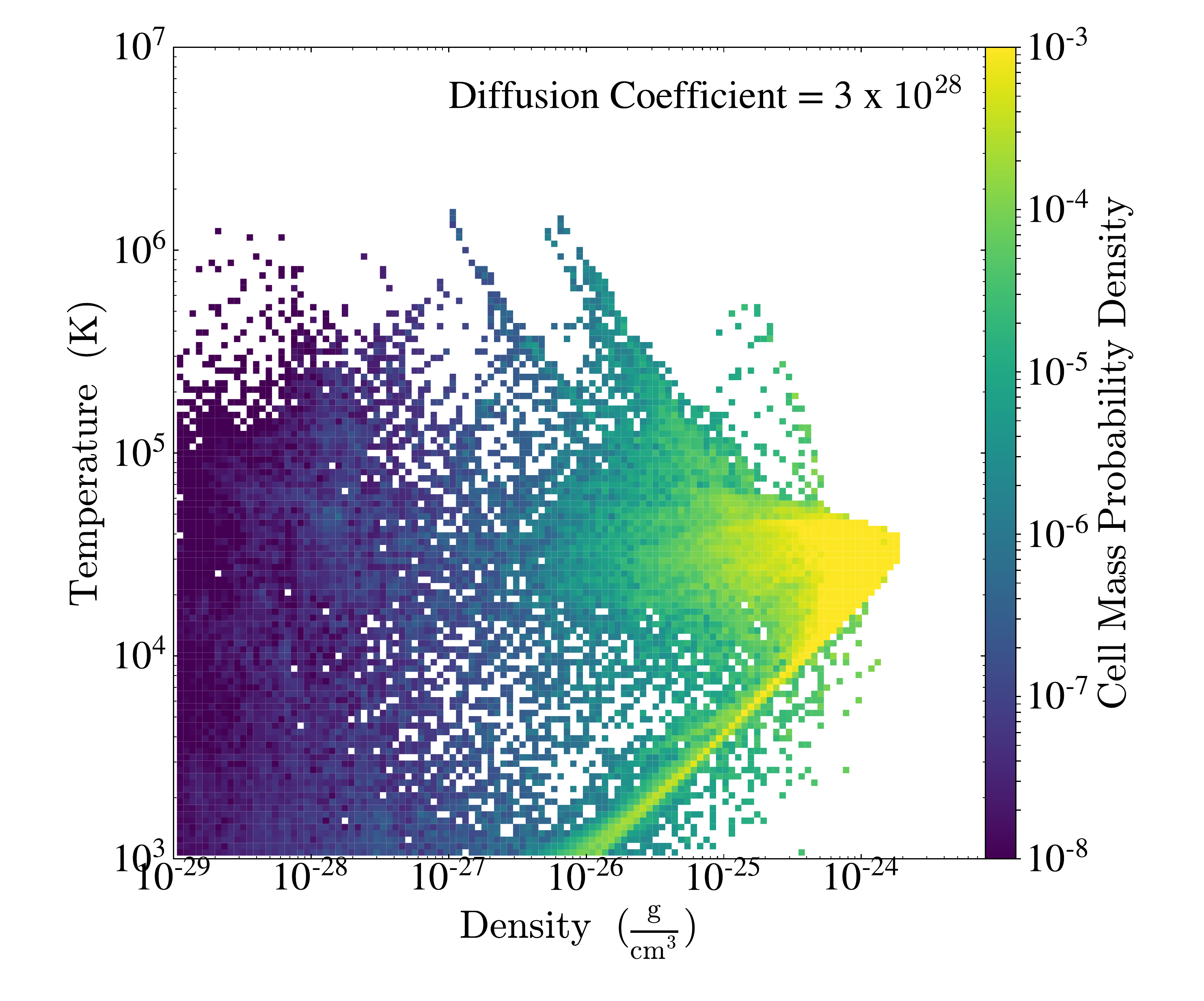}
\caption{Phase plots for our no cooling, diffusion coefficient $= 3 \times 10^{28}$ simulations run with a CFL number of 0.6 (left panel), fiducial CFL number of 0.2 (middle panel), and a CFL number of 0.2 but with a timestep further capped at $10^{10}$ s (right panel). The numerically unstable left panel shows much more high temperature gas owing to reconnection heating, while the middle and right panels are almost indistinguishable, as reconnection heating is suppressed by the smaller timesteps.}
\label{fig:phaseplots_cfl}
\end{figure}

\begin{figure}[t]
\centering
\includegraphics[width = 0.32\textwidth]{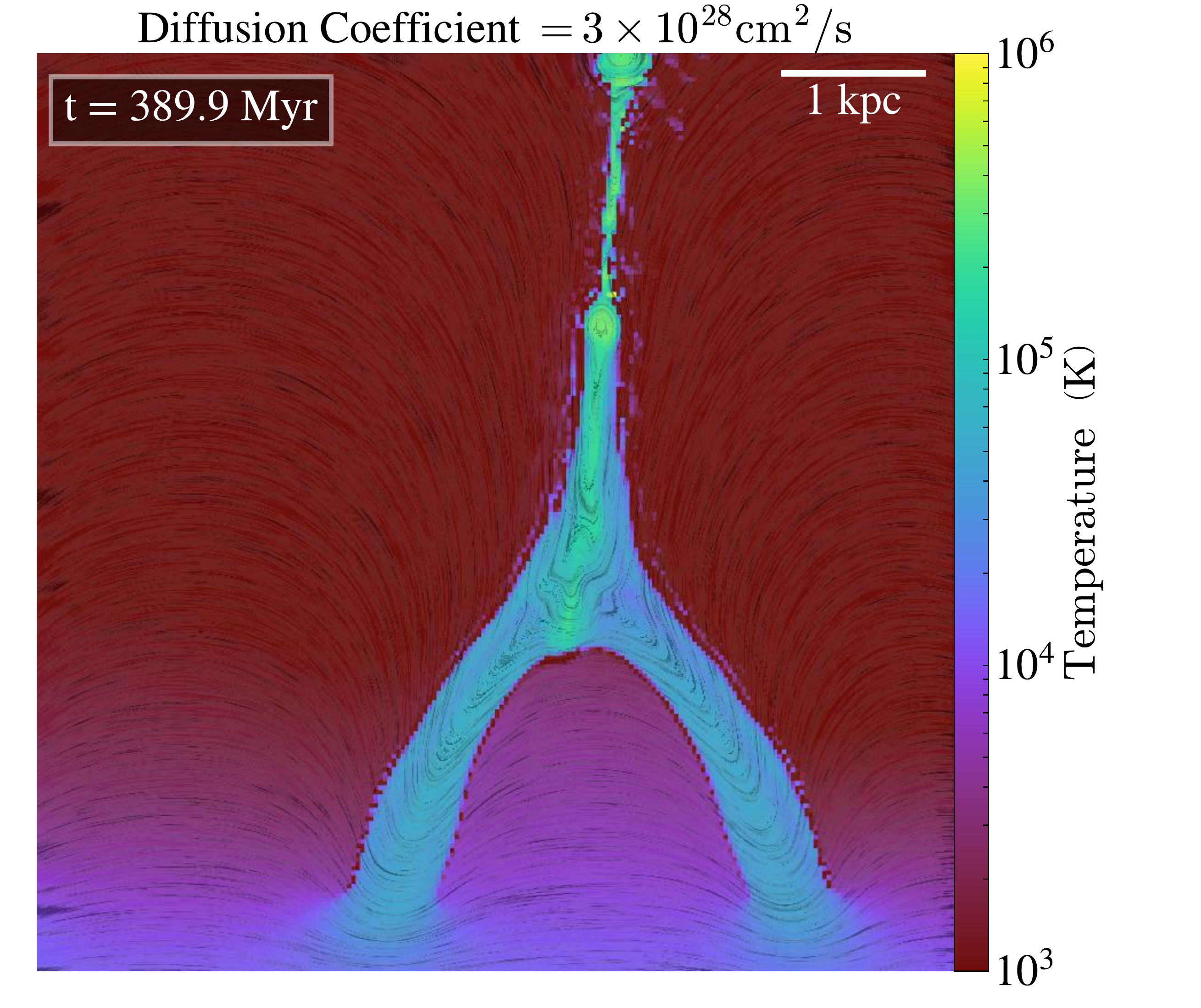}
\includegraphics[width = 0.32\textwidth]{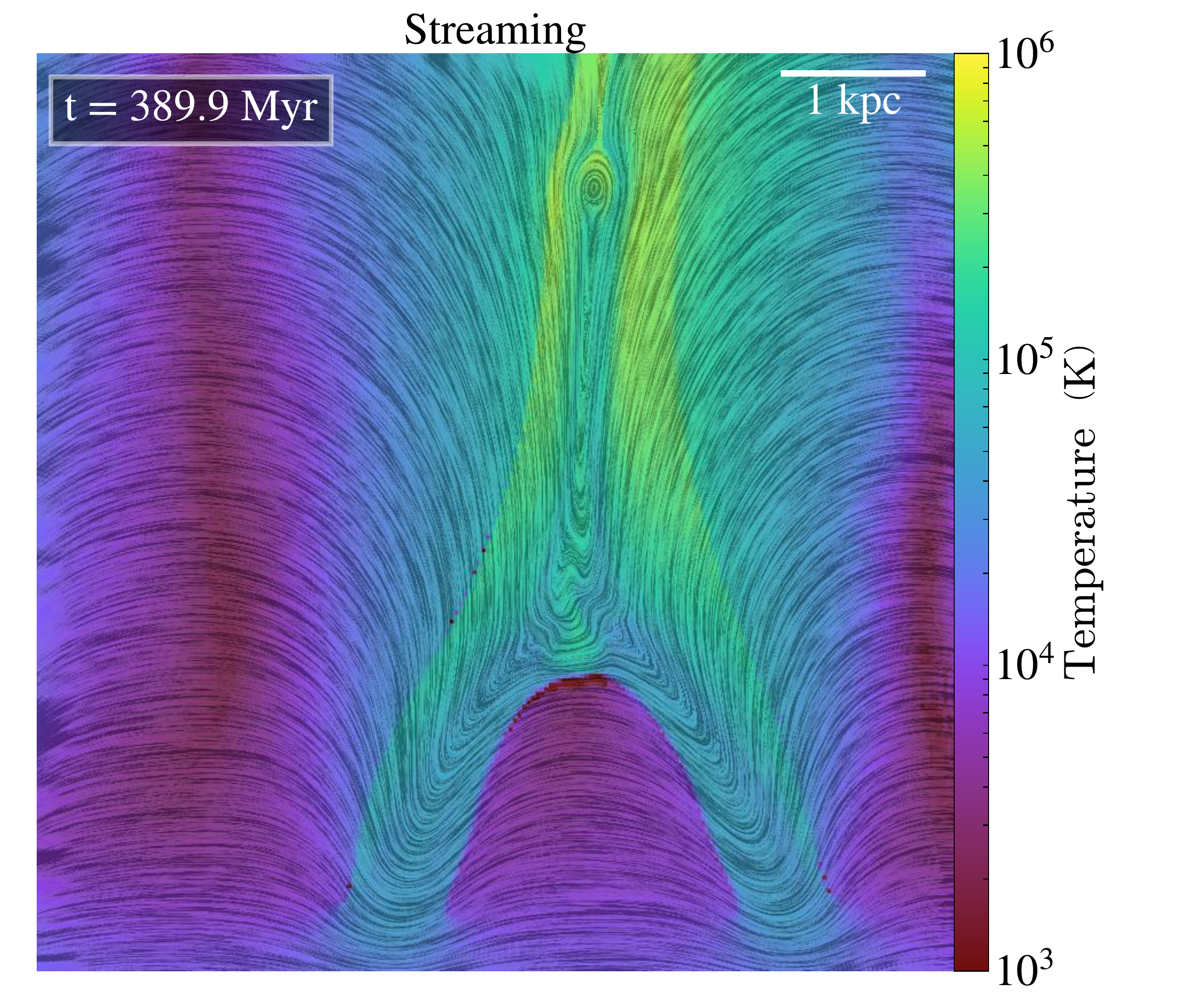}
\caption{These plots show a temperature zoom-in of two merging loops, where numerically-driven reconnection leads to a localized increase in gas pressure and the formation of two magnetic islands. Left panel: diffusion. Right panel: streaming. The formation of magnetic islands occurs in similar places at similar times. One can also see that the temperature increase is localized to the magnetic island, whereas cosmic ray heating due to streaming increases the gas temperature throughout the diffuse, extraplanar gas.}
\label{fig:reconnectionplots}
\end{figure}

To test that our simulations are well-converged and do not suffer from significant reconnective heating (compared to cosmic ray heating due to streaming), we re-ran our diffusion (coefficient = $3 \times 10^{28} cm^{2}/s$) and streaming simulations with a timestep capped at $10^{10}$ s, which is an order of magnitude below the timestep we observed during the nonlinear evolution of our diffusion simulations. Figure \ref{fig:reconnectionplots} shows a zoom-in of two merging loops and a break in magnetic topology, resulting in the formation of two magnetic islands and a localized increase in gas pressure. The two phase plots in Figure \ref{fig:phaseplots_cfl}  are for our fiducial diffusion simulation (right) and the same simulation with a capped timestep (left), showing no significant differences. We also observe a very similar evolution of total thermal energy in each simulation box and very similar plots of gas pressure, temperature, etc. at these two timesteps. Therefore, we consider these simulations to be well-converged, and most importantly, the gas phase is still starkly different from the streaming case, dominantly owing to collision-less heating due to streaming.

\section{Constant Gravity Simulations}
\label{sec:constantgrav}
For the linear stability analysis presented in HZ18, gravity is constant in the vertical direction, i.e. $g(y) = -g_{0} \hat{y}$. Assuming a constant temperature everywhere, the density, gas pressure, cosmic ray pressure, and magnetic pressure all drop exponentially with scale height H to enforce hydrostatic equilibrium. This setup is standard in the Parker instability literature as it makes the equations more tractable to solve analytically. However, it poses a few challenges when trying to compare simulations to analytic results.

It results in a discontinuity at the midplane (where g(y) abruptly changes sign) that complicates numerical studies; it is never possible to fully resolve this sharp transition. This leads to an initial adjustment in the simulation that flattens the exponential profile and sends a steepening sound wave outwards from the midplane. We use outflow or diode boundary conditions in the y-direction far away from the midplane in order to mitigate this effect. These boundary conditions allow shocks to leave the computational domain, as opposed to a reflecting boundary that would more accurately conserve total energy in the simulation box but would reflect this shock back towards our region of interest. However, the discontinuity at the midplane still creates numerical artifacts such as a jump in the values of $m$ and $c$ to slightly larger quantities.
    
Formally, in order to keep $\rho (\delta u)^{2}$ constant, $\delta u \propto e^{y/2H}$. This means the velocity perturbation approaches larger and larger values above the midplane. This seems like it would present an issue because we ideally want the dominant vertical wavelength to be as close to infinite as possible, as $k_{y} = 0$ has been shown to give the maximum growth rate and is easiest to compare to. Therefore, we extend our simulation box to be as tall as possible, while also maintaining a high enough resolution to resolve the scale height by at least a few cells. Because we expect perturbations in the disk itself to drive the instability, we tried simulation perturbations that either drop off as $e^{-|y|/H}$ or that have no vertical dependence at all. We found no difference in the resulting growth rates, and we find that they match the smooth gravity growth rates very well. 
    
Because of the constant gravity setup, the analysis of HZ18, for instance, filters out the midplane warping modes and only allows modes that are symmetric about the midplane. The midplane warping modes, however, typically have faster growth times than the odd modes due to the extra converstion of potential energy to kinetic energy when gas can cross the midplane.
 
\begin{figure}
    \centering
    \includegraphics[width=0.49\textwidth]{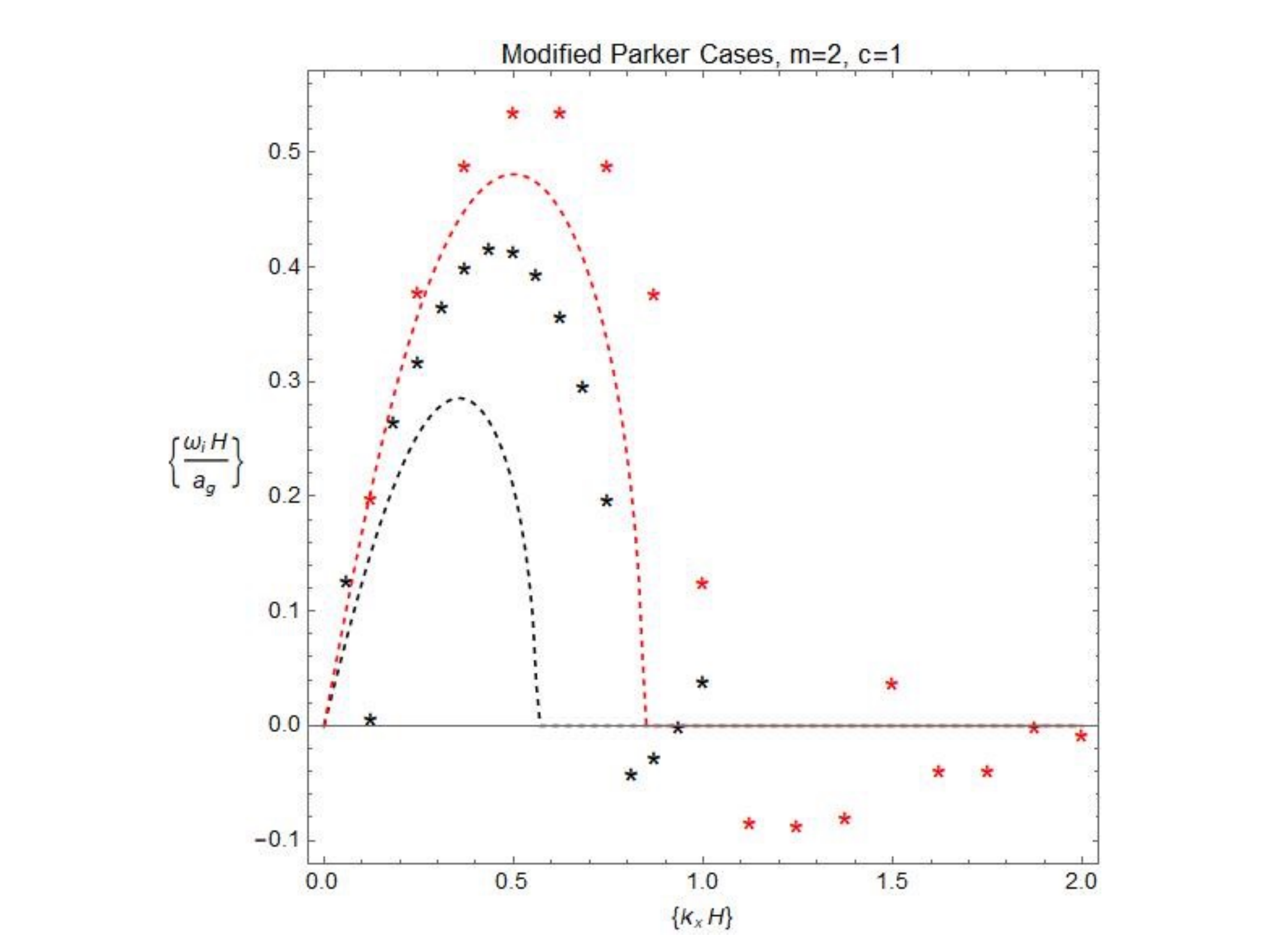}
    \caption{Example of a mismatch between constant gravity simulations (black stars are for Modified Parker, red stars are for Modified Parker with cooling) compared to the linear stability analysis (dashed curves).}
    \label{fig:constantGravFigure}
\end{figure}

We will show an example of our results using a full domain, with gravity instantaneously switching from positive to negative at $y = 0$. This setup allows for midplane warping modes, whereas a similar simulation cut at the midplane keeps the midplane untouched. A constant temperature of 8000 K, m = 2, c = 1, and gravitational force of $g_{0} = 3 \times 10^{-9}$ N are used in this case. We perturb the initial state using both the ``leading the horse to water" and ``horse race" types of perturbations, and we come to a similar conclusion: it is very hard to match linear theory in the constant gravity case. In almost every case we tried (varying m and c values, Modified Parker vs streaming, etc.), our simulated growth rates were \emph{higher} than that predicted by the linear stability analysis, as shown in Figure \ref{fig:constantGravFigure}. We believe this is partially due to a mismatch between the assumptions of the linear theory -- whereby the perturbations should be zero at the midplane -- and our simulation setup -- which allows the faster growing midplane crossing mode. This behavior was most obvious for cases with generally slow mode growth, which is hard to match even for our more reliable smooth gravity simulations. 

We also tried cutting the simulation box at the midplane and enforcing the pertubations to go to zero at lower boundary (y = 0); however, we encountered a variety of numerical artifacts caused by gravity being non-zero within one cell of the boundary. We also checked convergence with respect to resolution, box size, perturbation amplitude, etc. and our results consistently over-estimated the linear analysis growth rates, despite in most cases reproducing the general growth curve behavior shown in HZ18. We conclude, then, that our simulations are picking out the midplane crossing mode, which is not what the linear stability analysis gives us. Therefore, since a constant gravity is unrealistic anyway, we carried on with a smooth gravity profile. If you are a fellow Parker instability simulator, and you've read to the end of this Appendix, we caution you to do the same, or else you may suffer the same fate as these tired authors!

\bibliographystyle{aasjournal}
\bibliography{citations}

\end{document}